\documentclass[11pt]{article}
\usepackage[body={16cm,23cm}]{geometry}

\usepackage{cite}
\usepackage{amsmath,amssymb}
\usepackage{amsfonts}
\usepackage[mathscr]{eucal}
\usepackage{epic}
\usepackage{eepic}
\usepackage{enumerate}
\usepackage{graphicx}

\usepackage[hang,small,bf
]{caption}
\newcounter{app}
\newcounter{sapp}[app]
\newcounter{ssapp}[sapp]
\def\theapp{\Alph{app}}
\newcommand{\app}[1]{
\refstepcounter{app}{\vspace{7mm}
\noindent\Large\bf Appendix
\theapp.
 \ #1 \par \vspace{5mm}}
\setcounter{equation}{0}
\def\theequation{\Alph{app}.\arabic{equation}}}
\def\thesapp{\Alph{app}.\arabic{sapp}}
\newcommand{\sapp}[1]{\par \refstepcounter{sapp}{\vspace{2mm} 
\noindent\large\bf \thesapp
\ #1 \par \vspace{2mm}}
\def\theequation{\Alph{app}.\arabic{equation}}}
\def\thessapp{\Alph{app}.\arabic{sapp}.\arabic{ssapp}}
\newcommand{\ssapp}[1]{\par \refstepcounter{ssapp}{\vspace{2mm} 
\noindent\bf \thessapp
\ #1 \par \vspace{2mm}}
\def\theequation{\Alph{app}.\arabic{equation}}}

\renewcommand\theequation{\thesection.\arabic{equation}}
\def\nsection#1{\setcounter{equation}{0}\section{#1}}

\newtheorem{conj}{Conjecture}
\newcommand{\ds}{\displaystyle}

\newcommand{\R}{\mathbf{R}}

\newcommand{\TT}{\mathbb{T}}

\newcommand{\be}{\begin{equation}}
\newcommand{\ee}{\end{equation}}

\renewcommand{\a}{\alpha}
\renewcommand{\b}{\beta}
\def\evx{\mathsf{Ev}_x}
\def\str{\mathsf{Str}}

\newcommand{\Q}{{\mathbb Q}}
\newcommand{\Qb}{\overline{\mathbb Q}}
\newcommand{\Qs}{{\mathsf Q}}

\newcommand{\p}{{\mathsf p}}
\newcommand{\A}{{\mathbb A}}
\newcommand{\As}{{\mathsf A}}
\newcommand{\Ss}{{\mathbb S}}
\newcommand{\f}{{\mathsf b}}
\newcommand{\ox}{{\,\overline{x}}}
\newcommand{\aq}{{a_q}}
\newcommand{\qbinom}[2]{\genfrac{[}{]}{0pt}{}{#1}{#2}}

\renewcommand{\Re}{\mathrm{Re}}

\newcommand{\YBE}{Yang-Baxter equation}
\def\SE{Schr\"odinger equation}
\def\hf{{\frac{1}{2}}}

\renewcommand{\author}[1]{\large\rm #1\\ \bigskip}
\newcommand{\address}[1]{{\normalsize\it #1\\}\bigskip}
\renewcommand{\title}[1]{\bigskip\bigskip\Large\bf #1\bigskip\bigskip\\}

\begin{document}

\vglue 2cm

\begin{center}

\title{
Baxter's $\bf Q$-operators for supersymmetric spin chains%
}

\vspace{1cm}

\author{        Vladimir V. Bazhanov\footnote[1]{email:
                {\tt Vladimir.Bazhanov@anu.edu.au}} and 
                Zengo Tsuboi\footnote[2]{email: 
                {\tt ztsuboi@yahoo.co.jp, \\
\phantom{mmm}
URL: http://www.pref.okayama.jp/kikaku/kouryoushi/english/kouryoushi.htm}}}

\address{${}^1$ Department of Theoretical Physics,\\
         Research School of Physical Sciences and Engineering,\\
    Australian National University, Canberra, ACT 0200, Australia.\\
\vspace{0.5cm}
         ${}^2$ Okayama Institute for Quantum Physics,\\
                Kyoyama 1-9-1, Okayama 700-0015, Japan.}

\end{center}

\noindent 
Journal reference:  Nucl. Phys. B 805 [FS] (2008) 451-516 \\
DOI: 10.1016/j.nuclphysb.2008.06.025 \\
Report number: OIQP-07-13
\vspace{1cm}

\begin{abstract}
We develop Yang-Baxter integrability structures connected with 
the quantum affine superalgebra $U_q(\widehat{sl}(2|1))$. 
Baxter's {\bf Q}-operators are explicitly constructed as supertraces of certain
monodromy matrices associated with ($q$-deformed) bosonic and
fermionic oscillator algebras. There are six different 
${\bf Q}$-operators in this case, obeying a few 
fundamental fusion relations, which imply all 
functional relations between various commuting transfer matrices.   
The results are universal in the sense that they do not depend on the
quantum space of states and apply both to lattice models and
to continuous quantum field theory models as well.
\end{abstract}

\newpage 
\tableofcontents 
\newpage
\nsection{Introduction}
The method of the ${\bf  Q}$-operator,  introduced by Baxter in his seminal 
paper \cite{Bax72} on the exact solution of the eight-vertex model, finds many
applications in the theory of integrable quantum systems. Its
relationship to the algebraic structure of quantum groups
\cite{Drinfeld,Jimbo} was unraveled in \cite{BLZ97,BLZ99}. This method 
does not require the existence of the ``bare'' vacuum state and
therefore has wider applicability than the traditional approaches to
integrable systems such the coordinate \cite{Bethe} or algebraic Bethe
Ansatz \cite{Sklyanin-Takhtajan-Faddeev}.

Here we consider integrable models of statistical mechanics and quantum
field theory  associated with the 
quantum affine superalgebra $U_q(\widehat{sl}(2|1))$. 
The fundamental $R$-matrix serving these models was found by
Perk and Schultz \cite{Perk:1981},
\be
\renewcommand\arraystretch{1.5}
R(x)=q
 \left(
 \begin{array}{lll}
\phantom{-}q^{-e_{11}}-xq^{e_{11}-\hf} & 
\quad\phantom{-}(q^{-1}-q)q^{-\hf}xe_{21} & \quad (q^{-1}-q)q^{-\hf}xe_{31} \\
\phantom{-}(q^{-1}-q)e_{12} & \quad\phantom{-}q^{-e_{22}}-xq^{e_{22}-\hf} 
& \quad(q^{-1}-q)q^{-\hf}xe_{32} \\
(q^{-1}-q)e_{13} & \quad (q^{-1}-q)e_{23} &
 \quad q^{e_{33}}-xq^{-e_{33}-\hf} 
 \end{array}
 \right),\label{3-stateR}
\ee
where $e_{ij}$ is a $3 \times 3$ matrix whose $(k,l)$
element is  
$\delta_{ik}\delta_{jl}$. 
It defines 
an ``interaction-round-a-vertex''  model on the square lattice with
three different states, $s=1,2,3$, for each lattice edge.   
The states ``$1$'' and ``$2$'' will be referred to as bosonic (even) 
states and the state ``$3$'' as a fermionic (odd) state. 
We will call this model as the 3-state $gl(2|1)$-Perk-Schultz model
or just as the ``3-state model''.
It is worth noting that the paper \cite{Perk:1981} 
contains more general $R$-matrices with an 
arbitrary number of states per edge associated with the 
$U_q(\widehat{sl}(m|n))$ superalgebras. 
We also remark that closely related $R$-matrices for the non-graded 
 case of $U_q(\widehat{sl}(m))$ were previously given by Cherednik
 \cite{Cherednik80}.

The edge configurations of the whole lattice in the
3-state $gl(2|1)$-Perk-Schultz model obey 
simple kinematic constraints, analogous to 
the ``arrow conservation law'' in the ordinary 6-vertex model.
Here we consider the periodic boundary conditions in the horizontal
direction. Then for every allowed edge configuration 
the numbers $m_1,m_2,m_3$, counting the edges
of the types  ``$1$'', ``$2$'' and ``$3$'' in a horizontal row are the
same for all rows of the lattice. 
The row-to-row transfer matrix of the model reduces to a  
block-diagonal form,  where the blocks 
are labeled by these conserved numbers. 
Note that, $m_1 +m_2 +m_3 =L$, where $L$ is the horizontal size of the
lattice. 

The spin chain Hamiltonian connected with the $gl(2|1)$-Perk-Schultz
model describes the trigonometric generalization 
\cite{Perk:1981,Foerster-Karowski,Gonzalez-Ruiz94} of the  
supersymmetric $t$-$J$ model \cite{Sutherland75}. Both models (rational and
trigonometric, and also their multicomponent analogs) 
were studied by many authors (see for example,
\cite{Lai74}-\cite{FWZ07}). Owing to the 
edge-type conservation properties discussed above these models can be
solved via the ``nested'' Bethe Ansatz \cite{Lai74, Sutherland75}.
The problem of the diagonalization of the Hamiltonian is then reduced
to the solution of certain algebraic equations, called the
Bethe Ansatz equations, where 
the number of unknowns depends
edge occupation numbers $m_1,m_2,m_3$.  
It is important to note that the integrability of the 
model is not affected by an introduction of two arbitrary 
``horizontal fields'' (or ``boundary twists'') which requires only simple 
modifications to the transfer matrix and to the Hamiltonian. 
We found this generalization to
be extremely useful. In the following we will always consider the 
non-zero field case.   
 
It was remarked many times 
\cite{EK92,BCFH92,T98,GS03,BKSZ05,BDKM06,KSZ07,GV07,Woynarovich83} 
that there are equivalent, but different, forms of the
Bethe Ansatz in the model.  In fact, it is easy to argue that 
there are precisely $3!=6$ different Bethe {\em Ans\"atze}
in this case. They are related by permutations of the occupation numbers 
$m_1,m_2,m_3$. 
Indeed, there are three ways choose the bare vacuum state 
and then two ways to proceed on the second  ``nested'' stage of the
Bethe Ansatz.  
Of course, the super-symmetry does not play any special role in this
respect. Exactly the same counting also takes place for all 
models related with $U_q(\widehat{sl}(3))$ algebra, see 
\cite{Pronko-Stroganov00, BHK}. It is important to realize, however, 
that the above arguments fully apply only to a generic non-zero field case. 
If the fields are vanishing (or take some special values) then only a few of 
these Bethe Ans\"atze are well defined, while the other suffer 
from the ``beyond the equator'' problem, first encountered in
\cite{Pronko99} for the XXX-model.

The three-state $gl(2|1)$-Perk-Schultz $R$-matrix 
is just one representative of an infinite 
set of $R$-matrices associated with the $U_q(\widehat{sl}(2|1))$ algebra. 
These $R$-matrices can be constructed as  
different specializations of the universal $R$-matrix.
In particular, there are the so-called {\em fusion} \cite{KRS81} 
$R$-matrices related 
to the matrix  
representations of the finite-dimensional algebra $U_q({gl}(2|1))$. 
Other important $R$-matrices connected with
the ($q$-deformed) oscillator algebras and continuous quantum field theory. 
There are two ways this variety could  be used. First, one can 
consider models with different quantum spaces of states. Second, for
the same quantum space there are ``higher'' or ``fusion'' transfer matrices 
corresponding to different finite-dimensional representation of
$U_q({gl}(2|1))$ in the ``auxiliary'' space. 
All these transfer-matrices are operator-valued functions of
a (multiplicative) spectral variable ``$x$''. We will call them the 
${\bf T}$-operators.  
The ${\bf T}$-operators with
different values of $x$ form a commuting family of operators.  
They satisfy a number of important 
functional relations, called the {\em fusion relations} (see
eqs.\eqref{tfr} below). For the case 
of $U_q(\widehat{sl}(2|1))$ related models 
a complete set of 
these relations 
was proposed in \cite{T97,T98,T98-2} 
(see also, \cite{KNS93,Maassarani95,PF96,KSZ07,Zabrodin07,KV07}). 
However, a direct algebraic proof of these relations in full
generality, i.e., for an arbitrary quantum space and a generic value
of the deformation parameter $q$, was not hitherto known.
Here we fill this gap. 

The precise form of the functional relations,
obviously, depends on the normalization of the ${\bf
T}$-operators. Here we use a distinguished normalization determined by the 
universal $R$-matrix (see Eqs. \eqref{R-expan} and \eqref{t-uni2} below). 
The functional relations then take a universal form, which do not
contain various model-dependent scalar factor. Such factors  
are usually present in the transfer matrix relations in lattice
theory. To restore these factors in our approach 
one needs to explicitly calculate the 
specializations of the universal $R$-matrix for particular models. 
Here we compute these factors for the 3-state lattice model and 
in the case of the continuous conformal
field theory, arising in quantization of the AKNS soliton hierarchy
\cite{Fateev-Lukyanov2005}.

An important part in the theory of integrable quantum systems 
is played by the so-called $\bf Q$-operators, introduced by Baxter in
his pioneering work on the eight-vertex model of lattice statistics
\cite{Bax72}. 
The ${\bf Q}$-operators belong to the same commuting family of operators,
as the ${\bf T}$-operators.  
In this paper we present a complete 
algebraic theory of the ${\bf Q}$-operators for the models related
with $U_q(\widehat{sl}(2|1))$ algebra. There are six different  
${\bf Q}$-operators in this case. We denote them as $\Q_i(x)$ and 
$\overline{\Q}_i(x)$, $i=1,2,3$. They are single-valued functions  
in the whole complex plane of the variable of $x$, except the origin 
$x=0$, where they have simple algebraic
branching points,
\be
\Q_k(e^{2\pi i}\, x)=e^{2\pi i\/ \Ss_k}\, \Q_k(x),
\qquad \overline{\Q}_k(e^{2\pi i}\, x)=e^{-2\pi i\/ \Ss_k}\,
\overline{\Q}_k(x),
\qquad k=1,2,3,
\ee
Here $\Ss_1$, $\Ss_2$ and $\Ss_3$, such that 
$\Ss_1+\Ss_2+\Ss_3=0$, are constant operators 
(acting in the quantum space) given by
certain linear combinations of the Cartan generators of 
$U_q(\widehat{sl}(2|1))$ and external field parameters. 
They commute among themselves and with all the other operators in the
commuting family. Their eigenvalues ${\mathsf S}_i$ are
conserved quantum 
numbers, which in the case of the 3-state lattice model reduce to 
the edge occupation $m_1$, $m_2$ and $m_3$, mentioned above (see
Eq.\eqref{s-m} below). 
 
{}From an algebraic point of view the ${\bf Q}$-operators are very similar to
the transfer matrices. They are constructed as traces of certain 
(in general, infinite-dimensional) monodromy matrices,
arising as specializations of the universal $R$-matrix to
the representations of the fermionic and bosonic q-oscillator algebras. 
Using some special decomposition properties of products of these
infinite-dimensional representations we show that the ${\bf Q}$-operators 
satisfies a few fundamental functional relations. 
There are four independent Wronskian-type relations between the ${\bf
  Q}$-operators,      
\be
\renewcommand\arraystretch{2.0}
\begin{array}{rcl}
c_{12} &=&c_{13} \,\Q_1(q^{+\hf} x)\,
 \overline{\Q}_1(q^{-\hf}x)-  
c_{23} \,\Q_2(q^{+\hf} x) \,
\overline{\Q}_2(q^{-\hf}x)\\
c_{12} &=&c_{13} \,\Q_1(q^{-\hf} x)\, \overline{\Q}_1(q^{+\hf}x)-  
c_{23} \,\Q_2(q^{-\hf} x)\,
 \overline{\Q}_2(q^{+\hf}x),\\
 c_{21}\,\Q_{3}(x)&=&\overline{\Q}_{1}(qx)\,\overline{\Q}_{2}(q^{-1}x)-
\overline{\Q}_{1}(q^{-1}x)\,\overline{\Q}_{2}(qx),\\
c_{12}\,\overline{\Q}_{3}(x)&=&\Q_{1}(qx)\,\Q_{2}(q^{-1}x)-
\Q_{1}(q^{-1}x)\,\Q_{2}(qx),
\end{array}
\renewcommand\arraystretch{1.0}\label{Q-rel1}\\[.2cm]
\ee
where 
$c_{ij}=(z_i-z_j)/\sqrt{z_i z_j}$, with $z_1=q^{2\Ss_1}$,
$z_2=q^{2\Ss_2}$ and $z_3=z_1\,z_2=q^{-2\Ss_3}$.  

The fusion transfer matrices are expressed as polynomial combinations 
of the $\Q$-operators. In particular, the fundamental transfer matrix 
(corresponding to the three-dimensional representation in the
auxiliary space) is given by 
\be
c_{12}\,{{\mathbb T}}(x)=
c_{13} \Q_1(q^{\frac{3}{2}}x) \overline{\Q}_1(q^{-\frac{3}{2}}x) 
-c_{23} \Q_2(q^{\frac{3}{2}}x)
 \overline{\Q}_2(q^{-\frac{3}{2}}x)\label{atyp0} 
\ee
With an account of \eqref{Q-rel1} the last formula can be transformed to  
any of the six equivalent forms 
\be
{\mathbb T}(x)=
p_i\,\frac{\overline{\Q}_i(q^{-p_i-\frac{1}{2}}x)}
{\overline{\Q}_i(q^{p_i-\frac{1}{2}}x)}+
p_j\,\frac{\overline{\Q}_i(q^{p_i+2p_j-\frac{1}{2}}x)}
{\overline{\Q}_i(q^{p_i-\frac{1}{2}}x)}
\frac{\Q_k(q^{p_i-p_j-\frac{1}{2}}x)}
{\Q_k(q^{p_i+p_j-\frac{1}{2}}x)}
+p_k\,\frac{\Q_k(q^{p_i+p_j+2p_k-\frac{1}{2}}x)}
{\Q_k(q^{p_i+p_j-\frac{1}{2}}x)}, \label{six}
\ee
where $p_1=p_2=-p_3=1$ and $(i,j,k)$ is any permutation of $(1,2,3)$,
which are the standard Bethe Ansatz type expressions for the
transfer matrix. 
All the above functional relations are written in the normalization
of the universal $R$-matrix (used for both ${\bf T}$ and ${\bf Q}$
operators). They can be easily adjusted for the traditional 
normalization in the lattice theory, where the corresponding eigenvalues  
${\mathsf T}(x)$ and ${\mathsf Q}_i(x)$ and $\overline{\mathsf Q}_i(x)$ 
become finite polynomials of $x$. 
For instance, for the 3-state lattice model 
\begin{eqnarray}
&& {\mathsf T}(x)=
p_i\,{\mathsf f}(q^{2p_{i}-\frac{1}{2}}x)\,
\frac{\overline{\Qs}_i(q^{-p_i-\frac{1}{2}}x)}
{\overline{\Qs}_i(q^{p_i-\frac{1}{2}}x)}+
p_j\,{\mathsf f}(q^{-\frac{1}{2}}x)\,
\frac{\overline{\Qs}_i(q^{p_i+2p_j-\frac{1}{2}}x)}
{\overline{\Qs}_i(q^{p_i-\frac{1}{2}}x)}
\frac{\Qs_k(q^{p_i-p_j-\frac{1}{2}}x)}
{\Qs_k(q^{p_i+p_j-\frac{1}{2}}x)}
\nonumber \\ 
&& \hspace{40pt} +p_k\,{\mathsf f}(q^{-\frac{1}{2}}x)\,
\frac{\Qs_k(q^{p_i+p_j+2p_k-\frac{1}{2}}x)}
{\Qs_k(q^{p_i+p_j-\frac{1}{2}}x)}\label{Teig}
\end{eqnarray}
where ${\mathsf f}(x)=(1-x)^L$ and 
\be
{\mathsf Q}_i(x)
=x^{{\mathsf S}_i}\prod_{\ell=1}^{{m}_i}
(1-x/x^{(i)}_\ell) ,
\qquad \overline{{\mathsf Q}}_i(x)=x^{-{\mathsf S}_i}\prod_{\ell=1}^{L-{m}_i}
(1-x/{\overline x}^{(i)}_\ell).  \label{q-factor}
\ee
The  zeroes $\{x^{(i)}_\ell\}$, $\ell=1,2,\ldots,m_i$ and
$\{\overline{x}^{(i)}_\ell\}$, $\ell=1,2,\ldots,L-m_i$,\ $i=1,2,3$,
satisfy the Bethe Ansatz equations,
\begin{eqnarray}
-\frac{p_i}{p_j}\frac{{\mathsf f}(q^{+p_i}\,\overline{x}_\ell^{(i)})}
{{\mathsf f}(q^{-p_i}\,\overline{x}_\ell^{(i)})}&=&
\frac{\overline{\Qs}_i(q^{+2p_j}\overline{x}^{(i)}_\ell)}
{\overline{\Qs}_i(q^{-2p_i}\overline{x}^{(i)}_\ell )}
\frac{\Qs_k(q^{-p_j}\overline{x}^{(i)}_\ell)}
{\Qs_k(q^{+p_j}\overline{x}^{(i)}_\ell)},\qquad
\ell=1,2,\ldots,L-m_i \nonumber\\[.3cm]
-\frac{p_j}{p_k}&=&
\frac{\overline{\Qs}_i(q^{-p_j}x^{(k)}_\ell)}
{\overline{\Qs}_i(q^{+p_j}x^{(k)}_\ell)}
\frac{\Qs_k(q^{+2p_k}x^{(k)}_\ell)}
{\Qs_k(q^{-2p_j}x^{(k)}_\ell)}
,\qquad
\ell=1,2,\ldots,m_k \label{BAijk}
\end{eqnarray}
where, as before, $(i,j,k)$ denotes an arbitrary permutation of
$(1,2,3)$. Further, the eigenvalues of $\mathsf{S}_1$,  $\mathsf{S}_2$
and $\mathsf{S}_3$, entering the  exponents in \eqref{q-factor}, are given by 
\be
2\,{\mathsf S}_1=L-{m}_1+\f_1,\qquad
2\,{\mathsf S}_2=L-{m}_2+\f_2\qquad 2\,{\mathsf S}_3=
-L-{m}_3-\f_1-\f_2\ , \label{s-m}
\ee
where $m_1$ and $m_2$ and $m_3=L-m_1-m_2$ are the edge occupation
numbers, and $\f_{1}$ and $\f_{2}$ are arbitrary field parameters. 

Eqs.\eqref{BAijk} provide six 
self-contained sets of the Bethe Ansatz 
equations only involving zeros, belonging to
a pair of the eigenvalues $(\As_i(x),
\overline{\As}_j(x))$, \ $i\not=j$.  
Once any such pair is determined, the remaining zeroes can 
be found from the functional equations

The organization of the paper is as follows. The algebraic definitions
of the $R$-matrices, transfer matrices and ${\bf Q}$-operators are
given in Section~\ref{defin}. This section also contains necessary
information about the representation theory of $U_q(\widehat{sl}(2|1))$.
The functional relations are presented in
Section~\ref{funcrel}. 
Their applications in continuous quantum field theory and their
connections to the spectral theory of ordinary differential equations
are considered in Section~\ref{cqft}. A direct algebraic
proof of the functional relations is given in Section~\ref{proof}. 
Technical details of calculations are removed to four Appendices. 

Some of our results in Sect.~\ref{funcrel} partially overlap  with 
those in \cite{BDKM06} devoted to 
some models in the rational case $q=1$.
Our approach to the ${\bf Q}$-operators is different from that of
\cite{BDKM06}; in particular, it is applicable for an 
arbitrary quantum space and to generic values of $q$.

\nsection{Yang-Baxter equation, transfer matrices 
and ${\bf Q}$-operators}\label{defin}
\subsection{The universal $R$-matrix}
The quantum affine algebra ${\mathcal A}=U_q(\widehat{sl}(2|1))$
\cite{Y99} (see also \cite{KT94}) is 
generated by the elements $h_0, h_1, h_2$, $e_0, e_1, e_2$ and 
$f_0, f_1, f_2$, which are of two types: ``fermionic'' and ``bosonic''.
The elements $e_0,e_2,f_0,f_2$ are 
fermionic, while all the other generating elements are bosonic. 
It is convenient to assign the parity 
\be
\p(X)=\left\{\begin{array}{l}
1,\quad X=e_0,e_2,f_0,f_2,\cr
0, \quad X=e_1,f_1,h_0,h_1,h_2,\cr 
\end{array}\right.
\ee 
such that 
\be
\p(XY)=\p(X)+\p(Y) \pmod{2},\qquad X,Y\in {\mathcal A}, 
\ee 
and introduce the generalized commutator 
\be
[X,Y]_q=X Y-(-1)^{\p(X)\p(Y)}\,q\, YX. \label{gcom} 
\ee
Note, in particular, that $[X,Y]\equiv [X,Y]_1$ is reduced to the
ordinary commutator when at least one of the elements $X,Y$ is even 
and to the anti-commutator when both of them are odd. 
The algebra $U_q(\widehat{sl}(2|1))$ is defined by the following 
commutation relations
\be
[h_i,h_j]=0,\quad [h_i,e_j]=a_{ij}e_j,\quad [h_i,f_j]=-a_{ij}f_j,\quad
[e_i,f_j]=\delta_{ij}\,\frac{q^{h_i}-q^{-h_i}}{q-q^{-1}}\ , \label{acom}
\ee
where $i,j=0,1,2$, the Serre relations
\be
e_j^2=f_j^2=[e_{1},[e_{1},e_{j}]_{q^{-1}}]_{q}
=[f_{1},[f_{1},f_{j}]_{q^{-1}}]_{q}=0,  
\quad 
j=0,2,\label {aser1}
\ee
and the extra Serre relations 
\be
[e_{0},\,[e_{2},\,[e_{0},\,[e_{2},\,e_{1}]_{q^{-1}}]]]_{q}=
[e_{2},\,[e_{0},\,[e_{2},\,[e_{0},\,e_{1}]_{q^{-1}}]]]_{q},\label{aser2}
\ee
\be
[f_{0},\,[f_{2},\,[f_{0},\,[f_{2},\,f_{1}]_{q^{-1}}]]]_{q}=
[f_{2},\,[f_{0},\,[f_{2},\,[f_{0},\,f_{1}]_{q^{-1}}]]]_{q}\ .\label{aser3}
\ee
As usual, $(a_{ij})$ denotes the Cartan matrix
\begin{eqnarray}
(a_{ij})_{0 \le i,j \le 2}=
\left(
\begin{array}{ccc}
0 & -1 & 1 \\
-1 & 2 & -1 \\
1 & -1 & 0 
\end{array}
\right), \label{cartan}
\end{eqnarray}
Note, that the sum 
\be
k=h_0+h_1+h_2,
\ee
is a central element, commuting with all other elements of the algebra. 
In this paper, we consider the case $k=0$. 

The algebra ${\mathcal A}=U_q(\widehat{sl}(2|1))$ is a Hopf algebra
with the co-multiplication 
\be
\Delta: \quad {\mathcal A}\longrightarrow 
{\mathcal A}\otimes_s {\mathcal A}
\ee
defined as 
\begin{eqnarray}
&& \Delta(h_{i})=h_{i}\otimes_{s} 1+1\otimes_{s} h_{i}, \nonumber \\
&& \Delta(e_{i})=e_{i}\otimes_{s} 1 +q^{h_{i}} \otimes_{s} e_{i}, 
\label{comul1}\\
&& \Delta(f_{i})=f_{i}\otimes_{s} q^{-h_{i}} +1 \otimes_{s} f_{i},\nonumber
\end{eqnarray}
where $i=0,1,2$ and $\otimes_s$ denotes the graded tensor product, such
that 
\be
(A\otimes_{s}B)(C\otimes_{s}D)=(-1)^{\p(B)\p(C)}\,AC\otimes_{s}BD\label{gprod}
\ee
There is another co-multiplication $\Delta'$ obtained from \eqref{comul1}
by interchanging factors of the direct products,
\be
\Delta'=\sigma\circ \Delta,\qquad \sigma\circ (X\otimes_s Y)=(-1)^{\p(X)\p(Y)}
Y\otimes_s X,\qquad X,Y\in {\mathcal A}.\label{perm}
\ee

The Borel subalgebras ${\mathcal B}_{+}\subset {\mathcal A}$ 
and ${\mathcal B}_{-}\subset {\mathcal A}$ are generated by
$h_{0},h_{1},h_{2}, e_{0},e_{1},e_{2}$ and 
$h_{0},h_{1},h_{2}, f_{0},f_{1},f_{2}$, respectively.
There exists a unique element \cite{Dr85,KT92} 
\be
{\cal R}\in {\cal B}_+\otimes {\cal B}_-\ ,\label{imbed}
\ee
 satisfying the following relations 
\begin{eqnarray}
\Delta'(a)\ {\cal R}&=&{\cal R}\ \Delta(a)
\qquad (\forall\ a\in {\cal A})\, ,\nonumber\\
(\Delta\otimes_s 1)\, 
{\cal R}&=&{\cal R}^{13}\, {\cal R}^{23}\, ,\label{R-def}\\
(1\otimes_s\Delta)\, {\cal R}&=&{\cal R}^{13}\, {\cal R}^{12}\,\nonumber 
\end{eqnarray}
where 
${\cal R}^{12},\, {\cal R}^{13},\, {\cal R}^{23}\in
{\cal A}\otimes_s{\cal A}\otimes_s{\cal A}$ and
${\cal R}^{12}={\cal R}\otimes 1$, ${\cal R}^{23}=1\otimes {\cal R}$,
${\cal R}^{13}=(\sigma\otimes 1)\, {\cal R}^{23}$.
The element ${\cal R}$ is called the universal $R$-matrix. 
It satisfies the \YBE\ 
\be
{\cal R}^{12}{\cal R}^{13}{\cal R}^{23}={\cal R}^{23}{\cal
R}^{13}{\cal R}^{12}\ ,\label{YBE}
\ee
which is a simple corollary of the definitions \eqref{R-def}. 
The universal 
$R$-matrix is understood as a formal series in generators in ${\cal
B}_+ \otimes {\cal B}_-$. Its dependence on the Cartan elements 
can be isolated as a simple factor,
\be
{\cal R}=\overline{{\cal R}}\ q^{\mathcal K},\qquad {\mathcal
  K}=-h_0\otimes h_2 -h_2\otimes h_0,\label{R-red}
\ee
where the ``reduced'' universal $R$-matrix 
\be 
\overline{{\cal R}} = \hbox{series in\ \  } (e_j\otimes 1) \mbox{\ \ and\
  \ }(1 \otimes f_j)\ ,
\label{R-series}
\ee
is a series in 
$(e_j \otimes 1) \in {\cal B}_+\otimes 1$ \ and $(1\otimes f_j)\in
1\otimes{\cal B}_-$,\ \   $j=0,1,2$, \ and does not contain Cartan
elements. Remind that we assume $k=h_{0}+h_{1}+h_{2}=0$.
A few first terms in \eqref{R-series} can be readily calculated 
directly from the definitions \eqref{imbed} and \eqref{R-def}, 
\begin{eqnarray}
&& \overline{{\cal R}}=
1-(q-q^{-1})\sum_{j=0}^{2}(-1)^{p(j)}e_{j}\otimes_{s} f_{j}+
\frac{(q-q^{-1})^{2}}{q^{2}+1}(e_{1})^{2}\otimes_{s} (f_{1})^{2}  \nonumber \\
&& \qquad +(q-q^{-1})\sum_{i\ne j}
\left\{
e_{i}e_{j}\otimes_{s}
f_{j}f_{i}-(-1)^{{\mathsf p}(i){\mathsf p}(j)}
q^{-a_{ij}}e_{i}e_{j}\otimes_{s} f_{i}f_{j}   
\right\} +\cdots ,
\label{R-expan}
\end{eqnarray}
The symbol ${\mathsf p}(j)$ denotes the parity of the corresponding
element $e_j$, namely 
\be
{\mathsf p}(0)={\mathsf p}(2)=1,\qquad  {\mathsf p}(1)=0.\label{parity} 
\ee 
The higher terms in \eqref{R-expan} 
soon become very complicated and their
general form is unknown. 
This complexity  should not be surprising, since the universal $R$-matrix 
contains infinitely many nontrivial 
solutions of the \YBE\ associated with 
$U_q\big(\widehat{sl}(2|1)\big)$. 
Fortunately, for applications one only needs certain {\em specializations} of 
universal $R$-matrix, which can be calculated explicitly .  Almost all
these 
specializations are associated with the {\em evaluation homomorphisms}
from the infinite-dimensional algebra $U_q\big(\widehat{sl}(2|1)\big)$
(and from its Borel subalgebras) into finite-dimensional algebras.
The most important case is the {\em evaluation map} to 
the finite-dimensional quantum algebra $U_q\big({gl}(2|1)\big)$. 
This algebra is generated by the elements $E_{ii}$, $i=0,1,2$ and 
$E_{ij}$, $(i,j)=(1,2), (2,1), (2,3), (3,2)$, for which we also use
the notations 
\be
E_\alpha=E_{12},\quad 
E_\beta=E_{23},\quad 
F_\alpha=E_{21},\quad 
F_\beta=E_{32},\quad 
\ee
and 
\be
H_\alpha=E_{11}-E_{22},\qquad H_\beta=E_{22}+E_{33}, \qquad
H_{\alpha+\beta}=E_{11}+E_{33}.
\ee
The elements $E_\b$ and $F_\b$ are odd, ${\mathsf p}(E_\b)={\mathsf
  p}(F_\b)=1$, all other  
generators are even.
They satisfy the following relations (written with the generalized commutator
\eqref{gcom})
\be
\renewcommand\arraystretch{2.5}
\begin{array}{c}
[E_{ii},E_{jj}]=0,\quad [E_{ii},E_{kl}]=(\delta_{ik}-\delta_{il}) E_{kl},
\quad
[E_{\a_i},F_{\a_j}]=\ds
\delta_{\a_i,\a_j}\,{q^{H_{\a_i}}-q^{-H_{\a_i}}\over
q-q^{-1}},\\
E_\b^2=F_\b^2=[E_\a,[E_\a,E_\b]_{q^{-1}}]_q=[F_\a,[F_\a,F_\b]_{q^{-1}}]_q=0,
\end{array}\label{Uqgl21}
\renewcommand\arraystretch{1.0}
\ee
where  the Greek indices $\a_i$ and $\a_j$    
take two values $\a$ or $\b$. 
Introduce the following elements 
\begin{eqnarray}
&& E_{13}=q^{E_{22}+2E_{33}}[E_{12},E_{23}]_{q},\qquad 
E_{31}=[E_{32},E_{21}]_{q^{-1}}\, q^{-E_{22}-2E_{33}}
\\
&&\overline{E}_{13}=q^{-E_{22}-2E_{33}}[E_{12},E_{23}]_{q^{-1}},\qquad 
\overline{E}_{31}=[E_{32},E_{21}]_{q}\, q^{E_{22}+2E_{33}}
\end{eqnarray}

Let $x$ be a complex (spectral) parameter. Define the evaluation map
$$ \mathsf{Ev}_x: \qquad U_q(\widehat{sl}(2|1))\longrightarrow
U_q({gl}(2|1))$$ as follows,
\begin{eqnarray}
&&\evx(h_{0})=-E_{11} -E_{33}, \quad \evx(h_{1})=E_{11}-E_{22},
 \quad \evx(h_{2})=E_{22}+E_{33},\nonumber \\
&& \evx(e_{0})=-x\,E_{31}, \quad \evx(e_{1})=E_{12}, \quad
 \evx(e_{2})=E_{23}, \label{eval}\\ 
&& \evx(f_{0})=x^{-1}\,E_{13}, \quad \evx(f_{1})=E_{21}, \quad 
\evx(f_{2})=E_{32}. \nonumber
\end{eqnarray}
One can check that this map is an algebra homomorphism as all
the defining relations \eqref{acom}-\eqref{aser3} becomes 
corollaries  of \eqref{Uqgl21}\footnote{%
Note there is another (non-equivalent) 
evaluation map obtained from (\ref{eval}), 
if $E_{31}$ and $E_{13}$ are replaced with $\overline{E}_{31}$ and 
$\overline{E}_{13}$.}%
.

A brief introduction into the representation theory of $U_q(gl(2|1))$
is given in the Appendix~\ref{appA}. We also summarize some important
facts here. 
Let $\pi_{\mu}$, with $\mu=(\mu_1,\mu_2,\mu_3)$, such that $\mu_1-\mu_2\in
{\mathbb Z}_{\ge0}$, 
denotes the irreducible finite-dimensional 
representation of the $U_q(gl(2|1))$ with the highest weight $\mu$ and the
highest weight vector $|0>$  
defined as
\be
E_{12}\,|0>=E_{23}\,|0>=0,\qquad E_{ii}\,|0>=\mu_i\,|0>,\qquad i=1,2,3.
\ee
Any such representation 
is realized by linear transformations $\mathrm{End}(V)$ of
some graded vector space $V=V_0\oplus V_1$, where ${p}(V_0)=0$
and ${p}(V_1)=1$.  The latter is 
always a subspace of the vector space generated by a free action of the
elements $E_{21}$ and $E_{32}$ on the highest weight vector  
(note that the action of $E_{32}$ changes the
parity, while the action of $E_{21}$ leaves it unchanged).   
There are bases of $V$, called homogeneous, 
where all basis vectors $\{v_{i}\}$ have definite parities, i.e.,  
$v_{i} \in V_{0}$ or $v_{i} \in V_{1}$ for any $v_i$. 
Let $A$ be an arbitrary matrix 
$A \in \mathrm{End}(V)$, and $A_{ij}$ denote its 
matrix elements  in a homogeneous basis $A\,v_{k}=\sum_{j}v_{j}\,A_{jk}$. 
The supertrace of $A$ over $V$ is defined 
as ${\str}_{V}A=\sum_{j}(-1)^{p(v_{j})}A_{jj}$.

 Further, let $\pi_\mu(x)$ be
 the representation of $U_q(\widehat{sl}(2|1))$ obtained by the composition
of $\pi_\mu$ with the evaluation map \eqref{eval},
\be
\pi_\mu(x)=\pi_\mu\circ \evx\ .\label{evmap}
\ee
Then the
finite-dimensional $R$-matrices, obtained from the universal $R$-matrix, 
\be
   {\bf R}_{\mu_1\mu_2}(x_1/x_2)= 
(\pi_{\mu_1}(x_1)\otimes\pi_{\mu_2}(x_2)) [{\mathcal R}]\label{rmumu}
\ee 
satisfy  appropriate specializations of the Yang-Baxter equation
\eqref{YBE},
\be
   {\bf R}_{\mu_1\mu_2}(x_1/x_2)\,
   {\bf R}_{\mu_1\mu_3}(x_1/x_3)\,
   {\bf R}_{\mu_2\mu_3}(x_2/x_3)=
   {\bf R}_{\mu_2\mu_3}(x_2/x_3)\,
    {\bf R}_{\mu_1\mu_3}(x_1/x_3)\,
   {\bf R}_{\mu_1\mu_2}(x_1/x_2)\label{munu-ybe}
\ee
\subsection{${\bf T}$-operators (transfer matrices)}\label{tmatsec}
First, let us review standard definitions of the transfer matrices in
lattice models. 
The $R$-matrix ${\bf R}_{\mu\nu}(x)$, defined in \eqref{rmumu}, 
acts in the graded product of two representation
spaces $V_{\mu}\otimes_s V_{\nu}$.
It is convenient to consider the first of these spaces as an
``auxiliary space'' and the second one as a ``quantum space''.
The transfer matrix for a homogeneous periodic chain of the length
$L$ is defined as follows,  
\be
{\bf T}_{\mu}(x|\nu,y,L)=
{\str}_{\pi_\mu}\,\Big({\bf D}\,\underbrace{
{\bf R}_{\mu\nu}(x/y)\otimes_s {\bf R}_{\mu\nu}(x/y)\otimes_s \cdots \otimes_s 
{\bf R}_{\mu\nu}(x/y)}_{L-\mbox{\scriptsize{times}}} \Big)\ ,\label{tmat}
\ee
where the tensor product is taken with respect to the quantum 
spaces $V_\nu=\pi_\nu(y)$, while the matrix product and the supertrace
is taken with respect to the auxiliary space $V_\mu=\pi_\mu(x)$. 

The boundary
operator ${\bf D}$ reads  
\be
{\bf D}=q^{\f_1 E_{11}+ \f_2 E_{22} +(\f_1+\f_2) E_{33}}=
{\mathcal Ev}_x\,\big[\,q^{-\f_1\, h_0+\f_2 \,h_2}\,\big]\ ,
\label{Dboud-def}
\ee
where $E_{11},E_{22},E_{33}$ are defined in \eqref{Uqgl21} and 
$\f_1,\f_2$ denotes two arbitrary horizontal field parameters.   
The transfer matrix \eqref{tmat} acts in a Hilbert space  
\be
{\bf T}_{\mu}(x|\nu,y,L):
\qquad {\mathcal H}^{(\nu)}\to {\mathcal H}^{(\nu)},\qquad
{\mathcal H}^{(\nu)}= \underbrace{
V_{\nu}\otimes_s V_{\nu}\otimes_s \cdots \otimes_s V_{\nu}
}_{L-\mbox{\scriptsize{times}}}\ ,
\ee
which is the graded product of $L$ copies of the space $V_\nu=\pi_\nu(y)$. 

The symbols $\mu$ and $\nu$ in the 
notation for the transfer matrix $\TT_{\mu}(x|\nu,y,L)$, obviously,
refer to the auxiliary and quantum spaces respectively. 
For the same quantum space ${\mathcal H}^{(\nu)}$ there is an infinite
number of different 
transfer matrices, corresponding to different choices of the
representation $\mu$ in the auxiliary space.
The \YBE\ \eqref{munu-ybe} implies that these matrices form a commuting
family 
\be
[{\bf T}_{\mu_1}(x_1|\nu,y,L), 
{\bf T}_{\mu_2}(x_2|\nu,y,L)]=0, \qquad \mbox{for
  all} \quad \mu_1,\mu_2, x_1, x_2.\label {tcomm}
\ee
Note that due to the invariance property of the $R$-matrix (which 
trivially follows from  \eqref{comul1} and the first relation in
\eqref{R-def}) 
\be
({\bf D}_\mu\otimes_s {\bf D}_\nu)\, \R_{\mu\nu}(x)=
\R_{\mu\nu}(x) \, ({\bf D}_\mu\otimes_s {\bf D}_\nu),\qquad {\bf
  D}_\mu=
\pi_\mu\big[{\bf D}\big] , 
\ee
the commutativity \eqref{tcomm} is not
affected by the presence of non-zero fields in definition 
\eqref{tmat}. 

Below we will derive algebraic relation between different 
transfer matrices, using decomposition properties of 
products of representations of the quantum affine superalgebra
$U_q(\widehat{sl}(2|1))$ in the auxiliary space. We would like to
stress that our results are independent on the quantum space of the model. 
To facilitate these considerations it is useful to make a model-independent
definition of the transfer-matrices, 
\be
{\mathbb T}_\mu(x)=\big({\str}_{\pi_\mu(x)}\otimes 1\big) \,\Big[
 (q^{-\f_1 h_0+\f_2 h_2}\otimes 1 ) {\mathcal R}\Big]\label{t-uni1} 
\ee
where ${\cal R}$ is the universal $R$-matrix, and $\f_{1}$ and $\f_{2}$ are the
external field parameters.  This formula defines a ``universal'' ${\bf
  T}$-operator which is an element of the Borel subalgebra
${\mathcal B}_-$ associated with the quantum space.
To specialize it for a particular model one needs to choose an
appropriate representation of $\mathcal{B}_-$. For example, choosing
the latter to be the product $\pi_\nu(y)\otimes_s \pi_\nu(y)\otimes_s 
\cdots \otimes_s\pi_\nu(y)$, 
where $\pi_\nu(y)$ is defined by \eqref{evmap},
one immediately obtains\footnote{%
To evaluate the RHS of \eqref{tmat2} one 
needs to repeatedly use the third equation in \eqref{R-def}.} 
the lattice transfer
matrix \eqref{tmat}
\be
{\bf T}_\mu(x|\nu,y,L)=\big(
\underbrace{\pi_\nu(y)\otimes_s \pi_\nu(y)\otimes_s 
\cdots \otimes_s\pi_\nu(y)}_{L-\mbox{\scriptsize{times}}}
\big)\,\Big[{\mathbb T}_\mu(x)\Big]\ .\label{tmat2}
\ee
Another important example of the specialization of \eqref{t-uni1},
related with the continuous super-conformal field theory associated
with $U_q(\widehat{sl}(2|1))$ algebra, is considered in Section~\ref{cqft}. 

It is convenient to define new operators 
\be
z_1=q^{2\Ss_1}=q^{h_2+\f_1},\qquad z_2=q^{2\Ss_2}=q^{-h_0+\f_2},\qquad
z_3=q^{-2\Ss_3}=z_1z_2, \label{zs-def}
\ee
which are elements of the same Borel subalgebra ${\cal B}_-$. 
The definition \eqref{t-uni1} can be then rewritten as
\be
{\mathbb T}_\mu(x)=\big({\str}_{\pi_\mu(x)}\otimes 1\big)\Big[
 (1\otimes z_1)^{-(h_0\otimes1)}\,(1\otimes z_2)^{(h_2\otimes1)}\  
\overline{\mathcal R}\Big]\label{t-uni2} 
\ee
where $\overline{\cal R}$ is now the reduced universal $R$-matrix from 
\eqref{R-red}. The last formula looks a bit cumbersome, so in the
following we will use shorthand notations and simply write 
\be
{\mathbb T}_\mu(x)={\str}_{\pi_\mu(x)}
\Big[z_1^{-h_0}\,z_2^{h_2}\,\overline
{\mathcal R}\Big]\label{t-uni3} 
\ee
but assume the same meaning as in \eqref{t-uni2}.

Now define special notations for the most important ${\bf
  T}$-operators \eqref{t-uni3}, associated
with rectangular Young diagrams. We denote them as ${\mathbb
  T}_m^{(a)}(x)$, where $m \ge0$ is the length and $a\ge1$ is the 
height of the corresponding Young diagram, namely, 
\begin{itemize}
\item[(i)]
the operator ${\mathbb T}^{(1)}_m(x)$ 
corresponding to the $(2m+1)$-dimensional {\em class-2 atypical} 
representations with 
the highest weight 
$\mu=(m,0,0)$, 
\be
\TT^{(1)}_m(x)\equiv\TT_{(m,0,0)}(x),\qquad m\in{\mathbb Z}_{\ge0}.
\label{tm}
\ee 
\item[(ii)] the operator 
$\overline{\mathbb T}^{(1)}_m(x)$ corresponding to the 
$(2m+1)$-dimensional {\em
class-1 atypical} representations
 with the highest weight 
$\mu=(-1,-m,0)$,  
\be
{\overline \TT}^{(1)}_m(x)\equiv \TT^{(1)}_{-m-1}(x)
\equiv - \TT_{(-1,-m,0)}(x), \qquad m\in{\mathbb Z}_{\ge0}\ .
\label{tbarm}
\ee 
\item[(iii)]
special notations to the operators $\TT(x)$ and $\overline{\TT}(x)$
corresponding to  
the 3-dimensional representations ($m=0$ case in (i) and (ii) above)
\be
\TT(x)\equiv\TT^{(1)}_1(x)\equiv\TT_{(1,0,0)}(x),\qquad 
\overline{\TT}(x) \equiv {\overline \TT}^{(1)}_{1}(x) 
\equiv \TT^{(1)}_{-2}(x) \equiv -\TT_{(-1,-1,0)}(x)\label{3dim}
\ee
\item[(iv)] the operator ${\mathbb T}^{(2)}_c(x)$ corresponds to the
  4-dimensional {\em typical} representation ${\pi}_{\mu}(x)$
  with the highest weight $\mu=(c,c,0)$, where the parameter
  $c\in{\mathbb C}$ is not necessarily an integer   
\be
{\mathbb T}^{(2)}_c(x)=\TT_{(c,c,0)}(xq^{c+1}), \qquad
c\in{\mathbb C}\ .
\label{t2}
\ee
It is convenient to define also 
\be
\overline{{\mathbb T}}^{(2)}_c(x)={\mathbb T}^{(2)}_{-c-1}(x)\label{t2bar}
\ee
\end{itemize}
For $c=0$ or $c=-1$, the representation ${\pi}_{(c,c,0)}(x)$ 
becomes reducible (but still indecomposable). As a result one obtains
\be
{\mathbb T}^{(2)}_{0}(x)=\overline{{\mathbb T}}^{(2)}_{-1}(x)=
1-\TT^{(1)}_{1}(x),\qquad
\overline{{\mathbb T}}^{(2)}_{0}(x)={\mathbb T}^{(2)}_{-1}(x)
=1-{\overline \TT}^{(1)}_{1}(x).  \label{dec-typical}
\ee
Note  also that the case $m=0$ in \eqref{tm} and \eqref{tbarm} corresponds to
trivial one-dimensional representations, so that 
\be
\TT^{(1)}_0(x)\equiv1,\qquad  
\overline{\TT}^{(1)}_0(x)\equiv1.\label{tnot}
\ee

\subsection{${\bf Q}$-operators}\label{qopsec}
An important part in the theory of integrable quantum systems 
is played by the so-called ${\bf Q}$-operators, introduced by Baxter in
his pioneering work on the eight-vertex model of lattice statistics
\cite{Bax72}. From the algebraic point of view these operators are
not much different from the ${\bf T}$-operators. They are also
constructed as supertraces of certain (in general, infinite-dimensional) 
monodromy matrices. 
This is essentially the idea of Baxter's original work \cite{Bax72},
which has been further developed in \cite{BLZ97,BLZ99}.
In order to construct the ${\bf Q}$-operators this way one needs 
to find alternatives to the evaluation map \eqref{eval}. 
Each side of the \YBE\ \eqref{YBE} is
an element of ${\mathcal B}_+\otimes_s{\mathcal A}\otimes_s{\mathcal
  B}_-$. Therefore to obtain a specialization of this relation it is
not necessary to have a realization of the full quantum affine superalgebra in all
three factors of the direct product. For example, in the first factor one
only needs to construct a realization of the Borel subalgebra
${\mathcal B}_+$. The evaluation map \eqref{eval} is a particular, but
not the only example of such map. 
Below we construct several maps from ${\mathcal B}_+$ into
the graded direct products \eqref{gprod} of oscillator algebras.

Define the bosonic $q$-oscillator algebra
${\mathsf H}_q$, 
\be
{\mathsf H}_q:\qquad 
[{\mathcal H}^b,b^\pm]=\pm b^\pm,\qquad q b^+ b^- -q^{-1} b^-
b^+=\frac{1}{q-q^{-1}}, \qquad 
[b^{\pm},b^{\pm}]=0 \label{Hq-def}
\ee
and the fermionic oscillator algebra ${\mathsf F}$,
\be
{\mathsf F}:\qquad [{\mathcal H}^f,f^\pm]=\pm f^\pm,\qquad 
f^{+}f^{-} +f^{-}f^{+}=1,\qquad (f^+)^2=(f^-)^2=0,\label{F-def}
\ee
The Fermi operators $f^+$ and $f^-$ are odd, ${\mathsf p}(f^+)={\mathsf
p}(f^-)=1$; all the other generators are even. 

Define the following three maps. The first one is  
\be 
\rho_1(x):\qquad {\mathcal B}_+\to {\mathsf H}_q
\otimes{\mathsf   F}\label{map1}
\ee
where $x$ is the spectral parameter,
\be
\renewcommand\arraystretch{1.1}
\rho_1(x):\qquad\left\{
\begin{array}{lll}
e_{0}\to x f_{2}^{-}, \quad 
&e_{1}\to b_{1}^{+}\,q^{{\mathcal H}_{2}}, 
&e_{2}\to -q^{\frac{1}{2}}\,q^{-{\mathcal H}_{2}}\,b_{1}^{-}\,f_{2}^{+},
\\[0.4 cm] 
h_{0}\to -{\mathcal H}_{1}, 
&h_{1}\to 2{\mathcal H}_{1}+{\mathcal H}_{2}, 
&h_{2}\to -{\mathcal H}_{1}-{\mathcal H}_{2}.
\end{array}\right.
\renewcommand\arraystretch{1.0}
\ee
The second one
\be 
\rho_2(x) :\qquad {\mathcal B}_+\to 
{\mathsf   F}\otimes {\mathsf H}_q\label{map2}
\ee
is given by 
\be
\rho_2(x) :\qquad\left\{
\begin{array}{lll}
e_{0}\to -q^{\frac{1}{2}}xq^{-{\mathcal H}_{2}}f_{1}^{-}b_{2}^{+}, 
&e_{1}\to b_{2}^{-}, 
&e_{2}\to f_{1}^{+}q^{{\mathcal H}_{2}}, \\[0.4 cm] 
 h_{0}\to {\mathcal H}_{1}+{\mathcal H}_{2}, 
&h_{1}\to -{\mathcal H}_{1}-2{\mathcal H}_{2},  
&h_{2}\to {\mathcal H}_{2}.
\end{array}\right.
\ee
And the third one
\be 
\rho_3(x):\qquad {\mathcal B}_+\to {\mathsf F}
\otimes_s{\mathsf   F}\label{map3}
\ee
is given by 
\be
\rho_3(x):\qquad\left\{
\begin{array}{lll}
e_{0}\to x f_{1}^{+}q^{-{\mathcal H}_{2}}, 
&\ds e_{1}\to {q^{-\frac{1}{2}}}
q^{{\mathcal H}_{2}}f_{1}^{-}f_{2}^{+}/(q-q^{-1}), 
&e_{2}\to f_{2}^{-},\\[.4cm] 
h_{0}\to -{\mathcal H}_{2}, 
&h_{1}\to -{\mathcal H}_{1}+{\mathcal H}_{2}, 
&h_{2}\to  {\mathcal H}_{1}.
\end{array}\right.
\ee
The indices $1$ and $2$ above refer respectively 
to the first and second factors in the tensor products \eqref{map1},
\eqref{map2} and \eqref{map3}.
 
The operators $\Q_i(x)$ are defined similarly to the ${\bf
  T}$-operators \eqref{t-uni3},
\be
\Q_i(x)=x^{\Ss_i}\,\A_i(x),\qquad 
{\mathbb A}_i(x)=Z_i^{-1}\,{\str}_{\rho_i(x)}\,\Big[
  z_1^{-h_0}z_2^{h_2}\,\overline{\mathcal R}\,\Big], \qquad i=1,2,3,
\label{Q-def}
\ee
where $\Ss_i$, $i=1,2,3$  are defined in \eqref{zs-def} and the 
normalization constants read 
\be
Z_i={\str}_{\rho_i(x)}\,\Big[
  z_1^{-h_0}z_2^{h_2}\Big] \label{normali1}\ .
\ee
The trace is now taken over the {\em Fock space representations} of 
the $q$-oscillator superalgebras involved in the maps $\rho_i(x)$.
An important property of the definition \eqref{Q-def} is that 
the supertrace therein (normalized by the constants $Z_i$) is completely
determined by the commutation relations \eqref{Hq-def} and
\eqref{F-def} and the cyclic property of the supertrace, so the specific
choice of the representations in \eqref{Q-def} is not important 
as long as the supertrace exist. (Notice that the representations of the 
bosonic $q$-oscillator algebra \eqref{Hq-def} are infinite-dimensional 
so the question of convergence should be kept in mind. 
There is no real problem here, the convergence
can always be achieved with a proper choice of the external field
parameters $\f_{1,2}$. See Sect.\ref{fockspaces} for further details). 

The remaining three operators $\overline{\Q}_i(x)$ are defined 
in a similar way 
\be
\overline{\Q}_i(x)=x^{-\Ss_i}\,\overline{\A}_i(x),\qquad 
\overline{{\mathbb A}}_i(x)=Z_i^{-1}\,
{\str}_{\overline{\rho}_i(x)}\,\Big[
  z_1^{-h_0}z_2^{h_2}\,\overline{\mathcal R}\,\Big], \qquad i=1,2,3,
\label{Qbar-def}
\ee
where normalization constants read 
\be
\overline{Z}_i={\str}_{\overline{\rho}_i(x)}\, \Big[
  z_1^{-h_0}z_2^{h_2}\Big], \label{normali2}
\ee
and the corresponding maps $\overline{\rho}_i(x)$ are defined as follows. 
The first one
\be 
\overline{\rho}_1(x):\qquad {\mathcal B}_+\to {\mathsf H}_q
\otimes{\mathsf   F}
\ee
is given by
\begin{eqnarray}
\overline{\rho}_1(x):\qquad\left\{ 
\begin{array}{lll}
e_{0} \to x f_{2}^{+},  & 
e_{1} \to b_{1}^{-}q^{{\mathcal H}_{2}}, & 
e_{2} \to -q^{-\frac{1}{2}}q^{-{\mathcal H}_{2}}b_{1}^{+}f_{2}^{-}, \\ 
[0.4cm]
h_{0} \to {\mathcal H}_{1}, & 
h_{1} \to -2{\mathcal H}_{1}-{\mathcal H}_{2}, & 
h_{2} \to {\mathcal H}_{1}+{\mathcal H}_{2}.
\end{array}\right.\label{bar1}
\end{eqnarray}
The second one 
\be 
\overline{\rho}_2(x): \qquad {\mathcal B}_+\to 
{\mathsf   F}\otimes {\mathsf H}_q
\ee
is given by
\begin{eqnarray}
\overline{\rho}_2(x): \qquad\left\{ 
\begin{array}{lll}
e_{0} \to -q^{-\frac{1}{2}}xq^{-{\mathcal H}_{2}}f_{1}^{+}b_{2}^{-}, & 
e_{1} \to b_{2}^{+}, & 
e_{2} \to f_{1}^{-}q^{{\mathcal H}_{2}}, \\ [0.4cm]
h_{0} \to -{\mathcal H}_{1}-{\mathcal H}_{2}, &
h_{1} \to {\mathcal H}_{1}+2{\mathcal H}_{2}, &
h_{2} \to -{\mathcal H}_{2}.
\end{array}\right.\label{bar2}
\end{eqnarray}
And the third one  
\be 
\overline{\rho}_3(x):\qquad {\mathcal B}_+\to {\mathsf F}
\otimes_s{\mathsf   F}
\ee
is given by
\begin{eqnarray}
\overline{\rho}_3(x):\qquad\left\{ 
\begin{array}{lll}
e_{0} \to x f_{1}^{-}q^{-{\mathcal H}_{2}}, &  
e_{1} \to -{q^{\frac{1}{2}}}q^{{\mathcal H}_{2}}f_{1}^{+}f_{2}^{-}/(q-q^{-1}), 
& e_{2} \to f_{2}^{+}, \\ [0.4cm]
 h_{0} \to {\mathcal H}_{2}, &
h_{1} \to {\mathcal H}_{1}-{\mathcal H}_{2}, & 
 h_{2} \to  -{\mathcal H}_{1}.
\end{array}\right.\label{bar3}
\end{eqnarray}

\subsection{Lattice ${R}$-matrices}\label{latR}
As an illustration of the above construction consider
all specializations of the universal ${R}$-matrix related to
the $3$-state model 
\eqref{3-stateR}.
We postpone details of calculations to a separate 
publication \cite{workinprogress}, but present the final results
here. 

Below we will fix the (local) quantum space at each lattice site  
to be the $3$-dimensional evaluation representation
$\pi_{(1,0,0)}(q^{\frac{1}{2}})$  
with the weight $\mu=(1,0,0)$ and the spectral parameter $q^{\frac{1}{2}}$,
as defined in \eqref{evmap}.
First consider the ${\bf L}$-operator, obtained as 
\be
{\bf   L}(x)={{N}}'(x)\ 
\big(\mathsf{Ev}_x \otimes_s \pi_{(1,0,0)}(q^{\frac{1}{2}}) 
\big)\big[\mathcal{R}\big]\ .\label{L-def}
\ee 
where $\evx$ is the evaluation map \eqref{eval} and ${N}'$ is
a normalization factor.
It is convenient to present this operator  
\be
{\bf L}=\sum_{i,j=1}^3 L_{ij}\otimes_{s}e_{ij},\qquad L_{ij}\in U_q({gl}(2|1))
\ee
where $(e_{ij})_{kl}=\delta_{ik}\delta_{jl}$ is the matrix unit, 
in the form of a $3\times3$ matrix, acting in the quantum space, 
\be
{\bf L}=\left(
\begin{array}{ccc}
L_{11} & L_{12} & L_{13} \\
L_{21} & L_{22} & L_{23} \\
-L_{31} & -L_{32} & L_{33} 
\end{array}
\right)\ ,\label{lmatrix}
\ee
whose entries are
operators belonging to the finite-dimensional algebra
$U_q({gl}(2|1))$ (the reader should pay
attention to the signs in \eqref{lmatrix}). 
Remind that the relevant parities are $p(L_{jk})=p(e_{jk})=p(j)+p(k)\pmod 2$, 
and $p(1)=p(2)=0$, $p(3)=1$. The operators $L_{ij}$ act in the
auxiliary space. 
With a suitable choice of the normalization factor ${N}'(x)$
one obtains (cf. \cite{Zhang91})
\be
{\bf L}(x)=\left(
\begin{array}{lll}
{\mathcal C}q^{-E_{11}}-q^{\frac{1}{2}}\,x \,q^{E_{11}}\quad
&
-q^{\frac{1}{2}}\, \aq\, x\,q^{E_{11}}\,E_{21}\quad
& 
-q^{-\frac{1}{2}} \aq\, x \,{\mathcal C} E_{31}\,q^{E_{33}}
\\[.4cm]
-\aq \, {\mathcal C} E_{12}\,q^{-E_{11}}
&
{\mathcal C}q^{-E_{22}}-q^{\frac{1}{2}}\,x\,q^{E_{22}}
&
-q^{\frac{1}{2}} \,\aq \,x \,q^{E_{22}}\,E_{32}
\\[.4cm]
-q \,\aq \,q^{-E_{33}}\,E_{13}
&
-\aq\, {\mathcal C} E_{23}\,q^{-E_{22}}
&
{\mathcal C} q^{E_{33}}-q^{\frac{1}{2}}\,x \,q^{-E_{33}}
\\[.3cm]
\end{array}\right)\label{L-lattice}
\ee
where $a_q=(q-q^{-1})$ and 
\be
{\mathcal C}=q^{E_{11}+E_{22}+E_{33}}
\ee
is a central element of the algebra $U_q({gl}(2|1))$.    
Note that matrix
elements of ${\bf L}(x)$ are first order {\em polynomials} in the spectral
parameter $x$. Such normalization is especially convenient for
lattice models.
The factor $N'(x)$ depends on central elements 
of $U_q({gl}(2|1))$, so it depends on the representation. 
It can be thought of as a diagonal operator also acting 
in the auxiliary space. In general it is a meromorphic
function of $x$. The formula \eqref{L-lattice} can be further
specialized by choosing some particular representation $\pi_\mu$ of 
$U_q({gl}(2|1))$ in the auxiliary space.
The resulting operator is, essentially, a particular case  
of \eqref{rmumu} with $\nu=(1,0,0)$ and $y=q^{1/2}$, differing from
the later merely by a scalar factor 
\be
{\bf   L}_\mu(x)=\pi_\mu\big[{\bf L}(x)\big]=
{N}_\mu(x)\,\R_{\mu\nu}(x\,q^{-\frac{1}{2}})\Big|_{\nu=(1,0,0)}
\ee
where
\be
N_{{(\mu_1,\mu_2,\mu_3)}}(x)=
\frac{
\psi(xq^{\frac{1}{2}+\mu_{1}-\mu_{2}-\mu_{3}})
\psi(xq^{-\frac{3}{2}-\mu_{1}+\mu_{2}-\mu_{3}})}
{\psi(xq^{-\frac{3}{2}-\mu_{1}-\mu_{2}-3\mu_{3}})},
\quad \psi(x)=1-x. 
\ee
Similarly to \eqref{tmat} it is convenient to define lattice 
transfer matrices constructed with this ${\bf L}$-operator,
\begin{eqnarray}
{\bf T}^{(L)}_\mu(x)&=&\mathsf{Str}_{\pi_\mu}\Big\{
{\bf D}\ \underbrace{{\bf L}(x)\otimes_s {\bf L}(x)\otimes_s
\cdots \otimes_s {\bf L}(x)}_{L-\mbox{\scriptsize{times}}}
\Big\}\\[.4cm]
&=&(N_\mu(x))^L\ 
\big(
\underbrace{\pi_\nu(q^{\frac{1}{2}})\otimes_s
  \pi_\nu(q^{\frac{1}{2}})\otimes_s   
\cdots \otimes_s\pi_\nu(q^{\frac{1}{2}})}_{L-\mbox{\scriptsize{times}}}
\big)\,\Big[\TT_\mu(x)\Big]\ .\label{t-norm}
\end{eqnarray}
which, to within the scalar factor $(N_\mu)^L$, coincide with 
${\bf T}_\mu(x|L,q^{\frac{1}{2}},\nu)$ for 
$\nu=(1,0,0)$, defined in \eqref{tmat2}.
It is important that the re-normalized transfer matrix ${\bf
  T}^{(L)}_\mu(x)$ is a polynomial
function of the spectral parameter $x$. Thus, due to the
commutativity \eqref{tcomm}, all its eigenvalues are polynomials (more
precisely, they are, at most, $L$-th degree polynomials).
Finally, define a special notation for the fundamental transfer matrix
of the $3$-state model, corresponding to the $3$-dimensional
representation in the auxiliary space,
\be
{\bf T}(x)\equiv{\bf T}^{(L)}_\mu(x)\Big|_{\mu=(1,0,0)},
\ee
and
\be
N(x)=N_\mu(x)|_{\mu=(1,0,0)}=\psi(xq^{\frac{3}{2}})=(1-xq^{\frac{3}{2}}) .
\ee

Now define additional ${\bf L}$-operator associated with the oscillator
algebras \eqref{Hq-def}, \eqref{F-def}. 
\be
{\bf   L}_j(x)=N_j(x)\ 
\big(\rho_j(x) \otimes_s \pi_{(1,0,0)}(q^{\frac{1}{2}}) 
\big)\big[\mathcal{R}\big]\ ,\qquad j=1,2,3 \label{LQ-def}
\ee 
and 
\be
\overline{\bf   L}_j(x)=\overline{N}_j(x)\ 
\big(\overline{\rho}_j(x) \otimes_s \pi_{(1,0,0)}(q^{\frac{1}{2}}) 
\big)\big[\mathcal{R}\big]\ ,\qquad j=1,2,3 \label{LQbar-def}
\ee 
where ${\mathcal R}$ is the universal $R$-matrix and 
the maps $\rho_j(x)$ and $\overline{\rho}_j(x)$ 
defined in the previous subsection.
Explicitly, one obtains \cite{workinprogress} 
\be
{\bf L}_1(x)=\left(
\begin{array}{lll}
q^{-{\mathcal H}_{1}-{\mathcal H}_{2}}
-x q^{{\mathcal H}_{1}+{\mathcal H}_{2}}
{\mathcal C}_{b_{1}}{\mathcal C}_{f_{2}}\
& 
-\aq x
q^{{\mathcal H}_{1}+{\mathcal H}_{2}}b_{1}^{-}
{\mathcal C}_{f_{2}}\ 
&
q^{-\frac{3}{2}}\aq xq^{-{\mathcal H}_{2}} f_{2}^{-}
\\[.4cm]
-\aq b_{1}^{+}q^{-{\mathcal H}_{1}}
&
q^{{\mathcal H}_{1}}
&
0
\\[.4cm]
q^{\frac{3}{2}} q^{{\mathcal H}_{1}-{\mathcal H}_{2}}f_{2}^{+}
{\mathcal C}_{b_{1}}
&
q^{\frac{3}{2}}\aq q^{{\mathcal H}_{1}-
{\mathcal H}_{2}}b_{1}^{-}f_{2}^{+}
&
q^{-{\mathcal H}_{2}}
\end{array}\right),\label{L1-lattice}
\ee

\bigskip
\be
{\bf L}_2(x)=\left(
\begin{array}{lll}
q^{{\mathcal H}_{2}}
&
-\aq x
q^{{\mathcal H}_{1}-{\mathcal H}_{2}}b_{2}^{+}
{\mathcal C}_{f_{1}}
&
-q^{-1}\aq x 
 q^{-{\mathcal H}_{1}-{\mathcal H}_{2}}f_{1}^{-}b_{2}^{+}
\\[.4cm]
-q\aq  q^{{\mathcal H}_{2}}b_{2}^{-}\ 
&
q^{-{\mathcal H}_{1}-{\mathcal H}_{2}}
-xq^{{\mathcal H}_{1}+{\mathcal H}_{2}}
{\mathcal C}_{f_{1}}{\mathcal C}_{b_{2}}\ 
&
-xq^{-1}q^{-{\mathcal H}_{1}+{\mathcal H}_{2}}f_{1}^{-}
{\mathcal C}_{b_{2}}
\\[.4cm]
0 
&
-q\aq q^{-{\mathcal H}_{1}}f_{1}^{+}
&
q^{-{\mathcal H}_{1}}
\end{array}\right),\label{L2-lattice}
\ee

\bigskip
\be
{\bf L}_3(x)=\left(
\begin{array}{lll}
q^{{\mathcal H}_{1}}
&
0
&
q^{-\frac{3}{2}}\aq xq^{{\mathcal H}_{1}}f_{1}^{+}
\\[.4cm]
-q^{\frac{1}{2}}q^{{\mathcal H}_{1}+{\mathcal H}_{2}}f_{1}^{-}f_{2}^{+}
&
q^{{\mathcal H}_{2}}
&
q^{-1}xq^{-{\mathcal H}_{1}+{\mathcal H}_{2}}f_{2}^{+}
{\mathcal C}_{f_{1}}^{-1}
\\[.4cm]
-q^{\frac{3}{2}}q^{{\mathcal H}_{1}-{\mathcal H}_{2}}f_{1}^{-}
{\mathcal C}_{f_{2}}^{-1}\ 
&
-q\aq  q^{{\mathcal H}_{2}} f_{2}^{-}\ 
&
q^{{\mathcal H}_{1}+{\mathcal H}_{2}}
-xq^{-{\mathcal H}_{1}-{\mathcal H}_{2}}
{\mathcal C}_{f_{1}}^{-1}{\mathcal C}_{f_{2}}^{-1}
\end{array}\right),\label{L3-lattice}
\ee
and 

\bigskip
\be
\overline{\bf L}_1(x)=\left(
\begin{array}{lll}
q^{{\mathcal H}_{1}+{\mathcal H}_{2}}
&
-\aq x
q^{-{\mathcal H}_{1}-{\mathcal H}_{2}}b_{1}^{+}
&
q^{-\frac{3}{2}}\aq x q^{{\mathcal H}_{2}}f_{2}^{+}
\\[.4cm]
-\aq b_{1}^{-}q^{{\mathcal H}_{1}+2{\mathcal H}_{2}}
&
q^{-{\mathcal H}_{1}}-xq^{{\mathcal H}_{1}}
{\mathcal C}_{b_{1}}
&
-q^{-\frac{3}{2}}\aq^{2}x
q^{2{\mathcal H}_{2}} b_{1}^{-}f_{2}^{+}
\\[.4cm]
-q^{\frac{3}{2}} q^{{\mathcal H}_{1}+{\mathcal H}_{2}}f_{2}^{-}\ 
&
q^{\frac{1}{2}}\aq q^{-{\mathcal H}_{1}-
{\mathcal H}_{2}}b_{1}^{+}f_{2}^{-}\ 
&
q^{{\mathcal H}_{2}}-xq^{-{\mathcal H}_{2}}
{\mathcal C}_{f_{2}}^{-1} 
\end{array}\right)\ ,\label{Lb1-lattice}
\ee

\bigskip
\be
\overline{\bf L}_2(x)=\left(
\begin{array}{lll}
q^{-{\mathcal H}_{2}}-
xq^{{\mathcal H}_{2}}
{\mathcal C}_{b_{2}}\ 
&
-\aq x
q^{{\mathcal H}_{1}+{\mathcal H}_{2}}b_{2}^{-}\ 
&
-q^{-2}\aq x
 q^{{\mathcal H}_{1}-{\mathcal H}_{2}}f_{1}^{+}b_{2}^{-}
\\[.4cm]
-q\aq  q^{-{\mathcal H}_{2}}b_{2}^{+}
&
q^{{\mathcal H}_{1}+{\mathcal H}_{2}}
&
xq^{-1}
q^{{\mathcal H}_{1}-{\mathcal H}_{2}}f_{1}^{+}
\\[.4cm]
q\aq^{2}f_{1}^{-}b_{2}^{+}
&
-q\aq q^{{\mathcal H}_{1}+2{\mathcal H}_{2}}f_{1}^{-}
&
q^{{\mathcal H}_{1}}-
xq^{-{\mathcal H}_{1}}
{\mathcal C}_{f_{1}}^{-1}
\end{array}\right),\label{Lb2-lattice}
\ee

\bigskip
\be
\overline{\bf L}_3(x)=\left(
\begin{array}{lll}
q^{-{\mathcal H}_{1}}-
 xq^{{\mathcal H}_{1}}
{\mathcal C}_{f_{1}}
&
-q^{\frac{1}{2}}\aq^{2}xq^{-2{\mathcal H}_{2}}
f_{1}^{-}f_{2}^{+}
&
q^{-\frac{3}{2}}\aq xq^{-{\mathcal H}_{1}-2{\mathcal H}_{2}}f_{1}^{-}
\\[.4cm]
q^{\frac{3}{2}} q^{-{\mathcal H}_{1}+{\mathcal H}_{2}}
f_{1}^{+}f_{2}^{-}\ 
&
q^{-{\mathcal H}_{2}}-
 xq^{{\mathcal H}_{2}}
{\mathcal C}_{f_{2}}\ 
&
-q^{-1}xq^{-{\mathcal H}_{1}-
{\mathcal H}_{2}}f_{2}^{-}
\\[.4cm]
q^{\frac{3}{2}}q^{-{\mathcal H}_{1}+{\mathcal H}_{2}}f_{1}^{+}
&
-\aq  f_{2}^{+}q^{-{\mathcal H}_{2}}
&
q^{-{\mathcal H}_{1}-{\mathcal H}_{2}}
\end{array}\right),\label{Lb3-lattice}
\ee
where where $a_q=(q-q^{-1})$ and 
the central charges for the bosonic and fermionic 
oscillator algebras \eqref{Hq-def} and \eqref{F-def} are given by
\begin{eqnarray}
{\mathcal C}_{b}&=&\ds q\,q^{-2{\mathcal
    H}_b}\,\big(1-(q-q^{-1})^{2}\,b^{+}\,b^{-}\big),\\[.4cm]
{\mathcal C}_{f}&=&\ds q^{-1}\,q^{-2{\mathcal H}_f}\,\big(1-f^{+}f^{-}\big)
\end{eqnarray}
The normalization factors read
\be\label{fan-fac}
\begin{array}{rclrclrcl}
N_{1}(x)&=&\psi(x),&N_{2}(x)&=&\psi(x),&N_{3}(x)&=&1/{\psi(x)}, \\[.4cm]
\overline{N}_{1}(x)&=&1, & 
\overline{N}_{2}(x)&=&1, &
\overline{N}_{3}(x)&=&\psi(xq)\psi(xq^{-1})\ .
\end{array}
\ee
where $\psi(x)=1-x$.
For the $3$-state model the eigenvalues of the diagonal 
operators \eqref{zs-def} 
\be
z_1=q^{2\mathsf{S}_1},\qquad z_2=q^{2\mathsf{S}_2},\qquad
z_3=q^{-2\mathsf{S}_3},  \label{z3-def}   
\ee
where
\be
2\,{\mathsf S}_1=L-{m}_1+\f_1,\qquad
2\,{\mathsf S}_2=L-{m}_2+\f_2\qquad 2\,{\mathsf S}_3=-2\mathsf{S}_1-
2\mathsf{S}_2 \label{s-m2}
\ee
are expressed in terms of the external field parameters $\f_{1,2}$ and
the edge occupation numbers (or the ``magnon
numbers''), 
\be
m_i=\sum_{\ell=1}^L e^{(\ell)}_{ii} ,\qquad m_1+m_2+m_3=L\ ,
\ee
where $e^{(\ell)}_{ij}$ are three by three matrices
$||e^{(\ell)}_{ij}||_{ab}=\delta_{ia}\delta_{jb}$,\  $a,b=1,2,3$
acting at the $\ell$-th site of the chain.

We can now explicitly define lattice ${\bf Q}$-operators for the
$3$-state model,
\be
{\bf A}_j(x)=Z_j^{-1}\str_{\rho_j(x)}\Big\{
q^{-\f_1\,h_0+\f_2\,h_2}
\underbrace{{\bf L}_j(x)\otimes_s {\bf L}_j(x)\otimes_s
\cdots \otimes_s {\bf L}_j(x)}_{L-\mbox{\scriptsize{times}}}
\Big\}\label{Q-lat}
\ee
and 
\be
\overline{\bf A}_j(x)=\overline{Z}_j^{-1}\str_{\overline{\rho}_j(x)}\Big\{
q^{-\f_1\,h_0+\f_2\,h_2}
\underbrace{\overline{\bf L}_j(x)\otimes_s \overline{\bf L}_j(x)\otimes_s
\cdots \otimes_s \overline{\bf L}_j(x)}_{L-\mbox{\scriptsize{times}}}
\Big\}\label{Qb-lat}
\ee
where $j=1,2,3$, the generators $h_0$, $h_1$ in the
exponents act in the auxiliary
space and the quantities $Z_j$ and $\overline{Z}_j$ are given
in \eqref{normali1}. A word caution: one has to use there the same
representations of the oscillator algebras \eqref{Hq-def} and
\eqref{F-def} as in the corresponding formula \eqref{Q-lat} or
\eqref{Qb-lat}. Then these definitions 
become independent 
on a choice of these representations . The quantities $Z_j$ and 
$\overline{Z}_j$, however, do depend on this choice (see
Sect.\ref{fockspaces} for further details). 

It is evident from (\ref{L1-lattice}-\ref{Lb3-lattice}) that
${\bf A}_j(x)$ and $\overline{\bf A}_j(x)$ are operator-valued   
polynomials in $x$ of the degree $L$. They are simply connected with
the specializations of the universal ${\bf
  Q}$-operators \eqref{Q-def} and \eqref{Qbar-def} to the $3$-state model,
namely, 
\be
{\bf A}_j(x)=(N_j(x))^L\ 
\big(
\underbrace{\pi_\nu(q^{\frac{1}{2}})\otimes_s \pi_\nu(q^{\frac{1}{2}})\otimes_s 
\cdots \otimes_s\pi_\nu(q^{\frac{1}{2}})}_{L-\mbox{\scriptsize{times}}}
\big)\,\Big[\A_j(x)\Big],\qquad \nu=(1,0,0),\label{AA}
\ee 
and
\be
\overline{\bf A}_j(x)=(\overline{N}_j(x))^L\ 
\big(
\underbrace{\pi_\nu(q^{\frac{1}{2}})\otimes_s \pi_\nu(q^{\frac{1}{2}})\otimes_s 
\cdots \otimes_s\pi_\nu(q^{\frac{1}{2}})}_{L-\mbox{\scriptsize{times}}}
\big)\,\Big[\overline{\A}_j(x)\Big],\qquad \nu=(1,0,0),\label{AbAb}
\ee 
where $j=1,2,3$.

\nsection{Functional relations}\label{funcrel}
As is well known \cite{Baxterbook}, 
the analyticity of the transfer matrices 
becomes an extremely powerful condition when combined with the
functional relations which the transfer matrices satisfy, and, in
principle, allows one to determine all their eigenvalues. Therefore
the functional relations for the ${\bf T}$- and ${\bf Q}$-operators
are of a primary interest. We present these relations here, but
postpone their proof (which is purely algebraic) to
Section~\ref{proof}. It is convenient to split these relations
 into three groups, 
(i) the Wronskian-type relations, (ii) the ${\bf T}$-${\bf Q}$ relations
 and (iii) the fusion relations.
We remark that some of our results in this section 
overlap with those obtained in \cite{BDKM06,KSZ07,Zabrodin07,GV07}
 for the rational case $q=1$. 
\subsection{Wronskian-type relations} 
These relations only involve the ${\bf Q}$-operators, defined in 
\eqref{Q-def} and \eqref{Qbar-def}. 
First, there are four independent
relations, quoted in the Introduction,
\be
\renewcommand\arraystretch{2.0}
\begin{array}{rcl}
c_{12}  &=&c_{13} \,\Q_1(q^{+\frac{1}{2}} x)\, 
\overline{\Q}_1(q^{-\frac{1}{2}}x)-  
c_{23}\, \Q_2(q^{+\frac{1}{2}} x)\, 
\overline{\Q}_2(q^{-\frac{1}{2}}x)\\
c_{12}  &=& c_{13} \,\Q_1(q^{-\frac{1}{2}}x)\, 
\overline{\Q}_1(q^{+\frac{1}{2}} x)-  
c_{23} \,\Q_2(q^{-\frac{1}{2}}x)\, \overline{\Q}_2(q^{+\frac{1}{2}} x)
\end{array}\label{Q-rel}
\renewcommand\arraystretch{1.0}
\ee
and
\begin{subequations}
\begin{eqnarray}
 c_{21}\Q_{3}(x)&=&\overline{\Q}_{1}(xq)\overline{\Q}_{2}(xq^{-1})-
\overline{\Q}_{1}(xq^{-1})\overline{\Q}_{2}(xq),
 \label{sl21-rel1} \\[.2cm]
c_{12}\overline{\Q}_{3}(x)&=&\Q_{1}(xq)\Q_{2}(xq^{-1})-
\Q_{1}(xq^{-1})\Q_{2}(xq), \label{sl21-rel2}
\end{eqnarray}\label{sl21-rel12}
\end{subequations}
where 
\be
c_{ij}=\frac{z_{i}-z_{j}}{(z_{i}z_{j})^{\frac{1}{2}}}\label{cij-def}
\ee 
with the
operators $z_1$, $z_2$ and $z_3$ defined in \eqref{zs-def}.
Combining these relations one easily obtains  
\begin{eqnarray}
c_{13}\Q_{3}(x)\Q_{1}(x)&=&\overline{\Q}_{2}(xq)-
 \overline{\Q}_{2}(xq^{-1}), \label{sl21-rel3} \\[.2cm]
 c_{23}\Q_{2}(x)\Q_{3}(x)&=&\overline{\Q}_{1}(xq)-
 \overline{\Q}_{1}(xq^{-1}), \label{sl21-rel4} \\[.2cm] 
 c_{13}\overline{\Q}_{3}(x)\overline{\Q}_{1}(x)&=&\Q_{2}(xq^{-1})-\Q_{2}(xq),
 \label{sl21-rel5} \\[.2cm]
 c_{23}\overline{\Q}_{2}(x)\overline{\Q}_{3}(x)&=&\Q_{1}(xq^{-1})-\Q_{1}(xq). 
\label{sl21-rel6}
\end{eqnarray}
For example, substituting \eqref{sl21-rel2} into the LHS of
\eqref{sl21-rel3} and using both relations \eqref{Q-rel} in the
resulting expression, one gets
\be
\renewcommand\arraystretch{2.0}
\begin{array}{l}
c_{13}\,\Q_3(x)\, \Q_1(x)= \ds\frac{c_{13}}{c_{21}}\Big\{
\overline{\Q}_1(qx)\,\overline{\Q}_2(q^{-1}x)\,\Q_1(x)
-\overline{\Q}_1(q^{-1}x)\,\overline{\Q}_2(qx)\,\Q_1(x)\Big\}\\
=\ds\frac{1}{c_{21}}\Big\{\overline{\Q}_2(q^{-1}x)\Big[
c_{12}+c_{23}\,\Q_2(x)\,\overline{\Q}_2(qx)\Big]-
\overline{\Q}_2(qx)\Big[c_{12}+c_{23}\,\Q_2(x)\,
\overline{\Q}_2(q^{-1}x)\Big] \Big\} \\
=\overline{\Q}_{2}(xq)-
 \overline{\Q}_{2}(xq^{-1}).
\end{array}
\renewcommand\arraystretch{1.0}
\ee

\subsection{${\bf T}$-${\bf Q}$ relations} 
The next group of relations connects the
${\bf T}$- and ${\bf Q}$-operators. An important part of 
this group consists of relations, which 
express the ${\bf T}$-operators as polynomial combinations
of the ${\bf Q}$-operators. 
For the operators \eqref{tm}, \eqref{tbarm} and \eqref{t2}, introduced
above, these relations read
\begin{subequations}\label{atyp} 
\begin{eqnarray}
\renewcommand\arraystretch{2.0}
c_{12}\,\TT^{(1)}_m(x)&=&
c_{13} \Q_1(x q^{+m+\frac{1}{2}}) \overline{\Q}_1(x q^{-m-\frac{1}{2}}) 
-c_{23} \Q_2(x q^{+m+\frac{1}{2}}) 
\overline{\Q}_2(x q^{-m-\frac{1}{2}})\label{atypa}\\[.3cm]
c_{12}\,\overline{\TT}^{(1)}_m(x)&=&
c_{13} \Q_1(x q^{-m-\frac{1}{2}}) \overline{\Q}_1(x q^{+m+\frac{1}{2}}) 
-c_{23} \Q_2(x q^{-m-\frac{1}{2}}) 
\overline{\Q}_2(x q^{+m+\frac{1}{2}})\label{atypb}
\end{eqnarray}
\renewcommand\arraystretch{1.0}
\end{subequations}
where $m\in{\mathbb Z}_{\ge0}$ and 
\begin{subequations}\label{dim4}
\begin{eqnarray}
\TT^{(2)}_c(x)&=&c_{23} \, c_{13}\, \Q_3(x q^{-c-\frac{1}{2}})
 \overline{\Q}_3(xq^{+c+\frac{1}{2}})\label{dim4a} 
\\[.3cm]
\overline{\TT}^{(2)}_c(x)&=&c_{23} \, c_{13}\, \Q_3(x q^{+c+\frac{1}{2}})
 \overline{\Q}_3(xq^{-c-\frac{1}{2}})\label{dim4b} 
\end{eqnarray}
\end{subequations}
Taking into account the normalization conditions
\eqref{tnot} it is easy to see that Eqs.\eqref{atyp} 
with $m=0$ \ simply reduce to the Wronskian-type relations \eqref{Q-rel}.
The relation \eqref{dim4b} is just a corollary of the definition
\eqref{t2bar}. 

Using (\ref{Q-rel})-(\ref{sl21-rel6}) one can transform the expressions 
\eqref{atyp} to a more familiar Bethe-Ansatz type form. For example
the operators 
$\TT(x)$ and $\overline{\TT}(x)$, defined in \eqref{3dim},
corresponding to the 3-dimensional representations in auxiliary space, 
can be transformed to any of the six equivalent forms 
\be
\renewcommand\arraystretch{2.0}
\begin{array}{l}
{\mathbb T}(x)\,=\,\ds
p_i\,\frac{\overline{\Q}_i(q^{-p_i-\frac{1}{2}}x)}
{\overline{\Q}_i(q^{+p_i-\frac{1}{2}}x)}+
p_j\,\frac{\overline{\Q}_i(q^{+p_i+2p_j-\frac{1}{2}}x)}
{\overline{\Q}_i(q^{+p_i-\frac{1}{2}}x)}
\frac{\Q_k(q^{+p_i-p_j-\frac{1}{2}}x)}
{\Q_k(q^{+p_i+p_j-\frac{1}{2}}x)}
+p_k\,\frac{\Q_k(q^{+p_i+p_j+2p_k-\frac{1}{2}}x)}
{\Q_k(q^{+p_i+p_j-\frac{1}{2}}x)}, \\[.8cm]
\overline{\TT}(x)\,=\,\ds
p_i\,\frac{\overline{\Q}_i(q^{+p_i+\frac{1}{2}}x)}
{\overline{\Q}_i(q^{-p_i+\frac{1}{2}}x)}+
p_j\,\frac{\overline{\Q}_i(q^{-p_i-2p_j+\frac{1}{2}}x)}
{\overline{\Q}_i(q^{-p_i+\frac{1}{2}}x)}
\frac{\Q_k(q^{-p_i+p_j+\frac{1}{2}}x)}
{\Q_k(q^{-p_i-p_j+\frac{1}{2}}x)}
+p_k\,\frac{\Q_k(q^{-p_i-p_j-2p_k+\frac{1}{2}}x)}
{\Q_k(q^{-p_i-p_j+\frac{1}{2}}x)} , 
\end{array}\label{six2}
\renewcommand\arraystretch{1.0}
\ee
where $p_1=p_2=-p_3=1$ and $(i,j,k)$ is an arbitrary permutation of $(1,2,3)$. 
The most general transfer matrix expression of this type 
is considered in the Appendix C (see, Eq.\eqref{DVF-tab} therein). 

\subsection{Fusion relations} 
These relations only involve the
${\bf T}$-operators ${\TT}^{(a)}_m(x)$, corresponding to the  
rectangular Young diagrams of the length $m\ge0$ and the height
$a\ge1$.  A complete set of the fusion relations  for
$U_q(\widehat{sl}(r+1|s+1))$ involving the ${\bf T}$-operators
\eqref{atyp} and \eqref{dim4} has been previously proposed in
\cite{T97,T98,T98-2}. These relations 
were deduced there from the Bethe Ansatz solution of
the (r+s+2)-state lattice model.  In particular for
$U_q(\widehat{sl}(2|1))$ case, they take the form   
\begin{eqnarray}
&& \hspace{-10pt} 
\TT^{(1)}_{m}(q^{-1}x)\TT^{(1)}_{m}(qx)=
\TT^{(1)}_{m-1}(x)\TT^{(1)}_{m+1}(x) +
\TT^{(2)}_{m}(x) 
\quad  \mbox{for} \quad m \in {\mathbb Z}_{\ge 1}, \nonumber\\[.2cm] 
&& \hspace{-10pt} 
\TT^{(2)}_{1}(q^{-1}x)\TT^{(2)}_{1}(qx)=
\TT^{(2)}_{2}(x) +
\TT^{(1)}_{1}(x)\TT^{(3)}_{1}(x),\nonumber\\[.2cm] 
&& \hspace{-10pt}
\TT^{(2)}_{m}(q^{-1}x)\TT^{(2)}_{m}(qx)=
\TT^{(2)}_{m-1}(x)\TT^{(2)}_{m+1}(x) 
\quad  \mbox{for} \quad m \in {\mathbb Z}_{\ge 2}, \label{tfr}\\[.2cm] 
&& \hspace{-10pt}
\TT^{(a)}_{1}(q^{-1}x)\TT^{(a)}_{1}(qx)=
\TT^{(a-1)}_{1}(x)\TT^{(a+1)}_{1}(x) \quad 
  \mbox{for} \quad a \in {\mathbb Z}_{\ge 3}, \nonumber
\end{eqnarray}
where $\TT^{(1)}_{0}(x)=1$. 
There is also a duality relation:
\begin{eqnarray} 
&& \hspace{-35pt}
\TT^{(2)}_{a}(x)=
(-1)^{a-1} 
\TT^{(1+a)}_{1}(x) \quad  {\rm for} \quad a \in {\mathbb Z}_{\ge 1},
\end{eqnarray}
which maps last two relation in \eqref{tfr} into one another. 
The fusion relations for conjugate representations 
have also the same form\footnote{See Appendix B in \cite{T97}.}
\begin{eqnarray}
&& \hspace{-10pt} 
{\overline \TT}^{(1)}_{m}(q^{-1}x){\overline \TT}^{(1)}_{m}(qx)=
{\overline \TT}^{(1)}_{m-1}(x){\overline \TT}^{(1)}_{m+1}(x) +
{\overline \TT}^{(2)}_{m}(x) 
\quad  \mbox{for} \quad m \in {\mathbb Z}_{\ge 1}, \nonumber\\[.2cm] 
&& \hspace{-10pt} 
{\overline \TT}^{(2)}_{1}(q^{-1}x){\overline \TT}^{(2)}_{1}(qx)=
{\overline \TT}^{(2)}_{2}(x) +
{\overline \TT}^{(1)}_{1}(x){\overline \TT}^{(3)}_{1}(x),\nonumber\\[.2cm] 
&& \hspace{-10pt}
{\overline \TT}^{(2)}_{m}(q^{-1}x){\overline \TT}^{(2)}_{m}(qx)=
{\overline \TT}^{(2)}_{m-1}(x){\overline \TT}^{(2)}_{m+1}(x) 
\quad  \mbox{for} \quad m \in {\mathbb Z}_{\ge 2}, \label{tfr-con}\\[.2cm] 
&& \hspace{-10pt}
{\overline \TT}^{(a)}_{1}(q^{-1}x){\overline \TT}^{(a)}_{1}(qx)=
{\overline \TT}^{(a-1)}_{1}(x){\overline \TT}^{(a+1)}_{1}(x) \quad 
  \mbox{for} \quad a \in {\mathbb Z}_{\ge 3}, \nonumber
\end{eqnarray}
where ${\overline \TT}^{(1)}_{0}(x)=1$. 
The corresponding duality relation reads 
\begin{eqnarray} 
&& \hspace{-35pt}
{\overline \TT}^{(2)}_{a}(x)=
(-1)^{a-1} 
{\overline \TT}^{(1+a)}_{1}(x) \quad  {\rm for} \quad a \in {\mathbb
  Z}_{\ge 1}, 
\end{eqnarray}
Note that our results \eqref{Q-rel}-\eqref{dim4} imply 
all the fusion relations \eqref{tfr} and \eqref{tfr-con} as 
trivial corollaries. 
Additional functional relations for ${\bf T}$-operators are discussed 
in the Appendix C (see also \cite{KV07}).

Note also that the operator 
(\ref{dim4}) satisfies the relation\footnote{Eq.(\ref{tfr-typ}) is a
  special case of more general functional relation given in (3.33) of
  \cite{T98-2}}  
\be
{\mathbb T}^{(2)}_{c}(q^{-d}x) {\mathbb T}^{(2)}_{c}(q^{d}x)=
{\mathbb T}^{(2)}_{c-d}(x) {\mathbb T}^{(2)}_{c+d}(x)
\quad 
\mbox{for} \quad c,d \in {\mathbb C}, \label{tfr-typ}
\ee
which generalizes the third relation in \eqref{tfr}.
Finally, quote again useful relations (\ref{dec-typical}) connecting two
types of the ${\bf T}$-operators for the atypical and typical
representations. 

\subsection{Eigenvalue equations}
The ${\bf T}$- and ${\bf Q}$-operators with different values of the
spectral parameter $x$ form a commuting family of operators and can be
simultaneously diagonalized by an $x$-independent similarity transformation.
Therefore, the above functional equations are satisfied by 
eigenvalues of these operators, corresponding to the same eigenstate.
The equations are the same for all eigenstates. 
Here we presented these equations in a universal
model-independent form. They are written in the 
normalization of the 
universal $R$-matrix, which is uniquely defined by its series
expansion \eqref{R-expan}. This is a distinguished normalization where 
the functional equations does not contain any  
$x$-dependent scalar factors (however, the eigenvalues in this case 
are, in general, meromorphic functions of $x$). From analytic point of view 
it is more convenient to work with a normalization where the
eigenvalues are entire functions of $x$ (it is 
commonly used in the lattice theory since Baxter's pioneering work
\cite{Bax72}). With this ``analytic'' normalization the functional
equations acquire some non-universal scalar factors, depending on the
quantum space of the model.  

As an example consider the $3$-state lattice model. All relevant definition 
are already given in Section~\ref{latR}.
First consider the operators 
${\bf A}_j(x)$ and $\overline{\bf  A}_j(x)$, defined in 
\eqref{Q-lat} and \eqref{Qb-lat}, respectively.  
Let  ${\mathsf A}_j(x)$ and $\overline{\mathsf
  A}_j(x)$ denote a set eigenvalues of their eigenvalues,
corresponding to the same eigenstate.
All these eigenvalues are $L$-th degree polynomials in $x$, where $L$
is the length of the chain. Moreover, the definitions \eqref{Q-lat} and
\eqref{Qb-lat} imply 
\be
{\mathsf A}(0)=\overline{\mathsf A}(0)=1. \label{Alatnorm}
\ee
Remind that the constants $z_{1},z_{2},z_{3}$, defined in \eqref{z3-def}, are 
expressed through the conserved edge occupation 
numbers $m_1$, $m_2$ and $m_3$ and the external field parameters $\f_{1,2}$.
Now write the eigenvalues in the product form (cf. \eqref{q-factor})
\be
{\mathsf A}_i(x)
=\prod_{\ell=1}^{{m}_i}
(1-x/x^{(i)}_\ell) ,
\qquad \overline{{\mathsf A}}_i(x)=\prod_{\ell=1}^{L-{m}_i}
(1-x/{\overline x}^{(i)}_\ell).  \label{q-factor1}
\ee
where the numbers of zeroes are uniquely 
determined from elementary considerations of the 
leading $x\to\infty$ asymptotics in \eqref{Q-lat} and \eqref{Qb-lat}. 
Define the scalar function 
\be
{\mathsf f}(x)=(1-x)^L\ .
\ee
Substituting \eqref{AA} and \eqref{AbAb} into
(\ref{Q-rel}-\ref{sl21-rel6}) and using the expressions 
\eqref{fan-fac} for the normalization factors, one obtains,
\begin{eqnarray}
c_{12}\,{\mathsf f}(q^{+\hf}x)  
&=&c_{13}\,z_1^{+\frac{1}{2}}\, \As_1(q^{+\frac{1}{2}} x) \,
\overline{\As}_1(q^{-\frac{1}{2}}x)-  
c_{23}\, z_2^{+\frac{1}{2}}\, 
\As_2(q^{+\frac{1}{2}} x)\, 
\overline{\As}_2(q^{-\frac{1}{2}}x),
\label{Q-rel1-A}\\[.2cm]
c_{12}\,{\mathsf f}(q^{-\hf}x)    &=& c_{13}\,z_1^{-\frac{1}{2}}\, 
 \As_1(q^{-\frac{1}{2}}x) \,\overline{\As}_1(q^{+\frac{1}{2}} x)-  
c_{23}\,z_2^{-\frac{1}{2}}\,  
\As_2(q^{-\frac{1}{2}}x)\, \overline{\As}_2(q^{+\frac{1}{2}} x)\label{Q-rel2-A}
\end{eqnarray}
\begin{eqnarray}
 c_{21}{\mathsf f}(x)\As_{3}(x)&=&\Big(\frac{z_2}{z_1}\Big)^{\frac{1}{2}}\,
\overline{\As}_{1}(xq)\overline{\As}_{2}(xq^{-1})-
\Big(\frac{z_1}{z_2}\Big)^{\frac{1}{2}}\,
\overline{\As}_{1}(xq^{-1})\overline{\As}_{2}(xq),
 \label{sl21-rel1-A} \\[.2cm]
c_{12}\overline{\As}_{3}(x)&=&\Big(\frac{z_1}{z_2}\Big)^{\frac{1}{2}}\,
\As_{1}(xq)\As_{2}(xq^{-1})-
\Big(\frac{z_2}{z_1}\Big)^{\frac{1}{2}}\,\As_{1}(xq^{-1})\As_{2}(xq), 
 \label{sl21-rel2-A} 
\end{eqnarray}
and 
\begin{eqnarray}
c_{13}\As_{1}(x)\As_{3}(x)&=&
 \left(\frac{z_{1}}{z_{3}}\right)^{\frac{1}{2}} \overline{\As}_{2}(xq)-
 \left(\frac{z_{3}}{z_{1}}\right)^{\frac{1}{2}}
 \overline{\As}_{2}(xq^{-1}), 
 \label{sl21-rel3-A} \\[.2cm]
c_{23}\As_{2}(x)\As_{3}(x)&=&
\left(\frac{z_{2}}{z_{3}}\right)^{\frac{1}{2}} \overline{\As}_{1}(xq)-
 \left(\frac{z_{3}}{z_{2}}\right)^{\frac{1}{2}}
 \overline{\As}_{1}(xq^{-1}), 
 \label{sl21-rel4-A}  \\[.2cm]  
c_{13}\overline{\As}_{1}(x)\overline{\As}_{3}(x)&=&
\left(\frac{z_{1}}{z_{3}}\right)^{\frac{1}{2}}{\mathsf f}(q x)
 \As_{2}(xq^{-1})-
\left(\frac{z_{3}}{z_{1}}\right)^{\frac{1}{2}}{\mathsf f}(q^{-1} x)
 \As_{2}(xq),   
 \label{sl21-rel5-A} \\[.2cm]
c_{23}\overline{\As}_{2}(x)\overline{\As}_{3}(x)&=&
\left(\frac{z_{2}}{z_{3}}\right)^{\frac{1}{2}} {\mathsf f}(q x)
\As_{1}(xq^{-1})-
\left(\frac{z_{3}}{z_{2}}\right)^{\frac{1}{2}} {\mathsf f}(q^{-1} x)
\As_{1}(xq),  
\label{sl21-rel6-A}
\end{eqnarray}
The eigenvalues ${\mathsf T}_m^{(1)}(x)$ of the ${\bf T}$-operators 
\eqref{atyp} can be written in a similar form 
\be
c_{12} \,{\mathsf T}^{(1)}_m(x)=
c_{13}\, z_1^{m+\frac{1}{2}}\,\As_1(x q^{m+\frac{1}{2}})\, 
\overline{\As}_1(x q^{-m-\frac{1}{2}}) 
-c_{23}\, z_2^{m+\frac{1}{2}}\, 
\As_2(x q^{m+\frac{1}{2}})\, 
\overline{\As}_2(x q^{-m-\frac{1}{2}})\label{atyp-e} \ .
\ee

The eight relations (\ref{Q-rel1-A})-- (\ref{sl21-rel6-A})
can be solved in different ways. For example, for the coefficients of
the polynomials \eqref{q-factor1}. Altogether there are exactly $3L$ 
unknown coefficients. Let us count the number of equations.  
Begin by dropping the last four relation among the eight, since they are
simple corollaries of the first four. The remaining four relations are
of the degrees $L$, $L$, $L+m_3$  and $L-m_3$, respectively. They  
are trivially satisfied at $x=0$, so we are left with only $4L$ equations. 
It is easy to see that $L$ of them are dependent. For example,
expressing ${\mathsf A}_1(x)$ and ${\mathsf A}_2(x)$ from
\eqref{Q-rel1-A} and \eqref{Q-rel2-A} and substituting them into
\eqref{sl21-rel2-A} one immediately concludes that the RHS of 
\eqref{sl21-rel1-A} is always divisible by $\mathsf{f}(x)$. So that
$L$ out of $4L$ remaining equation are automatically satisfied. Thus
the there are exactly $3L$ polynomial equations for $3L$ unknown
coefficients.

The equations (\ref{Q-rel1-A}-\ref{sl21-rel6-A})
can also be solved for the zeroes of the polynomials \eqref{q-factor1}.
Let us rewrite these equation in the Bethe Ansatz form. Substitute
$x=q^{-\hf}x^{(1)}_\ell$ into \eqref{Q-rel1-A}  and 
$x=q^{+\hf}x^{(1)}_\ell$ into \eqref{Q-rel2-A} and then divide the
resulting two equation, 
\be
c_{12}\mathsf{f}(x^{(1)}_\ell)=-c_{23}\, z_2^{+\hf}
\overline{\As}_2(q^{-1}x^{(1)}_\ell) \As_2(x^{(1)}_\ell),\qquad
c_{12}\mathsf{f}(x^{(1)}_\ell)=-c_{23}\, z_2^{-\hf}
\overline{\As}_2(q^{+1}x^{(1)}_\ell)\As_2(x^{(1)}_\ell),\qquad
\ee
by one another. It follows that 
\be
z_2^{-1}=\frac{\overline{\As}_2(q^{-1}x^{(1)}_\ell)}
{\overline{\As}_2(q^{+1}x^{(1)}_\ell)},\qquad \ell=1,\ldots,m_1\ .
\ee
where $x^{(1)}_\ell$ denotes the zeroes of $\As_1(x)$. Performing
similar manipulation for other zeroes one obtains the complete set of
the Bethe Ansatz type equations.

\smallskip
{Zeroes of $\As_1(x)$}:
\be
-z_1^{-2} z_3 = \frac{\As_1(q^{+2}x^{(1)}_\ell)}{\As_1(q^{-2}x^{(1)}_\ell)}
\frac{\overline{\As}_3(q^{-1}x^{(1)}_\ell)}{\overline{\As}_3(q^{+1}x^{(1)}_\ell)},\qquad 
z_2^{-1}=\frac{\overline{\As}_2(q^{-1}x^{(1)}_\ell)}
{\overline{\As}_2(q^{+1}x^{(1)}_\ell)},\qquad \ell=1,\ldots,m_1\ ,\label{B3-1}
\ee

\smallskip
{Zeroes of $\As_2(x)$}:
\be
-z_2^{-2} z_3 = \frac{\As_2(q^{+2}x^{(2)}_\ell)}{\As_2(q^{-2}x^{(2)}_\ell)}
\frac{\overline{\As}_3(q^{-1}x^{(2)}_\ell)}{\overline{\As}_3(q^{+1}x^{(2)}_\ell)},\qquad 
z_1^{-1}=\frac{\overline{\As}_1(q^{-1}x^{(2)}_\ell)}
{\overline{\As}_1(q^{+1}x^{(2)}_\ell)},\qquad \ell=1,\ldots,m_2\ ,\label{B3-2}
\ee

\smallskip
{Zeroes of $\As_3(x)$}:
\be
z_1^{-1} = \frac{\overline{\As}_1(q^{-1}x^{(3)}_\ell)}
{\overline{\As}_1(q^{+1}x^{(3)}_\ell)},
\qquad
z_2^{-1} = \frac{\overline{\As}_2(q^{-1}x^{(3)}_\ell)}
{\overline{\As}_2(q^{+1}x^{(3)}_\ell)},\qquad \ell =1,\ldots,m_3\ ,\label{B3-3}
\ee

\smallskip
{Zeroes of $\overline{\As}_1(x)$}:
\begin{eqnarray}
 -z_1^{2} z_3^{-1}\, \frac{{\mathsf f}(q^{+1}\ox^{(1)})}
{{\mathsf f}(q^{-1}\ox^{(1)})}
= \frac{\overline{\As}_1(q^{+2}\ox^{(1)}_\ell)}
{\overline{\As}_1(q^{-2}\ox^{(1)}_\ell)}
\frac{{\As}_3(q^{-1}\ox^{(1)}_\ell)}{{\As}_3(q^{+1}\ox^{(1)}_\ell)},
&&  
z_2\,\frac{{\mathsf f}(q^{-1}\ox^{(1)})}
{{\mathsf f}(q^{+1}\ox^{(1)})}=\frac{{\As}_2(q^{-1}\ox^{(1)}_\ell)}
{{\As}_2(q^{+1}\ox^{(1)}_\ell)}, \nonumber \\[.3cm]
&& \qquad \ell=1,\ldots,L-m_1\ ,\label{B3-4}
\end{eqnarray}

\smallskip
{Zeroes of $\overline{\As}_2(x)$}:
\begin{eqnarray}
-z_2^{2} z_3^{-1}\,
\frac{{\mathsf f}(q^{+1}\ox^{(2)})}{{\mathsf f}(q^{-1}\ox^{(2)})} =
 \frac{\overline{\As}_2(q^{+2}\ox^{(2)}_\ell)}
{\overline{\As}_2(q^{-2}\ox^{(2)}_\ell)}
\frac{{\As}_3(q^{-1}\ox^{(2)}_\ell)}{{\As}_3(q^{+1}\ox^{(2)}_\ell)},&& 
z_1 \,\frac{{\mathsf f}(q^{-1}\ox^{(2)})}{{\mathsf f}(q^{+1}\ox^{(2)})}
=\frac{{\As}_1(q^{-1}\ox^{(2)}_\ell)}
{{\As}_1(q^{+1}\ox^{(2)}_\ell)}, \nonumber \\[.3cm]
&& \qquad\ell=1,\ldots,L-m_2\ ,\label{B3-5}
\end{eqnarray}

\smallskip
{Zeroes of $\overline{\As}_3(x)$}:
\begin{eqnarray}
z_1 \,\frac{{\mathsf f}(q^{-1}\ox^{(3)})}{{\mathsf f}(q^{+1}\ox^{(3)})}
= \frac{{\As}_1(q^{-1}\ox^{(3)}_\ell)}
{{\As}_1(q^{+1}\ox^{(3)}_\ell)},
&& 
z_2\,\frac{{\mathsf f}(q^{-1}\ox^{(3)})}{{\mathsf f}(q^{+1}\ox^{(3)})}
 = \frac{{\As}_2(q^{-1}\ox^{(3)}_\ell)}
{{\As}_2(q^{+1}\ox^{(3)}_\ell)}, 
\nonumber \\[.3cm]
 && \qquad \ell =1,\ldots,L-m_3\ ,\label{B3-6}
\end{eqnarray}

All factors $z_{1},z_{2},z_{3},$ in the above equations 
can be absorbed into redefined eigenvalues 
\be
\Qs_i(x)=x^{+\mathsf{S}_i}\,\As_i(x),\qquad
\overline{\Qs}_i(x)=x^{-\mathsf{S}_i}\,\overline{\As}_i(x),
\ee
The equations (\ref{B3-1})-(\ref{B3-6}) then become identical to 
the system \eqref{BAijk}, quoted in the introduction. There are six 
self-contained sets of the Bethe Ansatz 
equations involving only subsets of zeros, belonging to
any of the six pairs of the eigenvalues $(\As_i(x),
\overline{\As}_j(x))$, \ $i\not=j$.  
Once any such pair is determined, the remaining eigenvalues can 
be easily found by using the above functional 
equations\footnote{
For instance, if $(\As_1(x),
\overline{\As}_2(x))$ is known then: (i) $\As_3(x)$ and
$\overline{\As}_3(x)$ are explicitly expressed from
\eqref{sl21-rel3-A} and \eqref{sl21-rel6-A}, (ii) 
\eqref{sl21-rel1-A} and \eqref{sl21-rel2-A} then become linear equations
for the coefficients of $\As_2(x)$ and
$\overline{\As}_1(x)$.}.  

Finally note that for small chains the 
eigenvalues can, of course, be found 
from direct diagonalization the of ${\bf Q}$-operators (thanks to that we
now have their explicit definitions \eqref{Q-lat} and \eqref{Qb-lat}).   
As an illustration consider the simplest, but still interesting, case 
of a $1$-site chain, $L=1$. Evaluating the trace in 
\eqref{Q-lat} and \eqref{Qb-lat} one obtains the operators 
$\As_i(x)$ and $\overline{\As}_i(x))$ in the form of diagonal 3x3
matrices. We list their eigenvalues below.

\begin{enumerate}
\item
$m_1=1$, $m_2=0$, $m_3=0$, 
$z_1=q^{\f_1}$, $z_2=q^{\f_2+1}$, $z_3=q^{\f_1+\f_2+1}$,
\be
\As_1(x)=1-x
\frac{(z_{1}-z_{2})(q^{-2}z_{2}-1)}
 {(z_{1}-q^{-2}z_{2})(z_{2}-1)},\qquad
 \As_2(x)=1,\qquad \As_3(x)=1,
\ee
\be
\overline{\As}_1(x)=1,\qquad 
\overline{\As}_2(x)=1-q^{-1}x
\frac{z_{1}-z_{2}}{z_{1}-q^{-2}z_{2}},
\qquad 
\overline{\As}_3(x)=1-q^{-1}x
\frac{z_{2}-q^{2}}{z_{2}-1},
\ee
\item
$m_1=0$, $m_2=1$, $m_3=0$, $z_1=q^{\f_1+1}$, $z_2=q^{\f_2}$, 
$z_3=q^{\f_1+\f_2+1}$,
\be
\As_1(x)=1,\qquad 
\As_2(x)=
1-x
\frac{(z_{1}-z_{2})(z_{1}-q^{2})}
 {(z_{1}-q^{2}z_{2})(z_{1}-1)},
\qquad \As_3(x)=1,
\ee
\be
\overline{\As}_1(x)=1-qx
\frac{z_{1}-z_{2}}{z_{1}-q^{2}z_{2}},
\qquad \overline{\As}_2(x)=1,\qquad 
\overline{\As}_3(x)=1-qx
\frac{q^{-2}z_{1}-1}{z_{1}-1},
\ee
\item
$m_1=0$, $m_2=0$, $m_3=1$, $z_1=q^{\f_1+1}$, $z_2=q^{\f_2+1}$, 
$z_3=q^{\f_1+\f_2+2}$,
\be
\As_1(x)=1,\qquad \As_2(x)=1,\qquad
 \As_3(x)=1-x
\frac{(q^{-2}z_{1}-1)(z_{2}-q^{2})}
 {(z_{1}-1)(z_{2}-1)},
\ee
\be
\overline{\As}_1(x)=1-qx
\frac{q^{-2}z_{2}-1}{z_{2}-1},\qquad 
\overline{\As}_2(x)=1-q^{-1}x
\frac{z_{1}-q^{2}}{z_{1}-1},\qquad 
\overline{\As}_3(x)=1,
\ee
\end{enumerate} 
It is easy to check that these eigenvalues 
satisfy all the functional and Bethe Ansatz type equations given
above, as they, of course, should do.

\nsection{Applications in continuous quantum field theory.}\label{cqft}

In this section, we explain how the general results of the previous
sections can be specialized to the problems of the continuous quantum
field theory in two dimensions. 

\subsection{${\bf T}$- and ${\bf Q}$-operators in conformal field theory}\label{conformal}
The Borel subalgebra ${\cal B}_-$ of $U_q(\widehat{sl}(2|1))$, defined
  after \eqref{perm}, can be 
realized with two chiral Bose fields
\be
\phi^{(k)}(u)= X^{(k)}+  P^{(k)} u +
\sum_{n\not=0}\frac{a^{(k)}_{-n}}{n}\,e^{i u  n},\qquad   
k=1,2,\label{bose}
\ee
where $ P^{(k)}$ and $ X^{(k)}$ and $a^{(k)}_n$\ 
($n=\pm1,\pm2, \pm3,\ldots$)
are operators which satisfy the commutation relations of the
Heisenberg algebra,
\be
[{X}^{(k)},{P}^{(l)}]=i\, \delta_{kl},\qquad [a_m^{(k)},a_n^{(l)}]
=n\,\delta_{kl}\,\delta_{m+n,0}\ .
\ee
The variable $u$ is interpreted as the coordinate on the 2D cylinder
of the circumference $2\pi$. The field $\boldsymbol
{\phi}(u)=(\phi^{(1)}(u),\phi^{(2)}(u))$ is a quasi-periodic
function of $u$,
\be
\boldsymbol\phi(u+2\pi)=\boldsymbol\phi(u)+2\pi{\bf P},\qquad 
{\bf P}=(P^{(1)},P^{(2)})\ .
\ee
Let ${\boldsymbol \alpha}_1,{\boldsymbol \alpha}_2,{\boldsymbol
  \alpha}_3$ be 2-dimensional vectors  
\be
{\boldsymbol \alpha}_0=(\beta,\gamma),\qquad
{\boldsymbol \alpha}_1=(-2\beta\,,0),\qquad
{\boldsymbol \alpha}_2=(\beta,-\gamma),\qquad {\boldsymbol
  \alpha}_0+{\boldsymbol \alpha}_1+{\boldsymbol \alpha}_2=0, 
\ee
where $\beta$ and $\gamma$ are real constants, such that
\be
\beta^2+\gamma^2=1\ . 
\ee
Introduce the vertex operators 
\be
V_j(u)=:\,e^{i\,{\boldsymbol \alpha}_j{\boldsymbol\phi}(u)}\,: \qquad j=0,1,2,
\ee
where 
${\boldsymbol \alpha}_{j}{\boldsymbol\phi}(u)=
\alpha^{(1)}_{j}\,\phi^{(1)}(u)+\alpha^{(2)}_{j}\,\phi^{(2)}(u)$ 
denote the Euclidean scalar product of two-dimensional vectors 
${\boldsymbol \alpha}_{j}=(\alpha^{(1)}_{j},\alpha^{(2)}_{j})$ and 
${\boldsymbol\phi}(u)$, 
and the symbol $:\,\ldots\,:$ denotes the following normal 
ordering 
\be
:\,e^{i\,{\boldsymbol \alpha}_j\,{\boldsymbol\phi}(u)}\,:\equiv\,
\exp\big(i\,\sum_{n=1}^\infty{{\boldsymbol \alpha}_j\, {\bf
    a}_{-n}\over n}e^{inu}\big) 
\exp\big(i\,{\boldsymbol \alpha}_j({\bf X}+{\bf P}u)\big)
\exp\big(-i\sum_{n=1}^\infty{{\boldsymbol \alpha}_j\, {\bf a}_n\over
  n}e^{-inu}\big), 
\ee
where ${\bf a}_n=(a^{(1)}_n,a_n^{(2)})$, ${\bf X}=(X^{(1)},X^{(2)})$. 
It is easy to show that 
\be
{\bf P}\, V_j(u)=V_j(u)\,({\boldsymbol \alpha}_j+{\bf P}),\label{PV-comm}\ee
and 
\be
V_{i}(u)V_{j}(v)=(-1)^{\mathsf{p}(i)\mathsf{p}(j)}q^{-a_{ij}}V_{j}(v)V_{i}(u) 
\qquad {\rm for} \quad u>v
\ee
where the Cartan matrix and the parity ${\mathsf p}(j)$ 
are the same as in \eqref{cartan} and \eqref{parity} and 
\be
q=e^{-2i\pi\beta^2}.\label{q-cft}
\ee
Let ${\cal F}_{\bf p}$ be the Fock space generated by the free action 
of the operators $a^{(k)}_n$,\ with $n<0$, $k=1,2$  on 
the vacuum state $|{\bf p}>$, \ ${\bf p}=(p^{(1)},p^{(2)})$, defined
as  
\be
{\bf P}\,|{\bf p}>={\bf p}\,|{\bf p}>,\qquad {\bf a}_n\,|{\bf p}>=0, \qquad 
n>0.\label{vac}
\ee
The vertex operators act invariantly in the extended Fock space,  
\be
V_j(u): \quad 
\hat{\cal F}_{{\bf  p}}\to \hat{\cal F}_{{\bf  p}},\qquad 
\hat{\cal F}_{{\bf  p}}=\mathop{\bigoplus}_{(n_1,n_2,n_3)\in
  {\mathbb Z}^3}\  {\cal F}_{{\bf  p+n_1{\boldsymbol \alpha}_1+n_2
{\boldsymbol \alpha}_2+n_3{\boldsymbol 
      \alpha}_3}}\ .
\label{ext-sp}
\ee  
This space support the action of the Borel subalgebra ${\cal
  B}_-\subset U_q(\widehat{sl}(2|1))$, if one realizes its generators
as
\be
h_0=\frac{P^{(1)}}{2\beta}-\frac{P^{(2)}}{2\gamma},\qquad
h_1=-\frac{P^{(1)}}{\beta},\qquad
h_2=\frac{P^{(1)}}{2\beta}+\frac{P^{(2)}}{2\gamma},
\qquad h_0+h_1+h_2=0,\label{hh-def}
\ee
\be
f_j=-\frac{(-1)^{{\mathsf p}(j)}}{q-q^{-1}}\, \int_0^{2\pi}\, V_j(u)
du,\qquad j=0,1,2.\label{ff-def}
\ee
The commutation relations \eqref{acom} between \eqref{hh-def}
and \eqref{ff-def} trivially follow from \eqref{PV-comm}. The
generators \eqref{ff-def} satisfy the Serre relations \eqref{aser1} and
\eqref{aser3}. To see this one needs to rewrite the products of
$f_j$'s  in the form of {\em ordered} integrals. An example of such
calculations is given in the Appendix~\ref{app-R-expan}. 

Consider now the  specialization of the reduced universal
$R$-matrix \eqref{R-expan} to the representation 
\eqref{ff-def} in the quantum space (it is the second space in
\eqref{imbed}). It has an extremely elegant form 
\be
\overline{{\cal L}}=\overline{{\cal R}}_{\vert_{f_j=\rm{\eqref{ff-def}}}}=
{\cal P}\ \exp\Big(\int_0^{2\pi}\ {\cal Z}(u)du\Big),\label{zz-def}
\ee
where 
\be
{\cal Z}(u)=e_0\otimes_s V_0(u)+e_1\otimes_s V_1(u)+e_2\otimes_s 
V_2(u)\ .
\ee
It was first discovered in \cite{FL93} for finite-dimensional quantized
algebras
and then generalized in \cite{BLZ96} for the case of 
the quantum affine algebra $U_q(\widehat{sl}(2))$. A
proof can be found in \cite{BHK}. For the consistency of the fermionic
grading we also assume\footnote{The only effect of the extra signs 
arising from this relation is the negation of the spectral
parameter $x$.}
\be
(e_{i}\otimes_{s} V_{i}(u))(e_{j}\otimes_{s} V_{j}(v))
=(-1)^{{\mathsf p}(i){\mathsf p}(j)}
(e_{i}e_{j}\otimes_{s} V_{i}(u) V_{j}(v))\ .\label{Zu-def}
\ee
It follows then 
that the following ``universal'' ${\bf L}$-operator 
\be
{\cal L}={\cal P}\ \exp\Big(\int_0^{2\pi}\ {\cal Z}(u)du\Big)
\, q^{\cal K},\qquad {\cal
K}=-\frac{P^{(1)}}{2\beta}(h_0+h_2) +
\frac{P^{(2)}}{2\gamma}(h_2-h_0),
\label{L-uni} 
\ee
obtained from \eqref{zz-def} and \eqref{R-red} 
satisfy the Yang-Baxter equation
\be
{\cal R}_{12}\,{\cal L}_1\ {\cal L}_2 =
{\cal L}_2 \,{\cal L}_1\,{\cal R}_{12}\  .\label{ybe-cft}
\ee
where ${\cal R}_{12}$ is the universal $R$-matrix.
 The operator \eqref{L-uni} is an element of the Borel subalgebra
 ${\cal B}_+$ whose coefficients are operators acting in the
 quantum space \eqref{ext-sp}. 

Now we can define commuting 
${\bf T}$- and ${\bf Q}$-operators, acting in the Fock space of the
Bose fields \eqref{bose}, using our universal formulae \eqref{t-uni3},
\eqref{q-def} and \eqref{normali2}.
For the reason explained below we make a special choice of the
external parameters $\f_{1,2}$ in \eqref{zs-def} such that 
\be
z_1=q^{P^{(2)}/\gamma+P^{(1)}/\beta},\qquad
z_2=q^{P^{(2)}/\gamma-P^{(1)}/\beta},\qquad z_3=z_1 z_2\label{zz-c}
\ee
The definition \eqref{t-uni3} then become%
\footnote{Note that the choice \eqref{zz-c} leads to an extra
  factor $q^{\cal K}$ in \eqref{tt-def}
in addition to the one which comes from universal
  $R$-matrix.} 
\begin{eqnarray}
\hspace{-20pt}
\TT^{(CFT)}_\mu(x) &=&
{\str}_{\pi_\mu(x)}\Big[\, z_1^{-h_0}\, z_2^{h_2} \overline{\cal
  L}\,\Big]
 \\[.3cm]
&=&
{\str}_{\pi_\mu(x)}\Big[q^{2{\cal K}}{\cal P} \exp
\Big\{ \int_0^{2\pi} 
(e_0\otimes_s V_0(u)+e_1\otimes_s V_1(u)+e_2\otimes_s V_2(u))d u
\Big\}\Big]\label{tt-def}  \nonumber 
\end{eqnarray}
where ${\cal K}$ is given by \eqref{L-uni}. 
An important case is the 3-dimensional representations
\be
\TT^{(CFT)}(x)\equiv\TT_{(1,0,0)}^{(CFT)}(x),\qquad 
\overline{\TT}^{(CFT)}(x) 
 \equiv -\TT_{(-1,-1,0)}^{(CFT)}(x).
\label{dim3-CFT}
\ee
Similarly define the corresponding ${\bf Q}$-operators 
\begin{eqnarray}
\A_i^{(CFT)}(x)&=&Z_i^{-1}{\str}_{\rho_i(x)}
\Big[q^{2{\cal K}}{\cal P} \exp
\Big\{ \int_0^{2\pi} {\cal Z}(u) d u \Big\}\Big],\nonumber\\[.3cm]
\overline\A_i^{(CFT)}(x)&=&\overline{Z}^{-1}_{i}{\str}_{\overline\rho_i(x)}
\Big[q^{2{\cal K}}{\cal P} \exp
\Big\{ \int_0^{2\pi} {\cal Z}(u) d u \Big\}\Big],\label{qq-def}
\end{eqnarray}
where $i=1,2,3$, ${\cal Z}(u)$ is defined in \eqref{Zu-def} and 
$Z_i, \overline{Z}_i$ are given by \eqref{normali1} with $z_1$, $z_2$
and $z_3$ defined in \eqref{zz-c}. 

For the choice \eqref{zz-c} all 
the operators \eqref{tt-def} and \eqref{qq-def} commute with 
an infinite series of the local integral of motion (LIM), ${\mathbb
  I}_n$, $n=1,2,3,\ldots,\infty$,
\be
[\,{\mathbb I}_n, \TT^{(CFT)}_\mu(x)]=
[\,{\mathbb I}_n, \A^{(CFT)}_\mu(x)]=
[\,{\mathbb I}_n, \overline{\A}^{(CFT)}_\mu(x)]=0
\ee
of a two-dimensional CFT which arise 
in quantization of the AKNS soliton hierarchy \cite{Fateev-Lukyanov2005}.
They are defined as integrals 
\be
{\mathbb I}_n=\int_0^{2\pi}\ W_{n+1}(u) \, du\ .\label{LIM}
\ee
where the local densities are $W_{n+1}(u)$ are some polynomials 
in the $u$-derivatives of the Bose fields \eqref{bose} (the $k$-th
density $W_k(u)$ is of the degree $k$ in the derivative $\partial_u$).
The first non-trivial densities read \cite{Fateev-Lukyanov2005},
\begin{eqnarray}
W_2(u)&=& -\frac{1}{2}(\partial_{u}\phi^{(1)}(u))^{2}
-\frac{1}{2}(\partial_{u}\phi^{(2)}(u))^{2}+
\frac{1}{\sqrt{2n}}\partial_{u}^{2}\phi^{(1)}(u), \\[.3cm]
W_3(u)&=&\frac{6n+4}{6 \sqrt{2}}(\partial_{u}\phi^{(1)}(u))^{3}+
\frac{n}{\sqrt{2}}(\partial_{u}\phi^{(1)}(u))^{2}\partial_{u}\phi^{(2)}(u)+
\frac{n\sqrt{n}}{2}\partial_{u}^{2}\phi^{(1)}(u)\partial_{u}\phi^{(2)}(u)
\nonumber \\
&& \hspace{30pt} -
\frac{(n+2) \sqrt{n}}{2}
\partial_{u}\phi^{(1)}(u)\partial_{u}^{2}\phi^{(2)}(u)+
\frac{n+2}{6\sqrt{2}}\partial_{u}^{3}\phi^{(2)}(u),
\end{eqnarray}
where $\sqrt{n}=i\sqrt{2}/\beta$ \ is a parameter used in  
\cite{Fateev-Lukyanov2005} instead of our $\beta$ \ in \eqref{q-cft}. 
Note also that their chiral Bose 
fields $X(u)$ and $Y(u)$
are related to \eqref{bose} as $X(u)=\phi^{(1)}(u)/\sqrt{2}$ and 
$Y(u)=\phi^{(2)}(u)/\sqrt{2}$.

The CFT versions of the ${\bf T}$-  and ${\bf Q}$-operators defined
above act invariantly in the
single Fock space  ${\cal F}_{\bf p}$ (even though the ${\bf
  L}$-operator \eqref{L-uni} acts in the extended space
\eqref{ext-sp}). They satisfy all
the functional relation given in Sections~3.1--3.3, without any
modifications\footnote{Let us stress that in the considered case of CFT 
no additional scalar factors arise in the functional relations
(unlike the lattice models).}. 
One just need to supply the superscript $(CFT)$ to all
the ${\bf T}$-  and ${\bf Q}$-operators and identify the quantities
$z_1,z_2,z_3$ therein with those defined in \eqref{zs-def}.
Note that the vacuum state  
\eqref{vac} is an eigenstate for all these operators 
\be
\TT^{(CFT)}_\mu(x)|{\bf p}>={\mathsf T}^{(vac)}_\mu(x)|{\bf p}>,
\ee
\be
\A^{(CFT)}_i|{\bf p}>={\mathsf A}^{(vac)}_i(x)|{\bf p}>,\quad
\overline{\A}^{(CFT)}_i(x)|{\bf p}>
=\overline{{\mathsf A}}^{(vac)}_i(x)|{\bf p}>\ .
\ee
The eigenvalues satisfy the functional relations 
\begin{eqnarray}
c_{12}  
&=&c_{13}\,z_1^{+\frac{1}{2}}\, \As^{(vac)}_1(q^{+\frac{1}{2}} x) \,
\overline{\As}^{(vac)}_1(q^{-\frac{1}{2}}x)-  
c_{23}\, z_2^{+\frac{1}{2}}\, 
\As^{(vac)}_2(q^{+\frac{1}{2}} x)\, 
\overline{\As}^{(vac)}_2(q^{-\frac{1}{2}}x),
\label{CFT-rel1}\\[.2cm]
c_{12}
&=& c_{13}\,z_1^{-\frac{1}{2}}\, 
 \As_1^{(vac)}(q^{-\frac{1}{2}}x) 
\,\overline{\As}^{(vac)}_1(q^{+\frac{1}{2}} x)-  
c_{23}\,z_2^{-\frac{1}{2}}\,  
\As_2^{(vac)}(q^{-\frac{1}{2}}x)\, 
\overline{\As}_2^{(vac)}(q^{+\frac{1}{2}} x)\label{CFT-rel2}
\end{eqnarray}
which are corollaries of \eqref{Q-rel}.

The operators \eqref{tt-def} and \eqref{qq-def} 
understood as series in spectral parameter $x$. 
The first nontrivial terms in their expansions can be obtained from
the third order term in the expansion of the universal $R$-matrix
given in the Appendix~\ref{app-R-expan},
\begin{eqnarray}
\TT^{(CFT)}(x)&=&z_1+z_2-z_3+xG_{1}+O(x^{2}),
\\[.2cm]
\overline{\TT}^{(CFT)}(x)
&=&-z_{3}^{-1}+z_{2}^{-1}+z_{1}^{-1}-
 x(z_{1}z_{2}z_{3})^{-1}\overline{G}_{1}+O(x^{2}),
 \nonumber 
\end{eqnarray}
where $\TT^{(CFT)}(x)$ and $\overline{\TT}^{(CFT)}(x)$ are defined in
\eqref{dim3-CFT}. The quantities $G_{1}$ and 
$\overline{G}_{1}$ are {\em non-local} integrals of
motion defined as linear
combinations of ordered integrals of the vertex operators,
\begin{eqnarray}
G_1&=& z_{1}J(1,2,0)+z_{2}J(2,0,1)-z_{3}J(0,1,2) \\[.3cm]
\overline{G}_1&=& -z_{3}z_{2}J(0,2,1)-z_{1}z_{3}J(1,0,2)+z_{2}z_{1}J(2,1,0),
\end{eqnarray}
where
\begin{eqnarray}
J(i_{1},i_{2},\cdots,i_{n})=
\int_{u_{1}\ge u_{2} \ge \cdots \ge u_{n}}
 V_{i_{1}}(u_{1}) V_{i_{2}}(u_{2}) \cdots V_{i_{n}}(u_{n})
du_{1}du_{2} \cdots du_{n}.
\end{eqnarray}

Similarly for the ${\bf Q}$-operators \eqref{qq-def} one has 
\be
\begin{array}{rcl}
 \overline{\A}^{(CFT)}_{1}(x)&=&\ds 1-
\frac{q^{\frac{3}{2}}x(q^{-1}z_{1}G_{1}+\overline{G}_{1})}
{(q-q^{-1}) (q^{2}z_{2}-z_{1})(z_{3}-z_{1})}+O(x^{2}),  \\[.3cm]
 \overline{\A}^{(CFT)}_{2}(x)&=&\ds 1-
\frac{q^{\frac{3}{2}}x(q^{-1}z_{2}G_{1}+\overline{G}_{1})}
{(q-q^{-1}) (q^{2}z_{1}-z_{2})(z_{3}-z_{2})}+O(x^{2}), \\[.3cm]
 \overline{\A}^{(CFT)}_{3}(x)&=&\ds 1-
\frac{q^{-\frac{1}{2}}x(qz_{3}G_{1}+\overline{G}_{1})}
{(q-q^{-1}) (z_{1}-z_{3})(z_{2}-z_{3})}+O(x^{2}), \\[.3cm]
 \A^{(CFT)}_{1}(x)&=&\ds 1+
\frac{q^{-\frac{3}{2}}x(qz_{1}G_{1}+\overline{G}_{1})}
{(q-q^{-1}) (q^{-2}z_{2}-z_{1})(z_{3}-z_{1})}+O(x^{2}), \\[.3cm]
 \A^{(CFT)}_{2}(x)&=&\ds 1+
\frac{q^{-\frac{3}{2}}x(qz_{2}G_{1}+\overline{G}_{1})}
{(q-q^{-1}) (q^{-2}z_{1}-z_{2})(z_{3}-z_{2})}+O(x^{2}), \\[.3cm]
\A^{(CFT)}_{3}(x)&=&\ds 1+
\frac{q^{\frac{1}{2}}x(q^{-1}z_{3}G_{1}+\overline{G}_{1})}
{(q-q^{-1}) (z_{1}-z_{3})(z_{2}-z_{3})}+O(x^{2}) \ .
\end{array}\label{A-def}
\ee

Note, that the ${\bf T}$-operators similar to \eqref{tt-def} but with
a different 
realization of the vertex operators (through two Bose and two Fermi
free fields) were introduced in \cite{KZ06} 
in connection with a CFT with an extended (super)symmetry
which arises in quantization
of a supersymmetric extension of the KdV hierarchy 
(also related 
with  the $U_q(\widehat{sl}(2|1))$ algebra). As stated in
\cite{KZ06} their analog of the ${\bf L}$-operator \eqref{L-uni} 
satisfy the same Yang-Baxter equation \eqref{ybe-cft}.
Our results then imply that their ${\bf T}$-operators obeys exactly the same 
functional relations as in Section~3, provided
one uses the same definitions \eqref{tt-def},
but with their realization of the vertex operators
and also define by
\eqref{qq-def} the corresponding ${\bf Q}$-operators, 
which were not considered in \cite{KZ06}.
It would be interesting to further clarify these connections.

\subsection{Connections with the spectral theory of differential
 equations.}
Here we briefly illustrate the remarkable correspondence between the spectral 
theory of the Schr\"odinger equation
and the integrable structure of the conformal field 
theory, which attracted much attention recently 
\cite{Voros94,DT,BLZ98,S01,DDMST06}.
This correspondence relates some spectral characteristics of certain ordinary  
differential equations (and, more generally, integro-differential
equations \cite{DDMST06}) 
to the eigenvalues of the continuous analogs of the Baxter's
$\bf Q$-operators in quantum field theory with the conformal symmetry 
(in general, with an extended (super)conformal symmetry). 

The relevant differential equation in our case is, in fact, a
one-dimensional \SE\ on the half-line,  
\begin{eqnarray}
\Big\{
-\frac{d^{2}}{dy^{2}} +\frac{\ell(\ell+1)}{y^{2}}+r\,y^{\alpha
  -1}+y^{2\alpha}-E\Big\}\,\Psi(y)&=&0, \qquad 0<y<+\infty. \label{SFL}
\end{eqnarray}
with real $0<\alpha<\infty$ and arbitrary $r$ and $\ell$.  
It was first considered in the case 
$l=0$ 
and $\alpha>0$ by Suzuki \cite{S01}, who pointed out its connection to
the quantum affine superalgebra $U_{q}(\hat{sl}(2|1))$ 
and to the corresponding Bethe Ansatz equations. 
Eq.(\ref{SFL}) with $l \ne 0$ appeared in \cite{DDT01}. 
The full
equation with arbitrary values of $\ell$ and $r$, in the regime $\alpha<-1$,  
was recently considered
in \cite{Fateev-Lukyanov2005} 
in connection with the quantization of the integrable AKNS soliton
hierarchy. 

Here we consider the case $\alpha>0$ with arbitrary values of $\ell$ and $r$.
To simplify our considerations we will assume that $\alpha>1$, however
the results apply to the full range $0<\alpha<\infty$. 
For $\Re\ \ell>-\frac{1}{2}$, \  Eq.\eqref{SFL} 
has a unique solution, satisfying the condition
\be
\psi(y,E,r,\ell)=\Big({{2\over 
\alpha+1}}\Big)^{\frac{2\ell+2+\alpha}{2(\alpha+1)}}\ 
\Gamma\Big(-\frac{2\ell+1}{\alpha+1}\Big)\ 
y^{\ell+1}
 +O(y^{\ell+3}),\ \ \ \
{\rm as}\ \ \ \ \ \ y\to 0\ . 
\ee
This solution can be analytically continued outside the domain
$\Re\ \ell>-\frac{1}{2}$. Obviously, the  
function $\psi(y,E,r,-\ell-1)$, defined in this way, satisfy the same 
equation \eqref{SFL} and for generic values of $\ell$ the two solutions 
\begin{equation} 
\psi_1(y)=\psi(y,E,r,\ell),\qquad
\psi_2(y)=\psi(y,E,r, -\ell-1),\label{basisdiff}
\end{equation}	
are linearly independent, since
\begin{equation}	
(4\pi i)^{-1}\,{\rm Wr} \left[\psi_1,\psi_2\right]
=\big(\,q^{\ell+\frac{1}{2}}-q^{-\ell-\frac{1}{2}}\,\big)^{-1}\ ,\label{psi-wr}
\end{equation}	
where ${\rm Wr}[f,g]=f\partial_y g-\partial_y f g$\   denotes the usual
Wronskian and 
\be
q=\exp\Big(\frac{2\pi i}{\alpha+1}\Big)\ .\label{q-def}
\ee
{}From now on we will make the $\ell$-dependence implicit,
considering $\ell$ as a fixed parameter.
Further, for all values of $E$ the equation \eqref{SFL} has a unique 
solution $\chi(y,E,r)$ 
which decays at $y\to+\infty$. We normalize this solution as
\be
\chi(y,E,r) \to 
\Big({{2\over\alpha+1}}\Big)^{-\frac{r}{2(\alpha+1)}}\ 
y^{-\frac{\alpha +r}{2}}\exp (-\frac{y^{\alpha +1}}{\alpha +1}),\qquad
y\to+\infty\ .
\ee
It can be expanded in the basis \eqref{basisdiff}
\begin{equation}	
\chi(y,E,r)=D_2(E,r)\,\frac{\psi_1(y)}
{\Gamma\Big(\frac{\alpha+r-2\ell}{2(\alpha+1)}\Big)}
\,+ 
D_1(E,r)\,\frac{\psi_2(y)}
{\Gamma\Big(\frac{\alpha+r+2\ell+2}{2(\alpha+1)}\Big)}\, 
, \label{chi}
\end{equation}
where the connection coefficients $D_{1,2}(E,r)$, which are entire
functions of $E$, are of our primary
interest. They normalized by the condition 
\be
D_{1,2}(E,r)=1+O(E),\qquad E\to0\ .\label{R-norm}
\ee
This follows from the fact that for $E=0$ the substitution 
\be
\Psi(y)=y^{\ell+1}\, \exp\Big({-\frac{y^{\alpha+1}}{\alpha+1}}\Big)\,
w\Big(\,\frac{2 \,y^{\alpha+1}}{\alpha+1}\,\Big)
\ee
brings Eq.\eqref{SFL} to the Kummer equation 
\be
z\,\frac{d^2}{dz^2}\,w(z)+(b-z)\,\frac{d}{dz}\,w(z)-a \,w(z)=0
\ee
where 
\be
a=\frac{\alpha+r+2\ell+2}{2(\alpha+1)},\qquad
b=\frac{\alpha+2\ell+2}{\alpha+1}\ .
\ee
Eq.\eqref{chi} with $E=0$ reduces to the 
relation between Kummer's functions given in $\S$13.1.3 of
ref.\cite{Abramowitz}. 

The connection coefficients $D_{1,2}(E,r)$  in \eqref{chi} can be
interpreted as the spectral determinants. Indeed, at certain isolated 
values of $E$ one of the solutions \eqref{basisdiff} will decay for
$x\to+\infty$ and, thus, becomes proportional to $\chi(y,E,r)$. 
One of the terms in the RHS of \eqref{chi} then vanish.
Let $\{E_n^{(i)}(r)\}_{n=1}^\infty$, $i=1,2$ denotes
ordered spectral sets such that
\be
\psi_1(y,E_n^{(1)}(r))\to 0,\qquad \psi_2(y,E_n^{(2)}(r))\to0,\qquad
y\to\infty\ .\label{s-cond}
\ee
It is easy to see then that
\be
D_{i}(E,r)=\prod_{n=1}^\infty
\Big(1-\frac{E}{E_n^{(i)}(r)}\Big),\qquad i=1,2\ .\label{Det-def}
\ee
Simple WKB analysis shows that at large $n$ the eigenvalues 
$E_n^{(1,2)}(r)$ \ accumulate along positive real axis and that
\be
E_n^{(1,2)}(r)\sim n^{\frac{2\alpha}{\alpha+1}},\qquad n\to\infty \ .
\ee
Therefore for $\alpha>1$ the infinite products \eqref{Det-def}
converge as written. It follows then
\be
\log D_{1,2}(E,r)\simeq \mbox{const} (-E)^{\frac{\alpha+1}{2\alpha}}, 
\qquad E\to\infty, \qquad |\arg (-E)|<\pi\ .\label{d-ass}
\ee
Strictly speaking, the spectral conditions \eqref{s-cond} 
only define \eqref{Det-def} up to the multiplication by an 
entire function without zeroes.
However, comparing \eqref{d-ass} with 
the large $E$ asymptotics, which follows from the 
quasi-classical approximation to \eqref{chi}, one concludes 
that this function is a constant and then from \eqref{R-norm} that it is equal
to one. 

The spectral determinants $D_{1,2}(E,r)$  
satisfy certain functional equation, which we will now derive. 
The key observation is that Eq.\eqref{SFL} is invariant under the
transformation  
\be
\hat\Omega: \qquad y\to q^{1/2}\, y,\quad r\to -r, 
\quad \ell\to \ell,\quad E\to
q^{-1} E,\qquad \label{trans}
\ee
where $q$ is the same as in \eqref{q-def}. Therefore the functions
\be
\chi_k(y)=\big(i q^{-\frac{1}{4}-\frac{r}{4}}\big)^k\ 
\hat\Omega^k\, \big[\,\chi(y,E,r)\,\big],\qquad k=0,1,2,\ldots,\infty , 
\label{chik-def}
\ee
also satisfy \eqref{SFL}. It is easy to check that 
\be
{\rm Wr} \left[\chi_0,\chi_1\right]=2\ .
\ee
The solutions \eqref{basisdiff} are simply transformed under
\eqref{trans},
\be
\hat\Omega\big[\psi_1(y)]=q^{(\ell+1)/2}\, \psi_1(y),
\qquad
\hat\Omega\big[\psi_2(y)]=q^{-\ell/2}\, \psi_2(y)
\ee
Introduce the constants,
\be
(z_1)^{\frac{1}{2}}=i\,
q^{\frac{\ell}{2}+\frac{1}{4}-\frac{r}{4}},\qquad
(z_2)^{\frac{1}{2}}=-i\,
q^{-\frac{\ell}{2}-\frac{1}{4}-\frac{r}{4}},\qquad
(z_3)^{\frac{1}{2}}=(z_1\,z_2)^{\frac{1}{2}}=q^{-\frac{r}{2}}\ ,\label{z-def}
\ee
and also $c_{ij}=(z_i-z_j)/(z_i\,z_j)^{\frac{1}{2}}$,
\be
c_{12}=q^{-\ell-\frac{1}{2}}-q^{+\ell+\frac{1}{2}},\quad
c_{13}=iq^{+\frac{\ell}{2}+\frac{1}{4}+\frac{r}{4}}
+iq^{-\frac{\ell}{2}-\frac{1}{4}-\frac{r}{4}},\quad
c_{23}=-iq^{+\frac{\ell}{2}+\frac{1}{4}-\frac{r}{4}}
-iq^{-\frac{\ell}{2}-\frac{1}{4}+\frac{r}{4}},\quad\label{c-def}
\ee
It follows from \eqref{chi} and \eqref{chik-def}-\eqref{c-def} that 
\be
c_{12}=c_{13}\,z_1^{\frac{1}{2}}\,D_1(q\,E,r)\,D_2(E,-r)-
c_{23}\,z_2^{\frac{1}{2}}\,D_1(E,-r)\,D_2(q\,E,r)\ .\label{D-rel1}
\ee
Negating $r$ in the last relation one also gets 
\be
c_{12}=c_{13}\,z_1^{-\frac{1}{2}}\,D_1(E,r)\,D_2(q\,E,-r)-
c_{23}\,z_2^{-\frac{1}{2}}\,D_1(q\,E,-r)\,D_2(E,r)\ .\label{D-rel2}
\ee
We now want to identify the functions $D_{1,2}(E,\pm r)$ with the vacuum 
eigenvalues $\As_{1,2}^{(vac)}(x)$ and
$\overline{\As}_{1,2}^{(vac)}(x)$ 
of the ${\bf Q}$-operators of the super-conformal field theory considered
in Section~\ref{conformal}. First let us identify the parameters $q$
and $z_1,z_2,z_3$ of this Section defined in \eqref{q-def} and
\eqref{z-def} with those in \eqref{q-cft} and \eqref{zz-c}. 
We expect
the following exact correspondence 
\be
D_i(E,r)=\As_i^{(vac)}(\rho E),\qquad  
D_i(E,-r)=\overline{\As}_{3-i}^{(vac)}(\rho E),\qquad i=1,2,\label{conj}
\ee
where $\rho$ is a scalar factor depending on $\alpha$. 
For an elementary consistency check one can easily verify that these
quantities obey the same normalization conditions \eqref{A-def} and
\eqref{R-norm} and satisfy the identical functional relations
\eqref{CFT-rel1}, \eqref{CFT-rel2} 
and \eqref{D-rel1}, \eqref{D-rel2}. A complete proof 
of \eqref{conj} (and, in particular, the calculation of the constant
$\rho$) requires much deeper considerations which (hopefully) 
will be presented elsewhere. Here we only mention that the vacuum eigenvalues 
of some other commuting operators plays the role of the
Stokes multipliers describing the monodromy properties of the
differential equation near its irregular singular point
$y=\infty$. 
Every three solutions of \eqref{SFL} satisfy a
linear relation, in particular,
\be
\chi_n(y,E,r)=X_n(E,r)\, \chi_0(y,E,r)+ Y_n(E,r) \, \chi_1(y,E,r), \qquad
n \in{\mathbb Z}\ ,
\ee
where 
$\chi_n$ is defined in \eqref{chik-def}. Using \eqref{psi-wr} and
\eqref{chi} and assuming the correspondence \eqref{conj} it is not
difficult to show that 
\be
X_{2k}(E,r)=(z_3)^{\frac{k}{2}}\, 
{\mathsf T}_{-k}^{(1)(vac)}(q^{-k}\,x),\qquad
Y_{2k+1}(E,r)={\mathsf T}_{k}^{(1)(vac)}(q^{-k}\,x),\qquad k\in
{\mathbb Z}
\ee
and (omitting unimportant factors here)
\begin{eqnarray}
X_{2k+1}(E,r)&\sim&
\Big(\frac{z_1}{z_2}\Big)^{\frac{k}{2}} 
\,\overline{\As}_1^{(vac)}(q^{-2k-1}\,x)\, \overline{\As}_2^{(vac)}
(q^{-1}\,x ) 
-
\Big(\frac{z_2}{z_1}\Big)^{\frac{k}{2}} 
\,\overline{\As}_1^{(vac)}(q^{-1}\,x)\, \overline{\As}_2^{(vac)}(q^{-2k-1}\,x) 
\nonumber \\[.3cm]
Y_{2k}(E,r)&\sim&
\Big(\frac{z_1}{z_2}\Big)^{\frac{k}{2}} 
\,{\As}_1^{(vac)}(x)\, {\As}_2^{(vac)}(q^{-2k}\,x ) 
-
\Big(\frac{z_2}{z_1}\Big)^{\frac{k}{2}} 
\,{\As}_1^{(vac)}(q^{-2k}\,x)\, {\As}_2^{(vac)}(x) 
\end{eqnarray}
where $x\equiv\rho E$. It is interesting to note that
\be
Y_2(E)\sim
\overline{\As}_3^{(vac)}(q^{-1}x), \qquad X_3(E)\sim
{\As}_3^{(vac)}(q^{-2}\,x)\ .
\ee


\nsection{Algebraic proof of the functional relations}\label{proof}

In this section we will prove all the functional relations among the
${\bf Q}$-operators and ${\bf T}$-operators, given in
Sect.\ref{funcrel}.  Fortunately, due to symmetry transformations
(see below) a set of functional relations, which require a separate
proof, reduces to only three relations \eqref{sl21-rel1},
\eqref{atypa} and \eqref{dim4a}. Their proof is presented below. 

\subsection{Symmetry transformations}
\label{sect-sym}
The ${\bf T}$- and ${\bf Q}$-operators possess a
number of simple, but important symmetry relations.  Consider the
following automorphism of the algebra $U_q(\widehat{sl}(2|1))$, 
\be
\begin{array}{ll}
\sigma_{02}: \qquad &
\left\{\begin{array}{lll}
e_0\to e_2,\qquad &e_1\to e_1,\qquad &e_2\to e_0,\\[.3cm]
f_0\to f_2,\qquad &f_1\to f_1,\qquad &f_2\to f_0,\\[.3cm]
h_0\to h_2,\qquad &h_1\to h_1,\qquad &h_2\to h_0\ .
\end{array}\right.
\end{array}\label{sigma02a}
\ee
Note that this is an involution $(\sigma_{02})^{2}=1$.
It is easy to see that all the defining relations \eqref{acom},
\eqref{aser1}, \eqref{aser2} are invariant with respect to
\eqref{sigma02a}. 
This transformation also preserves the parities of
the elements of the algebra and the
co-multiplication \eqref{comul1}, 
\be
\Delta(\sigma_{02})=\sigma_{02}\otimes_s\sigma_{02}\ .\label{sig-del}
\ee  
It follows then that the universal $R$-matrix is
invariant with respect to the diagonal action of $\sigma_{02}$,
\be
(\sigma_{02}\otimes_s \sigma_{02})[{\mathcal R}]={\mathcal R}\ .\label{sig-R}
\ee
Further, let the external field parameters $\f_1$ and $\f_2$ 
in \eqref{t-uni1} are also replaced 
\be
\sigma_{02}:\qquad \f_1\to -\f_2,\qquad \f_2\to -\f_1,\label{sigma02b}
\ee
simultaneously with \eqref{sigma02a}. Note that the combined transformation 
\eqref{sigma02a}, \eqref{sigma02b} acts on the operators $z_{1,2,3}$
from \eqref{zs-def} as follows
\be
\sigma_{02}:\qquad z_1\to 1/z_2,\qquad z_2\to 1/z_1,\qquad z_3\to
1/z_3\ .\label{sig-z}
\ee

Further, the substitution $\sigma_{02}$ is also an automorphism of the
Borel subalgebra ${\mathcal B}_+$ (${\mathcal B}_-$) 
and, therefore, transforms its representations into each
other. Namely, for the maps $\rho_i(x)$ and $\overline{\rho}_i(x)$,
introduced in Section~\ref{qopsec}, such transformation 
leads to the following relations\footnote{%
The stated equivalences hold up to certain similarity
transformations of the products of the oscillator algebras
\eqref{Hq-def}, \eqref{F-def}, which does affect the (super)trace.}
\be
\begin{array}{ll}\begin{array}{rclrcl}
\rho_i(x)&\to& \rho_i(x)\cdot\sigma_{02}\simeq
\overline{\rho}_{3-i}(x),\quad&
\rho_3(x)&\to& \rho_3(x)\cdot\sigma_{02}\simeq
\overline{\rho}_{3}(x), 
\\[.3cm] 
\overline{\rho}_i(x)&\to &\overline{\rho}_i(x)\cdot\sigma_{02}\simeq
\rho_{3-i}(x),\quad& 
\overline{\rho}_3(x)&\to &\overline{\rho}_3(x)\cdot\sigma_{02}\simeq
\rho_{3}(x),
\end{array}&\quad i=1,2.\label{sig-rho}
\end{array}
\ee
By definition, the ${\bf Q}$-operators \eqref{Q-def} and
\eqref{Qbar-def} are elements of the Borel subalgebra ${\mathcal
  B}_-$, associated with the quantum space. It is easy to see that the
action of the automorphism $\sigma_{02}$ on this subalgebra (together
with the substitution \eqref{sigma02b}) 
will induce some permutation of the ${\bf
  Q}$-operators. Namely, from \eqref{sig-R}, \eqref{sig-z} and
\eqref{sig-rho} it follows that this action is 
\be\begin{array}{ll}\begin{array}{rclrcl}
\Q_i(x)&\to& \sigma_{02}\big[\Q_i(x)\big]=
\overline{\Q}_{3-i}(x),\quad&
\Q_3(x)&\to& \sigma_{02}\big[\Q_3(x)\big]=
\overline{\Q}_{3}(x), 
\\[.3cm] 
\Qb_i(x)&\to& \sigma_{02}\big[\Qb_i(x)\big]=
{\Q}_{3-i}(x),\quad&
\Qb_3(x)&\to& \sigma_{02}\big[\Qb_3(x)\big]=
{\Q}_{3}(x), 
\end{array}&\quad i=1,2.\label{sig-Q}
\end{array}
\ee
Similarly, taking into account 
\eqref{atyp}, \eqref{sig-z} and \eqref{sig-Q} one gets
\be
\TT_m^{(1)}(x)\to \sigma_{02}\big[\TT^{(1)}_{m}(x)\big]=\TT^{(1)}_{-m-1}(x), \qquad
m\in {\mathbb Z}\ .\label{sig-T}
\ee
Finally, the proof of the functional relations, given below, 
is based on decomposition properties of products of the
representations of ${\mathcal B}_+(U_q(\widehat{sl}(2|1)))$ with
respect to the co-multiplication \eqref{comul1}. Owing to
\eqref{sig-del}, the whole set of the functional relations splits
into pairs of relations following from each other under the 
substitution \eqref{sig-Q}, \eqref{sig-T}.

This symmetry leaves only three independent functional relations: 
\eqref{sl21-rel1}, \eqref{atypa}
with $m\in {\mathbb Z}_{\ge0}$ and \eqref{dim4a}, while all other relations
become their simple corollaries.  
To proceed further with an algebraic 
proof of the remaining three relations we need
some new notations. 

\subsection{Additional notations}
\subsubsection{Shifted modules}
For any $j_{0},j_{2} \in
{\mathbb C}$, let $p_{[j_{0},j_{2}]}$ be 
 a shift automorphism of the Borel subalgebra ${\mathcal B}_{+}\subset
 U_q(\widehat{sl}(2|1))$ such that 
\begin{eqnarray}
p_{[j_{0},j_{2}]}(e_{i})=e_{i}, \quad 
p_{[j_{0},j_{2}]}(h_{0})=h_{0}+j_{0}, \quad  
p_{[j_{0},j_{2}]}(h_{1})=h_{1}-j_{0}-j_{2}, \quad 
p_{[j_{0},j_{2}]}(h_{2})=h_{2}+j_{2} .
\end{eqnarray}
For any representation $\pi$ of ${\mathcal B}_{+}$
define {\em shifted representation}
$$\pi[j_{0},j_{2}]=\pi \cdot p_{[j_{0},j_{2}]}.$$
 We will often use the following identity
\begin{eqnarray}
{\str}_{\pi[j_{0},j_{2}]} (z_{1}^{-h_{0}}z_{2}^{h_{2}} \overline{{\mathcal R}})
={\str}_{\pi} (z_{1}^{-h_{0}-j_{0}}z_{2}^{h_{2}+j_{2}} \overline{{\mathcal R}})
=z_{1}^{-j_{0}}z_{2}^{j_{2}}
{\str}_{\pi} (z_{1}^{-h_{0}}z_{2}^{h_{2}} \overline{{\mathcal R}}),
\label{shift-id} 
\end{eqnarray}
where the super trace is understood as ${\str}_{\pi[j_{0},j_{2}]}
\otimes 1$ or as ${\str}_{\pi} \otimes 1$, see the note after \eqref{t-uni3}.
Here we used the fact that the reduced universal $R$-matrix 
$\overline{{\mathcal R}} \in {\mathcal B}_{+} \otimes {\mathcal
  B}_{-}$, defined in \eqref{R-red} and  \eqref{R-series},  
 does not contain powers of the Cartan elements ${h_{i}}\otimes 1$. 
\subsubsection{Modified versions of the maps $\rho_i$ and $\overline{\rho}_i$}
{}From now on and to the rest of the paper (including all appendices)
we will use a slightly modified version of the maps introduced 
in Sect.~\ref{qopsec}
\be
\begin{array}{rclrcl}
\rho_{1}^{\prime}(x)&=&x^{+(1\otimes {\mathcal H}^{f})}\cdot\rho_{1}(x)
\cdot x^{-(1\otimes{\mathcal H}^{f})}, 
\qquad 
&\overline{\rho}_{1}^{\prime}(x)&=& 
x^{-(1\otimes{\mathcal H}^{f})}\cdot \overline{\rho}_{1}(x)\cdot 
x^{+(1\otimes{\mathcal H}^{f})}, \\ [.3cm]
\rho_{2}^{\prime}(x)&=&\rho_{2}(x), 
\qquad 
&\overline{\rho}_{2}^{\prime}(x)&=&\overline{\rho}_{2}(x), \\[.4cm]
\rho_{3}^{\prime}(x)&=&x^{-({\mathcal H}^{f}\otimes1)}\cdot \rho_{3}(x)\cdot 
x^{+({\mathcal H}^{f}\otimes1)}, 
\qquad 
&\overline{\rho}_{3}^{\prime}(x)&=&
x^{+({\mathcal H}^{f}\otimes1)}\cdot \overline{\rho}_{3}(x)
\cdot x^{-({\mathcal H}^{f}\otimes1)}\ .
\end{array}\label{modmaps}
\ee
They contain additional similarity transformation in some fermionic
Fock spaces. Obviously, these transformations do not affect 
the definition of the ${\bf Q}$-operators, 
which only involve the super-trace.

\subsubsection{Fock spaces for oscillator algebras} \label{fockspaces}
It was already remarked that the definitions \eqref{Q-def} and
\eqref{Qbar-def} does not depend on a choice of representations of
oscillator algebras \eqref{Hq-def} and \eqref{F-def}. 
For definiteness, assume 
that $|q|=1$, but not a root of unity $q^N\not =1$ (the reasonings can also
be repeated when $|q|\not=1$). 
Consider, for example, the simplest non-trivial trace,
$\mbox{Tr}\big(e^{-\omega  
\mathcal{H}^b} \,b^+   b^-\big)$,
for the bosonic algebra \eqref{Hq-def}. Using the commutation
relation and the cyclic property of the trace, one obtains 
\begin{eqnarray}
{\mbox{Tr}\big( e^{\omega \mathcal{H}^b} b^+
  b^-\big)}&=&q^{-2}\,{\mbox{Tr}\big( e^{\omega \mathcal{H}^b} b^-
  b^+\big)}+\frac{1}{q(q-q^{-1})}\, {\mbox{Tr}\big( e^{\omega
    \mathcal{H}^b} \big)}\\[.3cm]
&=&q^{-2} e^{-\omega} \,{\mbox{Tr}\big( e^{\omega \mathcal{H}^b} b^+
  b^-\big)}+\frac{1}{q(q-q^{-1})}\, {\mbox{Tr}\big( e^{\omega
    \mathcal{H}^b} \big)}. 
\end{eqnarray}
It follows then
\be\label{2btrace}
\frac{{\mbox{Tr}\big( e^{\omega \mathcal{H}^b} b^+
  b^-\big)}}
{\mbox{Tr}\big( e^{\omega
    \mathcal{H}^b}\big)}=\frac{1}{q(q-q^{-1})(1-q^{-2} e^{-\omega})}\ .
\ee
The only assumption made in this calculation is the existence of the trace. 

The same quantity \eqref{2btrace} can also be calculated by
using the highest weight representations (Fock representations) of the
algebra \eqref{Hq-def}. This algebra has only two non-equivalent Fock
representations $w_\pm(\mathcal{H}_q)$, acting on the bases \ 
$|k\rangle_\pm$, $k \in {\mathbb Z}_{\ge 0}$,
\be\label{tworep}
w_\pm(b^\mp)|k\rangle_\pm=|k+1\rangle_\pm,\quad
w_\pm(b^\pm)|k\rangle_\pm=\frac{1-q^{\mp 2k}}{(q-q^{-1})^2}\,|k-1\rangle_\pm,
\quad w_\pm(\mathcal{H}^b)|k\rangle_\pm=\mp k|k\rangle_\pm\ , 
\ee
where one needs to take all upper or all lower signs. Using these
definitions, one easily obtains two expressions 
\begin{equation}
\begin{array}{rcl}
\mbox{Tr}_{w_+}
\big( e^{\omega \mathcal{H}^b}\big)&=&\ds \sum_{k=0}^\infty
e^{-k\omega}=\frac{1}{1-e^{-\omega}} , \\[.4cm]
\mbox{Tr}_{w_+}
\big( e^{\omega \mathcal{H}^b} b^+
  b^-\big)&=&\ds \sum_{k=0}^\infty
e^{-k\omega}\,\frac{(1-q^{-2}q^{-2k})}{(q-q^{-1})^2}=\frac{1}{q(q-q^{-1})
(1-e^{-\omega})(1-q^{-2}e^{-\omega})},
\end{array}\label{traces}
\end{equation}
which imply formula \eqref{2btrace}.
For $|q|=1$ the above series converge for $\mbox{Re}\,\omega>0$. Thus, 
the last calculation only implies \eqref{2btrace} in the half-plane
$\mbox{Re}\,\omega>0$. Of course, the final answer is a meromorphic
function of $\omega$ and can be analytically continued to the whole complex
$\omega$-plane. 

One can perform a similar calculation using the
second Fock representation,\  $w_-$, which requires 
$\mbox{Re}\,\omega<0$. Note that a replacement of $w_+$ with $w_-$ 
in \eqref{traces} changes the values of the trace. 
However, the final result for the ratio
\eqref{2btrace} remains the same, as expected. 
To summarize, for explicit calculations one can use any of the Fock
representations \eqref{tworep}, depending on convenience. 

For the fermionic algebra \eqref{F-def} there is only one
two-dimensional Fock representation (up to the shifts of the
$\mathcal{H}^f$).  However, to streamline the 
notations we define two representations 
\be
\begin{array}{rclrcl}
w_\pm(f^\pm)|0\rangle_\pm&=&0,\qquad 
w_\pm(f^\pm)|1\rangle_\pm&=&|0\rangle_\pm,\\[.4cm] 
w_\pm(f^\mp)|0\rangle_\pm&=&|1\rangle_\pm,\qquad 
w_\pm(f^\mp)|1\rangle_\pm&=&0,\\[.4cm] 
w_\pm(\mathcal{H}^f)|k\rangle_{\pm} &=& \mp k |k\rangle_{\pm},\qquad k \in \{0,1\},&&
\end{array}
\ee
differing by an exchange of the basis vectors $|0\rangle$ and
$|1\rangle$ and a shift of $\mathcal{H}^f$. 

Each of the definitions \eqref{Q-def} and
\eqref{Qbar-def} involves the super-trace over some representation of the
direct product of two oscillator algebras, entering the  
corresponding map $\rho_{i}(x)$ or
$\overline{\rho}_{i}(x)$. According to the above discussion this
representation in each case can be chosen in four ways
$W^{\xi_1,\xi_2}=w_{\xi_1}\otimes_s w_{\xi_2}$, labeled by two
sign variables $\xi_{1},\xi_{2}=\pm$. 
For the map $\rho_{a}^{\prime}(x)$ (resp. $\overline{\rho}_{a}^{\prime}(x)$),  
we will denote such representation as 
$W_{a}^{\xi_{1},\xi_{2}}(x)$ (resp. $\overline{W}_{a}^{\xi_{1},\xi_{2}}(x)$).  
Explicit form of the action of the Borel subalgebra ${\mathcal B}_+$
in the basis of these Fock representation is given in 
Appendix~\ref{q-osc-appendix}.
Here we will use the following 6 representations: 
 $\overline{W}^{++}_{1}(x)$, $\overline{W}^{--}_{2}(x)$, 
$\overline{W}^{-+}_{3}(x)$, $W^{--}_{1}(x)$,
$W^{++}_{2}(x)$ and $W^{+-}_{3}(x)$, completely presented in 
(\ref{w1b++})-(\ref{w3+-}).
The  normalization constants (\ref{normali1}) and (\ref{normali2}) 
for these representations are  
\begin{eqnarray}
Z_{1}=\overline{Z}_{1}=\frac{z_{2}(z_{1}-z_{3})}{z_{3}(z_{1}-z_{2})},\quad 
Z_{2}=\overline{Z}_{2}=\frac{z_{3}-z_{2}}{z_{1}-z_{2}},\quad 
Z_{3}=\overline{Z}_{3}=\frac{(z_{3}-z_{1})(z_{2}-z_{3})}{z_{2}z_{3}}, 
\label{normali3}
\end{eqnarray}
where $z_{1}z_{2}z_{3}^{-1}=1$.

As an example, we give here the representation $W^{+-}_3(x)$. It is a
4-dimensional representation spanned on the vectors 
\be
|m,n>_{+-}=|m>_+\otimes_s \,|n>_-=(f_{1}^{-})^{m}\,|0>_+
\otimes_s \,(f_{2}^{+})^{n}\,|0>_-=
(f_{1}^{-})^{m}(f_{2}^{+})^{n}|0>_{+-} , 
\ee
where $m,n=0,1$, with the following action of the generators of
${\mathcal B}_+$ 
\be
\begin{array}{rcl}
e_{0}|m,n>_{+-}&=&q^{-n}\,|m-1,n>_{+-}, \\[.4cm]
e_{1}|m,n>_{+-}&=&\ds 
\frac{(-1)^{m}\,q^{n+\frac{1}{2}}x}{q-q^{-1}}\,|m+1,n+1>_{+-},
\\[.4cm]
e_{2}|m,n>_{+-}&=&(-1)^{m}\,|m,n-1>_{+-},\\[.4cm]
(h_{0},h_{1},h_{2})\ |m,n>_{+-}&=&(-n,m+n,-m)\,|m,n>_{+-},
\end{array}\label{w3bar}
\ee
where $|m,n>_{+-}$ vanishes if either of the indices $m,n$
take values $-1$ or $+2$.
The parities of the vectors (important for the super trace) 
are equal to \be
p\big(|m,n>_{+-}\big)=(m+n) \pmod 2\ .
\ee

\subsection{Wronskian-type relation (\ref{sl21-rel12}) for ${\bf Q}$-operators}
\label{wrtype}
Below we will prove the relation (\ref{sl21-rel1}). Using
\eqref{Q-def} and \eqref{Qbar-def} we will write it in the form 
\be
-c_{12}{\mathbb A}_{3}(x)=\left(\frac{z_{2}}{z_{1}}\right)^{\frac{1}{2}}
 \overline{{\mathbb A}}_{1}(xq)\overline{{\mathbb A}}_{2}(xq^{-1})-
 \left(\frac{z_{1}}{z_{2}}\right)^{\frac{1}{2}}
 \overline{{\mathbb A}}_{1}(xq^{-1})\overline{{\mathbb A}}_{2}(xq). 
\ee
Making now a particular 
choice of representations for oscillator algebras (which utilizes the
freedom explained in the 
previous subsection) one can rewrite this functional relation as 
\begin{eqnarray}
\begin{array}{l}
\ds \frac{z_{2}}{z_{2}-z_{1}}\,
{\str}_{W_{3}^{+-}(x)}
\big(z_{1}^{-h_{0}}z_{2}^{h_{2}}\overline{{\mathcal R}}
\big)
\label{st-rel0}  \\[.4cm]
\ds \qquad\qquad=
{\str}_{\overline{W}_{1}^{++}(xq)\otimes_{s}\overline{W}_{2}^{--}(xq^{-1})} 
\big(z_{1}^{-h_{0}}z_{2}^{h_{2}}\overline{{\mathcal R}}\big)-
\frac{z_{1}}{z_{2}}\,
{\str}_{\overline{W}_{1}^{++}(xq^{-1})\otimes_{s}\overline{W}_{2}^{--}(xq)} 
\big(z_{1}^{-h_{0}}z_{2}^{h_{2}}\overline{{\mathcal R}}\big).
\end{array}
\end{eqnarray}
We will split the proof into two steps.

\bigskip\noindent
{\bf Step 1}. First, consider the tensor product module 
 $\overline{W}_{1}^{++}(xq)\otimes_{s}\overline{W}_{2}^{--}(xq^{-1})$.  
Let us write its basis vectors 
as 
\be
w_{j_{1},j_{2},j_{3},j_{4}}=|j_{1},j_{2}>_{++}
\otimes_{s}|j_{3},j_{4}>_{--}, \qquad j_{1},j_{4} \in {\mathbb Z}_{\ge
 0},\qquad j_{2},j_{3}=0,1\label{wjjjj}
\ee
where $|j_{1},j_{2}>_{++}$ and $|j_{3},j_{4}>_{--}$ denote bases in 
$\overline{W}_{1}^{++}(xq)$ and $\overline{W}_{2}^{--}(xq^{-1})$,
 defined in \eqref{w1b++} and \eqref{w2b--}, respectively.
We will assume that $w_{j_{1},j_{2},j_{3},j_{4}}\equiv0$, if the
 indices $j_1,j_2,j_3,j_4$ lie outside the domain specified in \eqref{wjjjj}.
Note that the parity $p\big(w_{j_{1},j_{2},j_{3},j_{4}}\big)=
 (j_{2}+j_{3}) \pmod{2}$. 
Taking into account \eqref{w1b++}, \eqref{w2b--} and the formula for
the co-multiplication \eqref{comul1}, one can calculate the
action of the generators of ${\mathcal B}_{+}$ on the basis \eqref{wjjjj},
\be
\label{ehw}
\begin{array}{rcl}
e_{0}\,w_{j_{1},j_{2},j_{3},j_{4}}&=&\ds
w_{j_{1},j_{2}-1,j_{3},j_{4}}+
\frac{(-1)^{j_{2}}(q^{j_{4}}-q^{-j_{4}})q^{-j_{1}
    -\frac{1}{2}}x}{(q-q^{-1})^2}w_{j_{1},j_{2},j_{3}+1,j_{4}-1},    
\\[.3cm]
e_{1}\,w_{j_{1},j_{2},j_{3},j_{4}}&=&\ds
 q^{-j_{2}}w_{j_{1}+1,j_{2},j_{3},j_{4}}
+q^{2j_{1}+j_{2}}w_{j_{1},j_{2},j_{3},j_{4}+1},
\\[.3cm]
 e_{2}\,w_{j_{1},j_{2},j_{3},j_{4}}&=&\ds
-\frac{x(1-q^{-2j_{1}})q^{j_{2}
+\frac{3}{2}}}{(q-q^{-1})^2}w_{j_{1}-1,j_{2}+1,j_{3},j_{4}}+ 
(-1)^{j_{2}}q^{-j_{1}-j_{2}+j_{4}}w_{j_{1},j_{2},j_{3}-1,j_{4}},
\\[.3cm]
h_{0}\,w_{j_{1},j_{2},j_{3},j_{4}}&=&
-(j_{1}+j_{3}+j_{4})\,w_{j_{1},j_{2},j_{3},j_{4}}, \\[.3cm]
h_{1}\,w_{j_{1},j_{2},j_{3},j_{4}}&=&
(2j_{1}+j_{2}+j_{3}+2j_{4})\,w_{j_{1},j_{2},j_{3},j_{4}}, \\[.3cm]
h_{2}\,w_{j_{1},j_{2},j_{3},j_{4}}&=&
-(j_{1}+j_{2}+j_{4})\,w_{j_{1},j_{2},j_{3},j_{4}}. 
\end{array}
\ee
It is convenient to define vectors, with the same weights,  
\be
\begin{array}{l}
w^{(1)}_{m,j}=w_{j,0,0,m-j}, \qquad 0 \le j \le m, \\[.3cm]
w^{(2)}_{m,j}=w_{j,1,1,m-j-1},\quad  
w^{(3)}_{m,j}=w_{j,1,0,m-j-1},\quad
w^{(4)}_{m,j}=w_{j,0,1,m-j-1}, \qquad 0 \le j \le m-1,
\end{array}
\ee
and introduce the following vectors
\begin{eqnarray}
\begin{split}
v^{(m)}_{00}&=\sum_{j=0}^{m}q^{(m-j-2)j}
\qbinom{m}{j}\left(
w^{(1)}_{m,j}-\frac{xq^{-j-\frac{1}{2}}[m-j]}{q-q^{-1}}w^{(2)}_{m,j}
\right),   \\[.3cm]
v^{(m)}_{10}&=\sum_{j=0}^{m}q^{(m-j-2)j}
\qbinom{m}{j}
w^{(3)}_{m+1,j}, \\[.3cm]
v^{(m)}_{01}&=\sum_{j=0}^{m}q^{(m-j)j-m}
\qbinom{m}{j}
w^{(4)}_{m+1,j}, \\[.3cm]
v^{(m)}_{11}&=\sum_{j=0}^{m+1}q^{(m-j)j-m}
\left(
\frac{q^{-\frac{1}{2}}(q-q^{-1})^{2}}{x}
\qbinom{m}{j-1}w^{(1)}_{m+1,j}+
q^{-2j-1}
\qbinom{m}{j}
w^{(2)}_{m+1,j}
\right), 
\end{split}
\end{eqnarray}
where 
\be
\qbinom{m}{j}
=\frac{[m]!}{[j]!\,[m-j]!},\qquad  j=0,1,\dots, m,\qquad m\ge0,
\ee
are the $q$-binomial coefficients,
\be
[m]!=[1][2]\cdots [m], \qquad m \in {\mathbb Z}_{\ge 1}, \qquad
[0]!=1,
\ee
is the $q$-factorial and 
\be
[r]=(q^{r}-q^{-r})/(q-q^{-1})\ .\label{qnumber}
\ee 
Here we put $\qbinom{m}{-1}=0$ for $m \in {\mathbb Z}_{\ge 0}$.
The action of generators of $\mathcal{B}_+$ on 
these vectors is as follows  
\begin{subequations}\label{actions1}
\be
\begin{array}{rclrclrcl}
e_{0}v^{(m)}_{00}&=&0,& 
e_{1}v^{(m)}_{00}&=&\frac{q^{\frac{1}{2}}x}
{q-q^{-1}}v^{(m)}_{11}+v^{(m+1)}_{00},&
e_{2}v^{(m)}_{00}&=&0, \\[.3cm]
e_{0}v^{(m)}_{10}&=&v^{(m)}_{00}, &
e_{1}v^{(m)}_{10}&=&qv^{(m+1)}_{10},& 
e_{2}v^{(m)}_{10}&=&0, \\[.3cm]
e_{0}v^{(m)}_{01}&=&0,&
e_{1}v^{(m)}_{01}&=&qv^{(m+1)}_{01},&
e_{2}v^{(m)}_{01}&=&v^{(m)}_{00}, \\[.3cm]
e_{0}v^{(m)}_{11}&=&q^{-1}v^{(m)}_{01}, &
e_{1}v^{(m)}_{11}&=&q^{2}v^{(m+1)}_{11}, &
e_{2}v^{(m)}_{11}&=&-v^{(m)}_{10},\\[.3cm]
\end{array}
\ee
and 
\be
\begin{array}{rclrrrl}
(h_0,h_1,h_2)\ v^{(m)}_{00}&=&(&-m,&2m,&-m&)\ v^{(m)}_{00}\\[.3cm]
(h_0,h_1,h_2)\ v^{(m)}_{10}&=&(&-m,&2m+1,&-m-1&)\ v^{(m)}_{10}\\[.3cm]
(h_0,h_1,h_2)\ v^{(m)}_{01}&=&(&-m-1,&2m+1,&-m&)\ v^{(m)}_{01}\\[.3cm]
(h_0,h_1,h_2)\ v^{(m)}_{11}&=&(&-m-1,&2m+2,&-m-1&)\ v^{(m)}_{11}. \\[.1cm]
\end{array}
\ee
\end{subequations}
For each $ m \in {\mathbb Z}_{\ge 0} $, let $W^{(m)}$ be 
the vector space spanned by the vectors 
$\{v^{(k)}_{00},v^{(k)}_{10},v^{(k)}_{01},v^{(k)}_{11}\}_{k=m}^{\infty}$.
By construction,
$$
\overline{W}_{1}^{++}(xq)\otimes_{s}\overline{W}_{2}^{--}(xq^{-1})
\supset W^{(0)}\supset W^{(1)} \supset W^{(2)} \supset
\ldots\ .
$$ 
Examining \eqref{actions1}, it is easy to conclude that 
\begin{itemize}
\item[(i)] each $W^{(m)}$ is  
an invariant subspace with respect to the 
action of ${\mathcal B}_{+}$,
\item[(ii)] for each $ m \in {\mathbb Z}_{\ge 0} $\ 
the factor module $W^{(m)}/W^{(m+1)}$ is isomorphic to the shifted 
module $W_{3}^{+-}(x)[-m,-m]$. To see this one needs to drop all
vectors $v^{(m+1)}_{jk}$ in the RHS of \eqref{actions1} and identify
the vectors $v^{(m)}_{jk}$ therein with $|j,k>_{+-}$ in \eqref{w3bar}. 
\end{itemize}
Applying the identity \eqref{shift-id} one obtains 
\be
\begin{array}{rl}
\ds {\str}_{W^{(0)}}(z_{1}^{-h_{0}}z_{2}^{h_{2}}\overline{{\mathcal R}})
&\ds=\sum_{m=0}^{\infty}
{\str}_{W^{(m)}/W^{(m+1)}}(z_{1}^{-h_{0}}z_{2}^{h_{2}}\overline{{\mathcal
    R}}) 
=\sum_{m=0}^{\infty}
{\str}_{W_{3}^{+-}(x)[-m,-m]}(z_{1}^{-h_{0}}z_{2}^{h_{2}}\overline{{\mathcal
    R}}) 
\\[.4cm]
&\ds=
\sum_{m=0}^{\infty}
    {\str}_{W_{3}^{+-}(x)}(z_{1}^{-h_{0}+m}z_{2}^{h_{2}-m}
\overline{{\mathcal R}})=
\frac{z_{2}}{z_{2}-z_{1}}{\str}_{W_{3}^{+-}(x)}(z_{1}^{-h_{0}}z_{2}^{h_{2}}
\overline{{\mathcal R}}).
\end{array}
\label{st-rel1}
\ee

\bigskip\noindent
{\bf Step 2}. Next we want to show that the factor module 
 $\overline{W}_{1}^{++}(xq)\otimes_{s}\overline{W}_{2}^{--}(xq^{-1})/W^{(0)}$ 
is isomorphic to 
$\overline{W}_{1}^{++}(xq^{-1})\otimes_{s}\overline{W}_{2}^{--}(xq)$ up to 
a shift automorphism. 

Introduce additional vectors in 
$\overline{W}_{1}^{++}(xq)\otimes_{s}\overline{W}_{2}^{--}(xq^{-1})$,
\be
\begin{array}{rcll}
u_{m,j}^{(1)}&=&\ds
\frac{q^{j(j-m-2)+m}}{
[m+1]
\qbinom{m}{j}}
\sum_{k=j+1}^{m+1} q^{k(m+1-k)} 
\qbinom{m+1}{j}
w^{(1)}_{m+1,k}\,  
&m\ge 0, \quad 
 0 \le j \le m ,  \\[.6cm]
u_{m,j}^{(2)}&=&\ds
\frac{q^{j(j-m)+m}}{
[m-j]
\qbinom{m}{j}}
\sum_{k=j+1}^{m+1} 
q^{k(m-k)}
\Big\{
\frac{q^{-\frac{1}{2}} 
(q-q^{-1})^{2}[k-j-1]}{[k]\,x}
&\hspace{-1em}\ds\qbinom{m}{k-1}
w^{(1)}_{m+1,k} \\[.6cm]
&&\ds \phantom{MMMM} +q^{-2k+j} 
\qbinom{m}{k}
w^{(2)}_{m+1,k} 
\Big\}\, 
 &m\ge 1, \quad 
 0 \le j \le m-1, \\[.5cm]
u_{m,j}^{(3)}&=&\ds 
\frac{q^{j(j-m+1)+m+1}}{
[m-j]\qbinom{m}{j}}
\sum_{k=j+1}^{m} q^{k(m-2-k)} 
\qbinom{m}{k}
w^{(3)}_{m+1,k} 
&m\ge1, \quad 
 0 \le j \le m-1, \\[.6cm]
u_{m,j}^{(4)}&=&\ds
\frac{q^{j(j-m-1)+m-2}}{
[m-j]\qbinom{m}{j}}
\sum_{k=j+1}^{m} q^{k(m-k)} 
\qbinom{m}{k}
w^{(4)}_{m+1,k}, &m\ge1, \quad 
 0 \le j \le m-1 .
\end{array}\label{vec-u4}
\ee
Note that the union of the three sets of vectors  
$$
\{v^{(m)}_{00},v^{(m)}_{10},v^{(m)}_{01},v^{(m)}_{11}\}_{m \in
  {\mathbb Z}_{\ge 0}}\,\bigcup\, 
\{u_{m,j}^{(1)}\}_{0 \le j \le m,\ m \in {\mathbb Z}_{\ge 0}}\,\bigcup\, 
\{u_{m,j}^{(2)},u_{m,j}^{(3)},u_{m,j}^{(4)}\}_{0 \le j \le m-1, \ 
m \in {\mathbb Z}_{\ge
    1}}$$   
completely span the vector space 
$\overline{W}_{1}^{++}(xq)\otimes_{s}\overline{W}_{2}^{--}(xq^{-1})$.  
The action of $\mathcal{B}_+$ on the vectors \eqref{vec-u4}
is as follows,  
\begin{subequations}\label{action_u}
\be
\begin{array}{rclrcl}
e_{0}u^{(1)}_{m,j}&=&\ds\frac{q^{-j+\frac{3}{2}}[m-j]x}
{q-q^{-1}}u^{(4)}_{m,j},&
e_{0}u^{(2)}_{m,j}&=&\ds u^{(4)}_{m,j},\\[.4cm]
e_{1}u^{(1)}_{m,j}&=&\ds q^{2j}u^{(1)}_{m+1,j}+u^{(1)}_{m+1,j+1},&
e_{1}u^{(2)}_{m,j}&=&\ds 
q^{2j+1}u^{(2)}_{m+1,j}+q^{-1}u^{(2)}_{m+1,j+1}, \\[.4cm]
e_{2}u^{(1)}_{m,j}&=&\ds
-\frac{q^{-j-\frac{1}{2}}[j]x}{q-q^{-1}}u^{(3)}_{m,j-1} 
 -\delta_{j,0} \frac{q^{2m+\frac{1}{2}}x}{q-q^{-1}}v_{10}^{(m)},&
e_{2}u^{(2)}_{m,j}&=&\ds -q^{-2j+m-2}u^{(3)}_{m,j},\\
\end{array}
\ee
\be
\begin{array}{rcl}
e_{0}u^{(3)}_{m,j}&=&\ds
u^{(1)}_{m-1,j}+\frac{q^{-j+\frac{3}{2}}x[j-m+1]}{q-q^{-1}}u^{(2)}_{m-1,j},
\\[.3cm]
e_{1}u^{(3)}_{m,j}&=&\ds 
q^{2j+1}u^{(3)}_{m+1,j}+q^{-1}u^{(3)}_{m+1,j+1}, 
\\[.3cm]
e_{2}u^{(3)}_{m,j}&=&0, \\[.3cm]
e_{0}u^{(4)}_{m,j}&=&0, \\[.3cm]
e_{1}u^{(4)}_{m,j}&=&q^{2j}u^{(4)}_{m+1,j}+u^{(4)}_{m+1,j+1}, \\[.3cm]
e_{2}u^{(4)}_{m,j}&=&\ds q^{-2j+m-1}u^{(1)}_{m-1,j}
-\frac{q^{-j-\frac{1}{2}}x[j]}{q-q^{-1}}u^{(2)}_{m-1,j-1}
-\delta_{j,0} \frac{xq^{3m-\frac{5}{2}}}{q-q^{-1}}v_{11}^{(m-1)},
\end{array}
\ee
and 
\be
\begin{array}{rclrrrl}
(h_0,h_1,h_2)\ u^{(1)}_{m,j}&=&(&-m-1,&2m+2,&-m-1&)\ u^{(1)}_{m,j}\\[.25cm]
(h_0,h_1,h_2)\ u^{(2)}_{m,j}&=&(&-m-1,&2m+2,&-m-1&)\ u^{(2)}_{m,j}\\[.25cm]
(h_0,h_1,h_2)\ u^{(3)}_{m,j}&=&(&-m,&2m+1,&-m-1&)\ u^{(3)}_{m,j}\\[.25cm]
(h_0,h_1,h_2)\ u^{(4)}_{m,j}&=&(&-m-1,&2m+1,&-m&)\ u^{(4)}_{m,j}\\[.1cm]
\end{array}
\ee
\end{subequations}
where $u^{(2)}_{0,j}=u^{(2)}_{m,-1}=u^{(3)}_{0,j}=u^{(3)}_{m,-1}\equiv0$.

Similarly to \eqref{ehw} write down the action of ${\mathcal B}_+$ 
on the basis vectors 
$$u_{j_{1},j_{2},j_{3},j_{4}}
=|j_{1},j_{2}>_{++}\otimes_{s}|j_{3},j_{4}>_{--}$$  
of the tensor product module 
$\overline{W}_{1}^{++}(xq^{-1})\otimes_{s}\overline{W}^{--}_{2}(xq)$, 
\be
\begin{array}{rcl}
e_{0}\,u_{j_{1},j_{2},j_{3},j_{4}}&=&\ds
u_{j_{1},j_{2}-1,j_{3},j_{4}}+
\frac{(-1)^{j_{2}}q^{-j_{1}+\frac{3}{2}}x(q^{j_{4}}-q^{-j_{4}})}
{(q-q^{-1})^2}\, 
u_{j_{1},j_{2},j_{3}+1,j_{4}-1},
\\[.4cm]
e_{1}\,u_{j_{1},j_{2},j_{3},j_{4}}&=&\ds
 q^{-j_{2}}\,u_{j_{1}+1,j_{2},j_{3},j_{4}}
+q^{2j_{1}+j_{2}}\,u_{j_{1},j_{2},j_{3},j_{4}+1},
\\[.4cm]
e_{2}\,u_{j_{1},j_{2},j_{3},j_{4}}&=&\ds
-\frac{q^{j_{2}-\frac{1}{2}}x(1-q^{-2j_{1}})}
{(q-q^{-1})^2}\,u_{j_{1}-1,j_{2}+1,j_{3},j_{4}}+
(-1)^{j_{2}}q^{-j_{1}-j_{2}+j_{4}}\,u_{j_{1},j_{2},j_{3}-1,j_{4}},
\\[.4cm]
\end{array}\label{u_action}
\ee
$$
(h_0,h_1,h_2)\ u_{j_{1},j_{2},j_{3},j_{4}}=
(-j_1-j_3-j_4,\ 2j_1+j_2+j_3+2j_4,\ -j_1-j_2-j_4) u_{j_{1},j_{2},j_{3},j_{4}}
$$
where $j_{1},j_{4} \in {\mathbb Z}_{\ge 0}$ and $j_{2},j_{3}=0,1$, 
otherwise $w_{j_{1},j_{2},j_{3},j_{4}}\equiv 0$. Also define 
\be
\begin{array}{l}
\overline{u}^{(1)}_{m,j}=u_{j,0,0,m-j},\qquad 
0 \le j \le m, \\[.3cm]
\overline{u}^{(2)}_{m,j}=u_{j,1,1,m-j-1}, \quad
\overline{u}^{(3)}_{m,j}=u_{j,1,0,m-j-1}, \quad
\overline{u}^{(4)}_{m,j}=u_{j,0,1,m-j-1}, 
\qquad 0 \le j \le m-1.
\end{array}\label{over_u}
\ee

Comparing \eqref{action_u} with \eqref{u_action} one concludes that 
\begin{itemize}
\item[(iii)]
the 
action of generators $e_{k}$ on vectors
$\{\overline{u}^{(a)}_{m,j}\}$, defined in \eqref{over_u}, 
is exactly the same as that on the vectors $\{u^{(a)}_{m,j}\}$ in 
$\overline{W}_{1}^{++}(xq)\otimes_{s}\overline{W}_{2}^{--}(xq^{-1})/W^{(0)}$.
To obtain the latter one needs to omit the terms containing 
$v^{(m)}_{ij}$ in the RHS of \eqref{action_u}. 
\item[(iv)]
the action of $h_{k}$ on $\{\overline{u}^{(a)}_{m,j}\}$ 
coincides with that for the vector $\{ u^{(a)}_{m,j}\}$ 
in $\overline{W}_{1}^{++}(xq)\otimes_{s}\overline{W}_{2}^{--}(xq^{-1})/W^{(0)}$
up to the shift $m \to m+1$. 
\end{itemize}
Thus one concludes that 
\be
\overline{W}_{1}^{++}(xq)\otimes_{s}
\overline{W}_{2}^{--}(xq^{-1})/W^{(0)} 
\simeq (\overline{W}_{1}^{++}(xq^{-1})
\otimes_{s}\overline{W}_{2}^{--}(xq))[-1,-1]\ .
\ee
Using the this formula and applying the identity \eqref{shift-id} one obtains  
\be
\begin{array}{rcl}
{\str}_{\overline{W}_{1}^{++}(xq)\otimes_{s}
\overline{W}_{2}^{--}(xq^{-1})/W^{(0)}}
(z_{1}^{-h_{0}}z_{2}^{h_{2}}\overline{{\mathcal R}})
&=&\ds \frac{z_{1}}{z_{2}}
{\str}_{\overline{W}_{1}^{++}(xq^{-1})\otimes_{s}\overline{W}_{2}^{--}(xq)}
(z_{1}^{-h_{0}}z_{2}^{h_{2}}\overline{{\mathcal R}}). 
\end{array}\label{st-rel2}
\ee
Combining this with (\ref{st-rel1}) one obtains (\ref{st-rel0}). 
This completes the proof of the relation (\ref{sl21-rel1}).

\subsection{Factorization formula (\ref{dim4})
for a typical representation}
Here we will prove (\ref{dim4}), namely 
\be
{\mathbb T}_{c}^{(2)}(x)
=(z_{1}-z_{3})(z_{2}-z_{3})\,z_{3}^{c-1}
{\mathbb A}_{3}(xq^{-c-\frac{1}{2}})
\overline{{\mathbb A}}_{3}(xq^{c+\frac{1}{2}}),\label{typfactor}
\ee
where ${\mathbb T}_{c}^{(2)}(x)$ is defined in \eqref{t2},
\be
{\mathbb T}_{c}^{(2)}(x)={\str}_{{\pi}_{(c,c,0)}(xq^{c+1})}
(z_{1}^{-h_{0}}z_{2}^{h_{2}}\overline{{\mathcal R}}).
\ee
The 4-dimensional typical representation $\pi_{(c,c,0)}(x)$ of the
Borel subalgebra ${\mathcal B}_{+}$ is described in the 
Appendix \ref{app-typical}.

Our proof of \eqref{typfactor} 
is based on the decomposition of the 16-dimensional 
tensor product module 
$W_{3}^{+-}(xq^{-c-\frac{1}{2}}) \otimes_{s} 
\overline{W}_{3}^{-+}(xq^{c+\frac{1}{2}}) $ 
into four 4-dimensional modules. With respect to the actions of 
${\mathcal B}_{+}$ 
each of these four 4-dimensional modules is isomorphic 
to a shifted typical representation  
${\pi}_{(c,c,0)}(xq^{c+1})[s^{(a)}_{0},s^{(a)}_{2}]$ 
($a \in \{1,2,3,4\}$), where the shift constants 
$s^{(a)}_{0}$ and $s^{(a)}_{2}$ are given in (\ref{shift-4dim}).


Introduce a basis in the 16-dimensional tensor product module  
\be
w_{j_{1},j_{2},j_{3},j_{4}}=
|j_{1},j_{2}>_{+-}\otimes _{s}|j_{3},j_{4}>_{-+} 
\in W_{3}^{+-}(xq^{-c-\frac{1}{2}}) 
\otimes_{s} \overline{W}_{3}^{-+}(xq^{c+\frac{1}{2}}),
\ee 
where $j_{1},j_{2},j_{3},j_{4}=0,1$, otherwise 
$w_{j_{1},j_{2},j_{3},j_{4}}\equiv0$. 
The parity of these vectors is $p(w_{j_{1},j_{2},j_{3},j_{4}})=
j_{1}+j_{2}+j_{3}+j_{4}\pmod 2$. 
Using \eqref{w3+-}, \eqref{w3b-+} and the formula for the
co-multiplication \eqref{comul1} one can calculate the action 
of ${\mathcal B}_{+}$ in this  tensor product module
\be
\begin{array}{rcl}
e_{0}\ w_{j_{1},j_{2},j_{3},j_{4}}&=&\ds
 q^{-j_{2}}w_{j_{1}-1,j_{2},j_{3},j_{4}}
  +(-1)^{j_{1}+j_{2}}q^{j_{4}-j_{2}}
   w_{j_{1},j_{2},j_{3}-1,j_{4}}\\[.3cm]
e_{1}\ w_{j_{1},j_{2},j_{3},j_{4}}&=&\ds
{x \left(q^{-c}
   w_{j_{1}+1,j_{2}+1,j_{3},j_{4}}-
q^{c+j_{1}+j_{2}}
   w_{j_{1},j_{2},j_{3}+1,j_{4}+1}\right)}/
   ({q-q^{-1}}),\\[.3cm]
e_{2}\ w_{j_{1},j_{2},j_{3},j_{4}}&=&\ds
(-1)^{j_{1}+j_{2}+j_{3}}
   q^{-j_{1}}w_{j_{1},j_{2},j_{3},j_{4}-1}
   +(-1)^{j_{1}}
   w_{j_{1},j_{2}-1,j_{3},j_{4}}, \\[.3cm]
(h_0,h_1,h_2)\ w_{j_{1},j_{2},j_{3},j_{4}}&=&
(-j_2-j_4,\ j_1+j_2+j_3+j_4,\ -j_1-j_3)\  w_{j_{1},j_{2},j_{3},j_{4}}.
\end{array}
\ee
We shall change the basis of the module from 
$\{ w_{j_{1},j_{2},j_{3},j_{4}}\}$ to $\{v^{(a)}_{j}\}$, where $j,a
\in \{1,2,3,4\}$: 
\be
\begin{array}{rclrcl}
v_{1}^{(1)}&=& -q^{2 c-1}(q-q^{-1}) x\, w_{0,0,1,1},&
 v_{2}^{(1)}&=&q^{c-1}x w_{0,1,1,1},\\[.3cm] 
 v_{3}^{(1)}&=&-q^{-c-1}(q-q^{-1})w_{0,1,0,0},&
v_{4}^{(1)}&=&w_{0,1,0,1},\\[.3cm]
v_{1}^{(2)}&=&x q^{c-1}w_{0,0,1,0}, &
v_{2}^{(2)}&=&-{x \left(q^{2 c}
   w_{0,0,1,1}-w_{1,1,0,0}\right)}/({q^2-1}), \\[.3cm]
v_{3}^{(2)}&=&w_{0,0,0,0}, &
v_{4}^{(2)}&=&{q^{-c} \left(q^{2 c+2}
   w_{0,0,0,1}-w_{0,1,0,0}\right)}/({q^2-1}),\\[.3cm]
v_{1}^{(3)}&=&q^{-c-2}  x w_{1,1,1,0}, & 
v_{2}^{(3)}&=&-{x w_{1,1,1,1}}/({q^2-1}), \\[.3cm]
v_{3}^{(3)}&=&-w_{0,0,1,1}+q w_{1,0,0,1}, &
v_{4}^{(3)}&=&{q^{-c}(w_{0,1,1,1}+q
   w_{1,1,0,1})}/({q^2-1}), 
\end{array}
\ee
$$
\begin{array}{rclrcl}
v_{1}^{(4)}&=&-q^{c-1} x w_{1,0,1,0},& 
 v_{2}^{(4)}&=&-{x \left(q^{2c+1} w_{1,0,1,1} +w_{1,1,1,0}\right)}
/({q^2-1}),\\[.3cm]
v_{3}^{(4)}&=&w_{1,0,0,0}-w_{0,0,1,0}, &
v_{4}^{(4)}&=&\{q^{c+2}( w_{0,0,1,1}-q w_{1,0,0,1}) +
q^{-c}(w_{0,1,1,0}+w_{1,1,0,0})\}/({q^2-1}).
\end{array}
$$
The action of the generators $e_k$, $k=0,1,2$ in this basis is 
\be
e_k\, v^{(a)}_j = \sum_{i=1}^4 A^k_{ij}\, v^{(a)}_i+
\sum_{b=a+1}^4 \sum_{i=1}^4\, B^k_{ijab}\, v^{(b)}_i,\qquad i,j,a \in \{1,2,3,4\}
\label{ek4d}\ee
where the only non-zero coefficients $A^k_{ij}$ are
\be 
A^0_{31}=-A^0_{42}=q^{c-1} x,\quad 
A^1_{23}=q^{1-c},\quad A^2_{12}=[c],\quad A^2_{34}=[c+1]\ ,
\ee
where the symbol $[c]$ is defined in \eqref{qnumber}. The coefficients
$B^k_{ijab}$ are known explicitly, but their exact form is 
not essential in the following. Note that, if the second term in
the formula \eqref{ek4d} is omitted, then for any fixed value of $a$ 
it becomes identical to the corresponding formula for 
the 4-dimensional representation $\pi_{(c,c,0)}(xq^{c+1})$, given in
 \eqref{typical-eva-app} up to a similarity transformation.  
The action of the generators $h_k$ in this basis is  
\be
h_k\, v^{(a)}_j= (\nu_{j,k}+s^{(a)}_k)\, v^{(a)}_j \label{hk4d}
\ee
where $a \in \{1,2,3,4\}$ and the weights $\nu_{j,k}$ are exactly the same as 
in the action of $h_k$ in the 4-dimensional 
representation $\pi_{(c,c,0)}(xq^{c+1})$, see
\eqref{typical-eva-app} in the Appendix~\ref{repsB}. The shift constants,  
satisfy the relation $s^{(a)}_{0}+s^{(a)}_{1}+s^{(a)}_{2}=0$;
explicitly they read 
\be
(s^{(a)}_0,s^{(a)}_1,s^{(a)}_2)=
(c-\alpha^{(a)}, \alpha^{(a)}+\beta^{(a)},-c-\beta^{(a)}),\quad a \in \{1,2,3,4\}, 
\label{shift-4dim}
\ee
where 
$$\beta^{(1)}=\beta^{(2)}=1,\quad \beta^{(3)}=\beta^{(4)}=2,\quad 
\alpha^{(1)}=\alpha^{(3)}=1,\quad   \alpha^{(2)}=\alpha^{(4)}=0.$$

Let $W^{(a)}$, $a \in \{1,2,3,4\}$ denotes the vector space spanned by the vectors 
$\{v^{(j)}_{1},v^{(j)}_{2},v^{(j)}_{3},v^{(j)}_{4}\}_{j=a}^{4}$. 
By construction, the original tensor product module 
$${W_{3}^{+-}(xq^{-c-\frac{1}{2}})\otimes_{s}
  \overline{W}_{3}^{-+}(xq^{c+\frac{1}{2}})}\simeq W^{(1)}$$
is isomorphic to $W^{(1)}$.  
 Examining \eqref{ek4d} and \eqref{hk4d} one easily finds that 
\begin{itemize} 
\item[(i)]
each $W^{(a)}$
is an invariant space with respect to the action 
of ${\mathcal B}_{+}$,
\item[(ii)] 
the following isomorphisms with respect to the action of ${\mathcal
  B}_{+}$ take place 
\begin{eqnarray}
W^{(a)}/W^{(a+1)} &\simeq& 
{\pi}_{(c,c,0)}(xq^{c+1})[s^{(a)}_{0},s^{(a)}_{2}],\qquad
a=1,2,3 \\[.3cm]
W^{(4)}&\simeq& 
{\pi}_{(c,c,0)}(xq^{c+1})[s^{(4)}_{0},s^{(4)}_{2}]\ , 
\end{eqnarray}
except that for $a=2,3$ the parities of all vectors on one
side of the correspondence need to be inverted, see below.   
\end{itemize}
Using these results, the definitions \eqref{Q-def} and
\eqref{Qbar-def} and the identity \eqref{shift-id} one obtains 
\be
\begin{array}{l}
Z_{3}\,\overline{Z}_{3}\,{\mathbb A}_{3}(xq^{-c-\frac{1}{2}})\,
\overline{{\mathbb A}}_{3}(xq^{c+\frac{1}{2}})
={\str}_{W_{3}^{+-}(xq^{-c-\frac{1}{2}})
\otimes_{s} \overline{W}_{3}^{-+}(xq^{c+\frac{1}{2}})}
(z_{1}^{-h_{0}}z_{2}^{h_{2}}
\overline{{\mathcal R}})\\[.4cm]
\ds\qquad ={\str}_{W^{(1)}}(z_{1}^{-h_{0}}z_{2}^{h_{2}} 
\overline{{\mathcal R}}) 
=\sum_{a=1}^3{\str}_{W^{(a)}/W^{(a+1)}}(z_{1}^{-h_{0}}z_{2}^{h_{2}} 
\overline{{\mathcal R}})+
{\str}_{W^{(4)}}(z_{1}^{-h_{0}}z_{2}^{h_{2}} \overline{{\mathcal R}}) \\[.4cm]
\qquad  =\ds
\sum_{a=1}^4 \ (-1)^{\gamma_a}\,{\str}_{{\pi}_{(c,c,0)}(xq^{c+1})}
(z_{1}^{-h_{0}-s^{(a)}_{0}}z_{2}^{h_{2}+s^{(a)}_{2}} \overline{{\mathcal R}})
\\[.6cm] 
\qquad =z_{2}^{-2}z_{3}^{-c-1}(z_{1}-z_{3})(z_{2}-z_{3})
{\str}_{{\pi}_{(c,c,0)}(xq^{c+1})}
(z_{1}^{-h_{0}}z_{2}^{h_{2}}\overline{{\mathcal R}})
\\[.4cm]
\qquad  =z_{2}^{-2}z_{3}^{-c-1}(z_{1}-z_{3})(z_{2}-z_{3}){\mathbb
  T}_{c}^{(2)}(x). 
\end{array}
\ee
The sign factor $(-1)^{\gamma_a}$, $\gamma_1=\gamma_4=0$,  
$\gamma_2=\gamma_3=1$, takes into account the 
parity of the vectors $p(v^{(a)}_{1})=\gamma_a$ (which needs to be
compared with  
the (even) parity of the highest weight vector in the
representation $\pi_{(c,c,0)}$). 
Using the expressions for the constants \eqref{normali3}
one finally arrives to \eqref{typfactor}.
\subsection{Wronskian type ${\bf T}$-${\bf Q}$ relation (\ref{atyp})} 
In this section it will be more convenient to use the Fock representations 
$\overline{W}^{--}_{1}(x)$ , 
$\overline{W}^{++}_{2}(x)$  
$W^{++}_{1}(x)$ and  
$W^{--}_{2}(x)$ defined in (\ref{w1b--}), 
(\ref{w2b++}), (\ref{w1++}) and (\ref{w2--}).
Note that are different from those used in 
Sect. 5.2.3, 5.3. 
The normalization constants (\ref{normali1}) 
and (\ref{normali2}) for these new representations read
\begin{eqnarray}
Z_{1}=\overline{Z}_{1}=\frac{z_{1}(1-z_{2})}{z_{1}-z_{2}},\quad 
Z_{2}=\overline{Z}_{2}=\frac{z_{1}-1}{z_{1}-z_{2}}.
\label{normali4}
\end{eqnarray}

We will prove (\ref{atypa}), 
\be
{\mathbb T}_{m}^{(1)}(x)=
\frac{c_{13}}{c_{12}}
z_{1}^{m+\frac{1}{2}}
{\mathbb A}_{1}(xq^{m+\frac{1}{2}})
\overline{{\mathbb A}}_{1}(xq^{-m-\frac{1}{2}})
-\frac{c_{23}}{c_{12}}
z_{2}^{m+\frac{1}{2}}
{\mathbb A}_{2}(xq^{m+\frac{1}{2}})
\overline{{\mathbb A}}_{2}(xq^{-m-\frac{1}{2}})
\label{wrons1}.
\ee
for $m\in {\mathbb Z}_{\ge0}$. 
For this purpose, we introduce more general ${\bf T}$-operators,
corresponding to the infinite-dimensional representations of 
$U_{q}(\widehat{sl}(2|1))$,
\be
{\mathbb T}^{+}_{m}(x)={\str}_{\pi^{+}_{(m,0,0)}(x)}
(z_{1}^{-h_{0}}z_{2}^{h_{2}}\overline{{\mathcal R}}),\qquad
{\mathbb T}^{-}_{m}(x)={\str}_{\pi^{-}_{(-1,m+1,0)}(x)}
(z_{1}^{-h_{0}}z_{2}^{h_{2}}\overline{{\mathcal R}}),
\label{T-infi}
\ee
where $m \in {\mathbb C}$. 
The representations $\pi^{+}_{\mu}(x)$ and $\pi^{-}_{\mu}(x)$ 
are defined in the Appendix~\ref{repsB}. 
For integer values of $m\in
{\mathbb Z}$ one of these representations becomes reducible and can be
decomposed into a semi-direct sum of a finite-dimensional and an
infinite-dimensional atypical representations. These properties are
studied in details in the Appendix~\ref{repsB}. Here we quote just one  
relevant formula \eqref{cl2-inf-iso},
\begin{eqnarray}
\pi^{+}_{(m,0,0)}(x)/\pi_{(m,0,0)}^{[0]}(x) &\simeq &
\pi^{-}_{(-1,m+1,0)}(x), \qquad m \in {\mathbb Z}_{\ge 0}\,
\label{decomp} 
\end{eqnarray}
The finite-dimensional representations $\pi_{(m,0,0)}^{[0]}(x)$, which
appears above is precisely that entering in the definition of the 
${\bf T}$-operators \eqref{tm}\footnote{%
The superscript ``$p$'' in the notation $\pi_\mu^{[p]}(x)$ denotes the
parity of the highest weight vector. The representations
$\pi_\mu^{[p]}(x)$ with $p=0$ and $p=1$ only differ by an overall
sign of the supertrace, but otherwise equivalent.}, namely,
\begin{eqnarray}
{\mathbb T}^{(1)}_{m}(x)&=&{\str}_{\pi_{(m,0,0)}^{[0]}(x)}
(z_{1}^{-h_{0}}z_{2}^{h_{2}}\overline{{\mathcal R}}),\qquad 
m \in {\mathbb Z}_{\ge 0},\nonumber \\[.3cm]
{\mathbb T}^{(1)}_{m}(x)&=&-
{\str}_{\pi_{(-1,m+1,0)}^{[0]}(x)}
(z_{1}^{-h_{0}}z_{2}^{h_{2}}\overline{{\mathcal R}}),
\qquad m \in {\mathbb Z}_{\le -2},\\[.3cm]
{\mathbb T}^{(1)}_{-1}(x)&=&
-{\str}_{\pi_{(-1,-1,1)}^{[1]}(x)}
(z_{1}^{-h_{0}}z_{2}^{h_{2}}\overline{{\mathcal R}})
\nonumber
\end{eqnarray}
On the level of supertraces Eq.\eqref{decomp} implies 
\be
{\mathbb T}^{(1)}_{m}(x)= {\mathbb T}^{+}_{m}(x)- {\mathbb
  T}^{-}_{m}(x), \qquad m \in {\mathbb Z}_{\ge0}, \label{toprove-aty}
\ee
where the operators $\TT^\pm(x)$ are defined by \eqref{T-infi}. 
As we shall see below, these operators factorize into products of two 
${\bf Q}$-operators 
\begin{eqnarray}
{\mathbb T}^{+}_{m}(x)
&=&\frac{c_{13}}{c_{12}}
z_{1}^{m+\frac{1}{2}}
{\mathbb A}_{1}(xq^{m+\frac{1}{2}})\overline{{\mathbb
    A}}_{1}(xq^{-m-\frac{1}{2}}), 
\label{fac1} \\[.3cm] 
{\mathbb T}^{-}_{m}(x)
&=&\frac{c_{23}}{c_{12}}
z_{2}^{m+\frac{1}{2}}
{\mathbb A}_{2}(xq^{m+\frac{1}{2}})\overline{{\mathbb
    A}}_{2}(xq^{-m-\frac{1}{2}}). 
\label{fac2}
\end{eqnarray}
These factorization properties hold for an arbitrary value of
$m\in {\mathbb C}$ (even though for \eqref{toprove-aty} one needs only
integer values of $m$). Thus, 
the Wronskian-like formula (\ref{wrons1}) in question 
is a corollary of the factorization relations 
(\ref{fac1}) and (\ref{fac2}). Their proof is presented below. 
Actually, one needs to prove 
one of these relations, since they follow from one another under the 
symmetry transformation \eqref{sig-z}, \eqref{sig-Q}.

\subsubsection{Factorization formula (\ref{fac1}) for
  infinite-dimensional representations}
The formula \eqref{fac1} reflects rather special properties of 
tensor product module $W_{1}^{++}(xq^{m+\frac{1}{2}})
\otimes_{s}\overline{W}_{1}^{--}(xq^{-m-\frac{1}{2}})$. 
Below we will show that this infinite-dimensional module 
can be decomposed into an
infinite number of infinite-dimensional modules,   
each of which is isomorphic to a shifted evaluation module  
$\pi^{+}_{(m,0,0)}(x)[j_{0},j_{2}]$ 
with $j_{0},j_{2} \in {\mathbb Z}$. 

Let us write the basis in  
$W_{1}^{++}(xq^{m+\frac{1}{2}})
  \otimes_{s}\overline{W}_{1}^{--}(xq^{-m-\frac{1}{2}})$ as 
\begin{eqnarray}
w_{j_{1},j_{2},j_{3},j_{4}}=|j_{1},j_{2}>_{++}\otimes_{s}|j_{3},j_{4}>_{--}, 
\quad j_{1},j_{3} \in {\mathbb Z}_{\ge 0}, \quad 
j_{2},j_{4} \in \{0,1 \}, 
\label{wjjjj-fac1}
\end{eqnarray}
where $|j_{1},j_{2}>_{++}$ and $|j_{3},j_{4}>_{--}$ denote the basis
vectors in 
$W_{1}^{++}(xq^{m+\frac{1}{2}})$ and 
$\overline{W}_{1}^{--}(xq^{-m-\frac{1}{2}})$, defined in 
\eqref{w1b--} and \eqref{w1++}, respectively.
The parity of these vectors is 
$$p(w_{j_{1},j_{2},j_{3},j_{4}})= j_{2}+j_{4} \pmod 2.$$
As usual, we assume that $w_{j_{1},j_{2},j_{3},j_{4}} \equiv 0$, 
if the indices $j_{1},j_{2},j_{3},j_{4}$ lie outside the domain
specified in (\ref{wjjjj-fac1}).  
Taking into account \eqref{w1b--}, \eqref{w1++} and the formula for
the co-multiplication \eqref{comul1}, one can find the
action of the generators of ${\mathcal B}_{+}$,
\be
\begin{array}{rcl}
e_{0}w_{j_{1},j_{2},j_{3},j_{4}}
&=& \ds w_{j_{1},j_{2}+1,j_{3},j_{4}}+
(-1)^{j_{2}}q^{j_{1}}w_{j_{1},j_{2},j_{3},j_{4}+1}, \\[.4cm]
e_{1}w_{j_{1},j_{2},j_{3},j_{4}}
&=& \ds \frac{1}{q-q^{-1}}
\left(
q^{-j_{1}-j_{2}}[j_{1}]w_{j_{1}-1,j_{2},j_{3},j_{4}}-
q^{-2j_{1}-j_{2}+j_{3}+j_{4}}[j_{3}]w_{j_{1},j_{2},j_{3}-1,j_{4}}
\right), \\[.5cm]
e_{2}w_{j_{1},j_{2},j_{3},j_{4}}
&=& \ds -x
\left( q^{m+1}w_{j_{1}+1,j_{2}-1,j_{3},j_{4}}+
(-1)^{j_{2}}
q^{j_{1}+j_{2}-m-1}w_{j_{1},j_{2},j_{3}+1,j_{4}-1}\right), \\[.4cm]
(h_{0},h_1,h_2)\,
w_{j_{1},j_{2},j_{3},j_{4}}
&=& \ds 
(j_{1}+j_{3},-2j_{1}-j_{2}-2j_{3}-j_{4},
j_{1}+j_{2}+j_{3}+j_{4})\,w_{j_{1},j_{2},j_{3},j_{4}}, 
\end{array}
\ee
It is convenient to define vectors, with the same weights, 
\be
w^{(1)}_{n,j}=w_{j,0,n-j,0}\,, \quad
w^{(2)}_{n,j}=w_{j,0,n-j,1}\,, \quad
w^{(3)}_{n,j}=w_{j,1,n-j,0}\,, \quad
w^{(4)}_{n,j}=w_{j,1,n-j,1}\,,
\ee
where $ 0 \le j \le n $, $n \in {\mathbb Z}_{\ge 0}$.
Introduce the following vectors,
\begin{eqnarray}
\begin{array}{rcl}
 v^{(1)}_{n,j}&=& \ds q^{-{j}/{2}-m}\,\lambda_j\,x\,
\sum_{k=0}^{n} 
q^{-k(k-n-2j-2)}
 \qbinom{n}{k} 
w^{(1)}_{j+n,k}
, \\[.5cm]
v^{(2)}_{n,j}&=& \ds q^{-{3j}/{2}}(q-q^{-1})\,\lambda_j\,
\sum_{k=0}^{n} 
q^{-k(k-n-2j-3)}
 \qbinom{n}{k} 
w^{(2)}_{j+n,k}
, \\[.5cm]
v^{(3)}_{n,j} &=& \ds q^{+{j}/{2}-m}\,\lambda_j\,x\,
\sum_{k=0}^{n} 
q^{-k(k-n-2j-2)}
 \qbinom{n}{k} 
(w^{(3)}_{j+n,k}+q^{2m-2j+k}w^{(2)}_{j+n,k})
, \\[.5cm]
v^{(4)}_{n,j}&=& \ds -q^{-{j}/{2}}(q-q^{-1})\,\lambda_j\,
\sum_{k=0}^{n} 
q^{-k(k-n-2j-3)} 
 \qbinom{n}{k} 
w^{(4)}_{j+n,k}, 
\end{array}
\end{eqnarray}
where $ n,j \in {\mathbb Z}_{\ge 0}$ and 
$\lambda_j=q^{-{j^2}/{2}}\, (q^{-1}-q)^j$.
Their parities are given by 
$$p(v^{(2)}_{n,k})=p(v^{(3)}_{n,k})=1,\qquad
p( v^{(1)}_{n,k})=p(v^{(4)}_{n,k})=0.$$ 
The action of 
${\mathcal B}_{+}$ on these vectors is as follows 
\begin{subequations} \label{actions-w1w1b}
\be
\begin{array}{lll}
 e_{0}v^{(1)}_{n,j}=-x[m-j]v^{(2)}_{n,j}+q^{-j}v^{(3)}_{n,j},\quad 
& e_{1}v^{(1)}_{n,j}=[j]v^{(1)}_{n,j-1}, \quad  
&e_{2}v^{(1)}_{n,j}=0, \\[.3cm]
 e_{0}v^{(2)}_{n,j}=-q^{-j}v^{(4)}_{n,j}, 
& e_{1}v^{(2)}_{n,j}=[j]v^{(2)}_{n,j-1}, 
& e_{2}v^{(2)}_{n,j}=v^{(1)}_{n,j+1}, \\[.3cm]
 e_{0}v^{(3)}_{n,j}=-x[m-j]v^{(4)}_{n,j}, 
& e_{1}v^{(3)}_{n,j}=[j]v^{(3)}_{n,j-1}, 
& e_{2}v^{(3)}_{n,j}=-xq^{m-j-1}v^{(1)}_{n+1,j}, \\[.3cm]
 e_{0}v^{(4)}_{n,j}=0, 
& e_{1}v^{(4)}_{n,j}=[j]v^{(4)}_{n,j-1}, 
& e_{2}v^{(4)}_{n,j}=v^{(3)}_{n,j+1}+xq^{m-j-2}v^{(2)}_{n+1,j},\\[.3cm] 
\end{array}
\ee
and 
\begin{eqnarray}
\begin{array}{rclrrrl}
(h_{0},h_{1},h_{2}) \ v^{(1)}_{n,j}&=&
( & n+j, & -2n-2j, & n+j & ) \ v^{(1)}_{n,j}, \\[.25cm]
(h_{0},h_{1},h_{2}) \ v^{(2)}_{n,j}&=&
( & n+j, & -2n-2j-1, & n+j+1 & ) \ v^{(2)}_{n,j}, \\[.25cm]
(h_{0},h_{1},h_{2}) \ v^{(3)}_{n,j}&=&
( & n+j, & -2n-2j-1, & n+j+1 & ) \ v^{(3)}_{n,j}, \\[.25cm] 
(h_{0},h_{1},h_{2}) \ v^{(4)}_{n,j}&=&
( & n+j, & -2n-2j-2, & n+j+2 & ) \ v^{(4)}_{n,j}. \\[.1cm]
\end{array}
\end{eqnarray}
\end{subequations}
Introduce vector spaces 
\be\begin{array}{l}
W_{2n},\ \ \ \qquad \mbox{spanned by the vectors}\ \  
\Big\{v^{(1)}_{p,j},\, v^{(2)}_{p,j},\, 
v^{(3)}_{p,j},\, v^{(4)}_{p,j}\Big\}_{p=n,\,j=0}^{\infty\quad\,
  \infty}\ ,\\[.6cm]
W_{2n+1},\qquad \mbox{spanned by the vectors}\ \  
\Big\{v^{(1)}_{p+1,j},\, v^{(2)}_{p+1,j},\, v^{(3)}_{p,j},\, v^{(4)}_{p,j}
\Big\}_{p=n,\, j=0}^{\infty\quad\,\infty}\ ,
\end{array}
\ee
where $n \in {\mathbb Z}_{\ge 0}$.  
By construction, 
\begin{eqnarray}
W_{1}^{++}(xq^{m+\frac{1}{2}})\otimes_{s}\overline{W}_{1}^{--}(xq^{-m-\frac{1}{2}})
 = W^{(0)} \supset W^{(1)} \supset W^{(2)} \dots 
\end{eqnarray}
Examining (\ref{actions-w1w1b}), we find that 
\begin{itemize}
\item[(i)]
for any $n \in {\mathbb Z}_{\ge 0}$, $W^{(n+1)}$ is an invariant subspace 
of $W^{(n)}$ with respect 
to the action of ${\mathcal B}_{+}$. 
\item[(ii)] 
for any $n \in {\mathbb Z}_{\ge 0}$, 
the following isomorphisms with respect to the action of ${\mathcal B}_{+}$
 take place
\begin{eqnarray}
W^{(2n)}/W^{(2n+1)} & \simeq & \pi^{+}_{(m,0,0)}(x)[n+m,n], \\[.3cm]
W^{(2n+1)}/W^{(2n+2)}  & \simeq & \pi^{+}_{(m,0,0)}(x)[n+m,n+1], 
\label{iso-2}
\end{eqnarray}
except for that 
the parities of all the vectors on one side of the second formula,  
(\ref{iso-2}),
need to be inverted (note that $v^{(1)}_{n,j}$ is even and $v^{(3)}_{n,j}$ 
is odd). Remind that the representation $\pi^+_{(m,0,0)}(x)$ is
defined in \eqref{piplus}. 

\end{itemize}
Using these results and the identity (\ref{shift-id}), we obtain 
\be
\begin{array}{l}
\ds
{\str}_{W_{1}^{++}(xq^{m+\frac{1}{2}})\otimes_{s}\overline{W}_{1}^{--}(xq^{-m-\frac{1}{2}})}
(z_{1}^{-h_{0}}z_{2}^{h_{2}}\overline{{\mathcal R}})=
{\str}_{W^{(0)}}
(z_{1}^{-h_{0}}z_{2}^{h_{2}}\overline{{\mathcal R}}) \\[.5cm]
\ds
\qquad=\sum_{n=0}^{\infty}
{\str}_{W^{(n)}/W^{(n+1)}}
(z_{1}^{-h_{0}}z_{2}^{h_{2}}\overline{{\mathcal R}})  \\[.5cm]
\ds
\qquad=
\sum_{n=0}^{\infty}
{\str}_{\pi^{+}_{(m,0,0)}(x)[n+m,n]}
(z_{1}^{-h_{0}}z_{2}^{h_{2}}\overline{{\mathcal R}})
-
\sum_{n=0}^{\infty}
{\str}_{\pi^{+}_{(m,0,0)}(x)[n+m,n+1]}
(z_{1}^{-h_{0}}z_{2}^{h_{2}}\overline{{\mathcal R}})
\\[.5cm]
\ds
\qquad =
\sum_{n=0}^{\infty}
{\str}_{\pi^{+}_{(m,0,0)}(x)}
(z_{1}^{-h_{0}-n-m}z_{2}^{h_{2}+n}\overline{{\mathcal R}})
-
\sum_{n=0}^{\infty}
{\str}_{\pi^{+}_{(m,0,0)}(x)}
(z_{1}^{-h_{0}-n-m}z_{2}^{h_{2}+n+1}\overline{{\mathcal R}})  \\[.5cm]
\ds 
\qquad=
\frac{(1-z_{2})z_{1}^{1-m}}{z_{1}-z_{2}}
{\str}_{\pi^{+}_{(m,0,0)}(x)}
(z_{1}^{-h_{0}}z_{2}^{h_{2}}\overline{{\mathcal R}}).
\end{array}
\ee
Then using the definitions (\ref{Q-def}), (\ref{normali1}), 
 (\ref{Qbar-def}), (\ref{normali2}),(\ref{cij-def}), (\ref{normali4}), 
 (\ref{T-infi})  
one finally arrive to (\ref{fac1}). This completes the proof of the
relation \eqref{wrons1}.

\nsection{Concluding remarks}
The Baxter's ${\bf Q}$-operators find many important
applications in the theory of integrable quantum systems. 
In this paper we have developed an algebraic theory of the ${\bf
  Q}$-operators for solvable models associated with the quantized affine 
algebra $U_q(\widehat{sl}(2|1))$, extending previously known results 
for $U_q(\widehat{sl}(2))$ \cite{BLZ97,BLZ99}  and 
$U_q(\widehat{sl}(3))$ \cite{BHK} (see also \cite{Hikami01} and \cite{Kojima08}
for $U_q(\widehat{sl}(n))$ case).

Our general formalism has been 
illustrated by two representative cases: the 3-state lattice model
and a continuous quantum field theory, associated with the AKNS soliton
hierarchy. Here we assumed a generic value of the deformation
parameter $q\not=1$. The isotropic case $q=1$ can be obtained by a
more or less straightforward limiting procedure (though requires additional
considerations, similar to \cite{Baz08}) and will be considered
elsewhere. 

Finally, note very useful connections between functional relations 
in solvable models with the theory of the (super) characters 
and symmetric functions. Namely we refer to the Weyl {\em first} and {\em
  second} formulae for the Schur functions. Actually, there two
different, but related, ``second'' formulae, often called the 
Jacobi-Trudy and  Giambelli formulae, respectively. 
These formulae have super-symmetric generalizations. They are referred
to by adding the adjective ``super-symmetric'' to the respective name
(except that the super-symmetric analog of the first Weyl formula is usually 
called the Sergeev-Pragacz formula).   
These connections are discussed in the Appendix C.

\section*{Acknowledgments}  
We thank  M. Bortz, P.G. Bouwknegt, M. Jimbo,
S. M. Khoroshkin and H. Yamane  for useful discussions.
We also thank V. Kazakov and M. Staudacher for their interest to this
work and numerous stimulating discussions.  
ZT would like to thank the members of the integrable system group at 
Australian National University, especially, M.T.Batchelor and X.W. Guan, 
for their kind hospitality during his stay at the ANU, where a part of this 
work was done. ZT was supported by Grant-in-Aid for Scientific
Research from JSPS, \#16914018,   
Bilateral Joint Projects ``Solvable models and their thermodynamics in 
statistical mechanics and field theory" from JSPS, 
Grant-in-Aid for Young Scientists, B \#19740244. 
The main results of this paper have been previously reported on a
number of conferences%
\footnote{
``Workshop and Summer School: 
From Statistical Mechanics to Conformal and Quantum Field Theory'', 
the university of Melbourne, January, 2007 
[http://www.smft2007.ms.unimelb.edu.au/program/LectureSeries.html]; 
meeting of the Physical Society of Japan, March, 2007;
``Physics and Mathematics of interacting quantum systems 
in low dimensions'', the University of Tokyo (Kashiwa), 24-26 May, 2007
[http://oshikawa.issp.u-tokyo.ac.jp/pmiqs/poster.html];  
La 79eme Rencontre entre physiciens theoriciens et mathematiciens 
``Supersymmetry and Integrability", IRMA Strasbourg, June, 2007 [http://www-irma.u-strasbg.fr/article383.html];  
Annual Statistical Mechanics Meeting, Australian National University,
December, 2007.}.
%

\app{Highest weight representations of $U_q(gl(2|1))$}
\addcontentsline{toc}{section}{Appendices}
\addcontentsline{toc}{subsection}{A. Highest weight representations of
  $U_q(gl(2|1))$} 
\label{appA}
Here we briefly discuss the representation theory of $U_q(gl(2|1))$. 
Let $V=V_{0}\oplus V_{1}$ be a ${\mathbb Z}_{2}$ graded vector space 
with the parity $p$ such that $p(v)=0$ (even) for $v \in V_{0}$ and 
$p(v)=1$ (odd) for $v \in V_{1}$. There is a basis $\{v_{i}\}$ of $V$,
 called a homogeneous basis, such that $v_{i} \in V_{0}$ or $v_{i} \in V_{1}$. 
For any linear operator $A \in {\mathrm End}(V)$, its 
matrix expression in this basis reads $Av_{k}=\sum_{j}v_{j}A_{jk}$. 
The supertrace of $A$ is defined as ${\str}A=\sum_{j}(-1)^{p(v_{j})}A_{jj}$. 
\sapp{Finite-dimensional representations}
In this paper we only need the {\em highest weight representations} 
of $U_q(gl(2|1))$.  
Any such representation can be constructed 
by the so-called induced module construction, which was 
proposed by Kac \cite{Kac78} for the $q=1$ case 
and generalized to the $q$-generic case in \cite{Zhang93}. 
It is built on the highest weight vector $|0>$ 
\begin{eqnarray}
E_{12}\,|0>=E_{23}\,|0>=0,\qquad E_{ii}\,|0>=\mu_i\,|0>,\qquad i=1,2,3.
\label{hw-def}
\end{eqnarray}
with the weight $\mu=(\mu_1,\mu_2,\mu_3)$, where  
$\mu_{1},\mu_{2},\mu_{3} \in {\mathbb C}$.
The module is finite-dimensional when  $\mu_{1}-\mu_{2} \in
{\mathbb Z}_{\ge 0}$ and will be denoted, in this case,
 as $\widehat{\pi}_{\mu}$.

The pair of numbers $[b_{1},b_{2}]$, where $b_{1}=\mu_{1}-\mu_{2}$ and
 $b_{2}=\mu_{2}+\mu_{3}$, is called the Kac-Dynkin label 
of $\widehat{\pi}_{\mu}$. 
Note that two modules $\widehat{\pi}_{(\mu_1,\mu_2,\mu_3)}$ and
 $\widehat{\pi}_{(\mu_1+\eta,\mu_2+\eta,\mu_3-\eta)}$, where $\eta$ is
 arbitrary, have the same Kac-Dynkinlabel. 
In general, $\widehat{\pi}_{\mu}$ is not an irreducible representation. 
The corresponding irreducible representation, 
obtained by factoring out a maximal proper invariant subspace 
of $\widehat{\pi}_{\mu}$, will be denotes as $\pi_{\mu}$. 
Sometimes, we will use the notation $\pi_{\mu}^{[p]}$ to indicate  
the parity $p$ of the highest weight vector of $\pi_{\mu}$. 
There are three types of finite dimensional representations:
\begin{itemize}
\item[(i)] typical representations
\be\label{typ-rep}
\dim \pi_\mu=4(\mu_1-\mu_2+1),\qquad 
(\mu_2+\mu_3)(\mu_{1}+\mu_{3}+1) \ne 0,\qquad \mu_1-\mu_2\in {\mathbb
  Z}_{\ge0},  
\ee
\item[(ii)] 
class-1 atypical representation, 
\be\label{cl1-rep}
\dim \pi_\mu=2(\mu_1-\mu_2)+3,\qquad \mu_{1}+\mu_{3}+1=0,
\qquad \mu_1-\mu_2\in {\mathbb Z}_{\ge-1},
\ee
\item[(iii)]
class-2 atypical representation, 
\be\label{cl2-rep}
\dim\pi_\mu=2(\mu_1-\mu_2)+1,\qquad \mu_2+\mu_3 = 0,
\qquad \mu_1-\mu_2\in {\mathbb Z}_{\ge0}.
\ee
\end{itemize}
Note that the case $\mu_{1}-\mu_{2}=-1$ in \eqref{cl1-rep}
is special; see (\ref{special-one}) below.  
\ssapp{Typical representations}
Below we present the action of the generators of $U_q(gl(2|1))$
 on a basis of $\widehat{\pi}_{\mu}$. Let $2l=\mu_{1}-\mu_{2}  
\in {\mathbb Z}_{\ge 0}$. 
Here we use a (slightly modified) basis from \cite{Ky-Van03}.
There are 4 sets of vectors 
\be
\{ w^{(1)}_{j}\}_{j=0}^{2l}\subset V^{(1)}, \qquad 
\{w^{(2)}_{j}\}_{j=0}^{2l+1}\subset V^{(2)}, \qquad 
\{w^{(3)}_{j}\}_{j=0}^{2l-1}\subset V^{(3)}, \qquad 
\{w^{(4)}_{j}\}_{j=0}^{2l}\subset V^{(4)},
\label{cond-pihat}
\ee
where $V^{(a)}$, $a=1,2,3,4$  denote the vector spaces spanned by
these sets.  Below we allow the index $j$ to take arbitrary integer
values and assume that $w^{(a)}_{j} \equiv 0$, if $j$ 
lies outside the intervals specified in (\ref{cond-pihat}) for each
value of $a=1,2,3,4$. 
The parities are $p(V^{(1)})=p(V^{(4)})=0$, $p(V^{(2)})=p(V^{(3)})=1$.
For $j \in {\mathbb Z}_{\ge 0}$, 
the action of the generators of  $U_q(gl(2|1))$  on these vectors 
reads  
\begin{eqnarray}
&& 
\begin{array}{rclrrrl}
(E_{11},E_{22},E_{33})\ w^{(1)}_{j}&=&
(&\mu_{1}-j,& \mu_{2}+j,& \mu_{3}&)\ w^{(1)}_{j}, \\[.25cm]
(E_{11},E_{22},E_{33})\ w^{(2)}_{j}&=&
(&\mu_{1}-j,&\mu_{2}-1+j,&\mu_{3}+1&)\ w^{(2)}_{j},\\[.25cm]
(E_{11},E_{22},E_{33})\ w^{(3)}_{j}&=&
(&\mu_{1}-1-j,&\mu_{2}+j,&\mu_{3}+1&)\ w^{(3)}_{j},\\[.25cm]
(E_{11},E_{22},E_{33})\ w^{(4)}_{j}&=&
(&\mu_{1}-1-j,&\mu_{2}-1+j,&\mu_{3}+2&)\ w^{(4)}_{j}, \\[.1cm]
\end{array}
\nonumber \\
&&
\begin{array}{rclrcl}
E_{12}w^{(1)}_{j}&=& [j]w^{(1)}_{j-1}, &
E_{12}w^{(2)}_{j}&=&[j]w^{(2)}_{j-1}, \\[.2cm]
E_{12}w^{(3)}_{j}&=&[j]w^{(3)}_{j-1}, &
E_{12}w^{(4)}_{j}&=&[j]w^{(4)}_{j-1}, \\[.2cm]
E_{21}w^{(1)}_{j}&=&[2l-j]w^{(1)}_{j+1}, &
E_{21}w^{(2)}_{j}&=&[2l+1-j]w^{(2)}_{j+1}, \\[.2cm]
E_{21}w^{(3)}_{j}&=&[2l-1-j]w^{(3)}_{j+1}, &
E_{21}w^{(4)}_{j}&=&[2l-j]w^{(4)}_{j+1}, 
\end{array}
\nonumber \\[.1cm]
&& 
\begin{array}{rclrcl}
E_{23}w^{(1)}_{j}&=& \ds 0, &
E_{23}w^{(2)}_{j}&=& \ds \frac{[\mu_{2}+\mu_{3}][2l+1-j]}{[2l+1]}w^{(1)}_{j},\\[.3cm]
E_{23}w^{(3)}_{j}&=& \ds \frac{[\mu_{1}+\mu_{3}+1]}{[2l+1]}w^{(1)}_{j+1},&
E_{23}w^{(4)}_{j}&=& \ds  
\frac{[\mu_{1}+\mu_{3}+1]}{[2l+1]}w^{(2)}_{j+1}-
\frac{[\mu_{2}+\mu_{3}][2l-j]}{[2l+1]}w^{(3)}_{j}, \\[.3cm]
E_{32}w^{(1)}_{j}&=& \ds  w^{(2)}_{j}+[j]w^{(3)}_{j-1}, &
E_{32}w^{(2)}_{j}&=&\ds [j]w^{(4)}_{j-1}, \\[.2cm] 
E_{32}w^{(3)}_{j}&=& \ds  -w^{(4)}_{j}, & 
E_{32}w^{(4)}_{j}&=&\ds 0,
\end{array}
\nonumber \\[.1cm]
&& \hspace{5pt} E_{31}w^{(1)}_{j} \; \; =\; \;  
q^{-2\mu_{3}}
\Bigl\{-
q^{-1-\mu_{1}}w^{(2)}_{j+1}+q^{-\mu_{2}}[2l-j]w^{(3)}_{j}
\Bigr\}, \label{atyp-general}\\[.2cm]
&& \hspace{5pt}  E_{31}w^{(2)}_{j} \; \; =\; \;  
q^{-1-\mu_{2}-2\mu_{3}}[2l+1-j]w^{(4)}_{j},\nonumber \\[.1cm]
&&
\begin{array}{rclrcl}
E_{31}w^{(3)}_{j}&=&\ds
q^{-2-\mu_{1}-2\mu_{3}}w^{(4)}_{j+1},&
E_{31}w^{(4)}_{j}&=&\ds 0, \\[.2cm]
E_{13}w^{(1)}_{j}&=&\ds 0, & 
E_{13}w^{(2)}_{j}&=&\ds -
\frac{q^{1+\mu_{1}+2\mu_{3}}[j][\mu_{2}+\mu_{3}]}{[2l+1]}
w^{(1)}_{j-1}, 
\end{array}
\nonumber \\[.1cm]
&& \hspace{5pt}  E_{13}w^{(3)}_{j} \; \; =\; \; 
\frac{q^{\mu_{2}+2\mu_{3}}[\mu_{1}+\mu_{3}+1]}{[2l+1]}
w^{(1)}_{j}, \nonumber \\[.1cm]
&& \hspace{5pt}  E_{13}w^{(4)}_{j} \; \; =\; \; 
q^{1+2\mu_{3}} \Bigl\{
\frac{q^{\mu_{2}}[\mu_{1}+\mu_{3}+1]}
{[2l+1]}w^{(2)}_{j}+
\frac{q^{1+\mu_{1}}[j][\mu_{2}+\mu_{3}]}
{[2l+1]}w^{(3)}_{j-1} \Bigr\}. \nonumber
\end{eqnarray}
The vector $w^{(1)}_{0}$ is the highest weight vector
\footnote{The vector $w^{(2)}_{0}$ is the 
highest weight vector when $\mu_{1}-\mu_{2}=-1$.}
 (denoted as $|0>$ in
\eqref{hw-def}).  When the weights $(\mu_1,\mu_2,\mu_3)$ meet the
conditions \eqref{typ-rep} 
the above formulae define a typical irreducible representation 
$\pi_{\mu}$ of the dimension  
$\dim \pi_\mu=4(\mu_{1}-\mu_{2}+1)$. 
\ssapp{The 4-dimensional typical representation}
The typical representation ${\pi}_{(\mu_{1},\mu_{1},\mu_{3})}$ is 
4-dimensional; it is spanned by 
the vectors $w^{(1)}_{0},w^{(2)}_{0},w^{(2)}_{1},w^{(4)}_{0}$. From
\eqref{atyp-general}, one obtains
\begin{subequations}\label{4dt}
\be
\begin{array}{rclrrrl}
 (E_{11},E_{22},E_{33})\ w^{(1)}_{0}&=&
(&\mu_{1},&\mu_{1},&\mu_{3}&) \ w^{(1)}_{0},   \\[.25cm]
 (E_{11},E_{22},E_{33}) \ w^{(2)}_{0}&=& 
(&\mu_{1},& \mu_{1}-1, & \mu_{3}+1& ) \ w^{(2)}_{0}, \\[.25cm]
 (E_{11},E_{22},E_{33}) \ w^{(2)}_{1}&=& 
(& \mu_{1}-1,& \mu_{1},& \mu_{3}+1& ) \ w^{(2)}_{1}, \\[.25cm]
 (E_{11},E_{22},E_{33}) \ w^{(4)}_{0}&=& 
(& \mu_{1}-1,& \mu_{1}-1,& \mu_{3}+2& ) \ w^{(4)}_{0}, \\[.2cm]
\end{array}
\ee
\be
\begin{array}{rclrcl}
E_{12}\,w^{(2)}_{1}&=&w^{(2)}_{0},&
E_{21}\,w^{(2)}_{0}&=&w^{(2)}_{1}, \\[.2cm] 

E_{23}\,w^{(2)}_{0}&=&[\mu_{1}+\mu_{3}]w^{(1)}_{0},& 
E_{23}\,w^{(4)}_{0}&=&[\mu_{1}+\mu_{3}+1]w^{(2)}_{1},\\[.2cm]
E_{32}\,w^{(1)}_{0}&=&w^{(2)}_{0}, &
E_{32}\,w^{(2)}_{1}&=&w^{(4)}_{0},\\[.2cm]
E_{31}\,w^{(1)}_{0}&=&
-q^{-1-\mu_{1}-2\mu_{3}}\,w^{(2)}_{1},&
E_{31}\,w^{(2)}_{0}&=&q^{-1-\mu_{1}-2\mu_{3}}\,w^{(4)}_{0},\\[.2cm]
E_{13}\,w^{(2)}_{1}&=&-
q^{1+\mu_{1}+2\mu_{3}}[\mu_{1}+\mu_{3}]
w^{(1)}_{0} ,&
E_{13}\,w^{(4)}_{0}&=& q^{1+\mu_{1}+2\mu_{3}} 
[\mu_{1}+\mu_{3}+1]w^{(2)}_{0}. 
\end{array}
\ee
where all other matrix element vanish.
\end{subequations}
\ssapp{Class-1 atypical representation}
When the weights fall to the case \eqref{cl1-rep} the module 
$\widehat{\pi}_{(\mu_{1},\mu_{2},-1-\mu_{1})}$ has an invariant
subspace $(V^{(3)}+V^{(4)})$, where $V^{(a)}$ is defined in
\eqref{cond-pihat}.
The factor space
$\widehat{\pi}_{(\mu_{1},\mu_{2},-1-\mu_{1})}/(V^{(3)}+V^{(4)})$
corresponds to the $(2(\mu_{1}-\mu_{2})+3)$-dimensional class-1 
atypical representation $\pi_{(\mu_{1},\mu_{2},-1-\mu_{1})}^{[0]}$. 
The action of the generators of $U_q(gl(2|1))$ is obtained from 
\eqref{atyp-general} by dropping the vectors $w^{(3)}_j$ and $w^{(4)}_j$. 
For $j \in {\mathbb Z}_{\ge 0}$ one obtains 
\be
\begin{array}{l}
\begin{array}{rclrrrl}
(E_{11},E_{22},E_{33}) \ w^{(1)}_{j}&=&
(& \mu_{1}-j,& \mu_{2}+j,& -1-\mu_{1}& ) \ w^{(1)}_{j},  \\[.25cm]
(E_{11},E_{22},E_{33}) \ w^{(2)}_{j}&=& 
(& \mu_{1}-j,& \mu_{2}-1+j,& -\mu_{1}& ) \ w^{(2)}_{j},  \\[.2cm]
\end{array}\\[.5cm]
\begin{array}{rclrcl}
E_{12}w^{(1)}_{j}&=&[j]w^{(1)}_{j-1}, &
E_{12}w^{(2)}_{j}&=&[j]w^{(2)}_{j-1},  \\[.2cm]
E_{21}w^{(1)}_{j}&=&[2l-j]w^{(1)}_{j+1}, &
E_{21}w^{(2)}_{j}&=&[2l+1-j]w^{(2)}_{j+1}, \\[.2cm]
E_{23}w^{(2)}_{j}&=&-[2l+1-j]w^{(1)}_{j},&
E_{32}w^{(1)}_{j}&=&w^{(2)}_{j}, \\[.2cm]
E_{31}w^{(1)}_{j}&=&
-q^{1+\mu_{1}}
w^{(2)}_{j+1}, &
E_{13}w^{(2)}_{j}&=&q^{-1-\mu_{1}}[j]
w^{(1)}_{j-1},
\end{array}
\end{array}
 \label{class1-1}
\ee
where $2l=\mu_1-\mu_2$ and all other matrix elements vanish.
The basis of this representation consists of the vectors
\be
\{ w^{(1)}_{j}\}_{j=0}^{2l}\subset V^{(1)}, \qquad 
\{w^{(2)}_{j}\}_{j=0}^{2l+1}\subset V^{(2)}. \qquad 
\label{cl1-bas}
\ee
Note, that $w^{(2)}_{2l+1}$  
is the lowest weight vector 
\be
E_{21}\,w^{(2)}_{2l+1}=E_{32}\,w^{(2)}_{2l+1}=0, \qquad 
(E_{11},E_{22},E_{33})\,w^{(2)}_{2l+1}=
(\mu_{2}-1,\mu_{1},-\mu_{1})\,w^{(2)}_{2l+1}.
\ee
Further, for $\mu_{1}-\mu_{2} \in {\mathbb Z}_{\ge 1}$ 
the invariant subspace $V^{(3)}+V^{(4)}$ is 
isomorphic to $\pi_{(\mu_{1}-1,\mu_{2},-\mu_{1})}^{[1]}$,
(while in the special point $\mu_1-\mu_2=0$ it is 
isomorphic to 
$\pi_{(\mu_{1}-1,\mu_{1}-1,1-\mu_{1})}^{[0]}$). 
In this way one obtains 
\begin{eqnarray}
\widehat{\pi}_{(\mu_{1},\mu_{2},-1-\mu_{1})}/
\pi_{(\mu_{1}-1,\mu_{2},-\mu_{1})}^{[1]} &\simeq& 
\pi_{(\mu_{1},\mu_{2},-1-\mu_{1})}^{[0]}, 
\qquad \mu_{1}-\mu_{2} \in {\mathbb Z}_{\ge 1}, \\[.2cm]
\widehat{\pi}_{(\mu_{1},\mu_{1},-1-\mu_{1})}/
\pi_{(\mu_{1}-1,\mu_{1}-1,1-\mu_{1})}^{[0]} &\simeq& 
\pi_{(\mu_{1},\mu_{1},-1-\mu_{1})}^{[0]}, \qquad \mu_1-\mu_2=0.
\label{iso5}
\end{eqnarray}
In the special case $\mu_{1}-\mu_{2}=-1$ the 
representation $\widehat{\pi}_{(\mu_{1},\mu_{1}+1,-\mu_{1}-1)}$ is
one-dimensional. It is spanned by 
the only vector $w^{(2)}_{0}$ whose weight is equal to 
$(\mu_{1},\mu_{1},-\mu_{1})$ and the parity is odd,
\be
\widehat{\pi}_{(\mu_{1},1+\mu_{1},-1-\mu_{1})} \simeq 
\pi_{(\mu_{1},\mu_{1},-\mu_{1})}^{[1]}. 
 \label{special-one}
\ee
\ssapp{Class-2 atypical representation}

As before, let $V^{(a)}$, $a=1,2,3,4$, be the vector spaces, defined in
\eqref{cond-pihat}.  In the case \eqref{cl2-rep} the module
$\widehat{\pi}_{(\mu_{1},\mu_{2},-\mu_{2})}$ is reducible; it contains
an invariant subspace $V^{(2)}\oplus V^{(4)}$.  
The factor space
$\widehat{\pi}_{(\mu_{1},\mu_{2},-\mu_{2})}/(V^{(2)}\oplus V^{(4)})$
corresponds to the $(2(\mu_{1}-\mu_{2})+1)$-dimensional class-2
atypical representation $\pi_{(\mu_{1},\mu_{2},-\mu_{2})}^{[0]}$.
To get explicit expressions for its matrix elements, one simply drops
all vectors $w^{(2)}_{j}$ and $w^{(4)}_{j}$ from \eqref{atyp-general}.
For $j \in {\mathbb Z}_{\ge 0}$ one obtains 
\be
\begin{array}{l}
\begin{array}{rclrrrl}
(E_{11},E_{22},E_{33}) \ w^{(1)}_{j}&=&
(& \mu_{1}-j,& \mu_{2}+j,& -\mu_{2}& ) \ w^{(1)}_{j}, \\[.25cm]
(E_{11},E_{22},E_{33}) \ w^{(3)}_{j}&=&
(& \mu_{1}-1-j,& \mu_{2}+j,& -\mu_{2}+1& ) \ w^{(3)}_{j}, \\[.1cm]
\end{array}\\[1.cm]
\begin{array}{rclrcl}
E_{12}w^{(1)}_{j} &=& [j]w^{(1)}_{j-1}, &
E_{12}w^{(3)}_{j} &=& [j]w^{(3)}_{j-1}, \\[.2cm]
E_{21}w^{(1)}_{j} &=& [2l-j]w^{(1)}_{j+1}, &
E_{21}w^{(3)}_{j} &=& [2l-1-j]w^{(3)}_{j+1}, 
\\[.2cm]
E_{23}w^{(3)}_{j} &=& w^{(1)}_{j+1},&
E_{32}w^{(1)}_{j} &=& [j]w^{(3)}_{j-1}, 
\\[.2cm]
E_{31}w^{(1)}_{j} &=& 
q^{\mu_{2}}[2l-j]w^{(3)}_{j}, &
E_{13}w^{(3)}_{j} &=& q^{-\mu_{2}}w^{(1)}_{j},
\end{array}
\label{class2-1}
\end{array}
\ee 
where all other matrix elements vanish.  
The basis of this representation consists of the vectors 
\be
\{ w^{(1)}_{j}\}_{j=0}^{2l}\subset V^{(1)}, \qquad 
\{w^{(3)}_{j}\}_{j=0}^{2l-1}\subset V^{(3)}, \qquad 
\label{cl2-bas}
\ee
of the parities are $p(w^{(1)}_{j})=0$ and $p(w^{(3)}_{j})=1$.
For 
$2l=\mu_1-\mu_2\in {\mathbb Z}_{\ge1}$ 
the vector $w^{(3)}_{2l-1}$ is the lowest weight vector 
\be
E_{21}w^{(3)}_{2l-1}=E_{32}w^{(3)}_{2l-1}=0,\qquad 
(E_{11},E_{22},E_{33})w^{(3)}_{2l-1}=
(\mu_{2},\mu_{1}-1,1-\mu_{2})w^{(3)}_{2l-1}.\label{class2-LW2}
\ee
For $\mu_{1}-\mu_{2}=0$, 
the representation $\pi_{(\mu_{1},\mu_{1}, -\mu_{1})}^{[0]}$ becomes 
one dimensional (with the only vector $w^{(1)}_{0}$).

A simple inspection shows that the invariant subspace $(V^{(2)}\oplus
V^{(4)})$ of $\widehat{\pi}_{(\mu_{1},\mu_{2},-\mu_{2})}$ is 
isomorphic to $\pi_{(\mu_{1},\mu_{2}-1,1-\mu_{2})}^{[1]}$, just defined in
\eqref{class2-1}. Thus, one
obtains  
\begin{eqnarray}
\widehat{\pi}_{(\mu_{1},\mu_{2},-\mu_{2})}
/\pi_{(\mu_{1},\mu_{2}-1,1-\mu_{2})}^{[1]}
\simeq  
\pi_{(\mu_{1},\mu_{2},-\mu_{2})}^{[0]},
\qquad\mu_{1}-\mu_{2} \in {\mathbb Z}_{\ge 0}.
\label{iso4}
\end{eqnarray}
%
\sapp{Infinite-dimensional representations of $U_{q}(gl(2|1))$} 
\label{sect-inf-dim}
Below
we will not assume any integrality conditions for the difference
$\mu_{1}-\mu_{2}$, unless otherwise is explicitly stated.  Introduce
two types of infinite-dimensional representations by the formulae
(\ref{class1-1}) and (\ref{class2-1}) assuming that the weights
$\mu_1,\mu_2\in {\mathbb C}$ are now arbitrary and the index $j$ takes
an infinite number of integer values $j=0,1,2,\ldots,\infty$. With
these conventions 
\begin{itemize}
\item[(iv)] Eq.\eqref{class1-1} defines a 
class-1 infinite-dimensional representation     
$\pi_{(\mu_{1},\mu_{2},-1-\mu_{1})}^{-}$, 
\item[(v)] Eq.\eqref{class2-1} defines a 
class-2 infinite-dimensional representations  
 $\pi_{(\mu_{1},\mu_{2},-\mu_{2})}^{+}$.
\end{itemize}

\ssapp{Class-1 infinite-dimensional representation
  $\pi_{(\mu_{1},\mu_{2},-1-\mu_{1})}^{-}$}\label{cl1-inf-dim} 
The action of generators for this representation is defined by
  \eqref{class1-1}, the same formulae as for the finite dimensional
  representation $\pi_{(\mu_{1},\mu_{2},-1-\mu_{1})}^{-}$, except that
  the index $j$ now runs infinitely many values $j\in {\mathbb Z}_{\ge 0}$.
The basis consists of the vectors $\{w^{(1)}_{j}\}_{j=0}^\infty$
and $ \{w^{(2)}_{j}\}_{j=0}^\infty$ with 
the parities $p(w^{(1)}_{j})=0$, $p(w^{(2)}_{j})=1$.
The highest weight vector is $w^{(1)}_{0}$. 
We assume also that $w^{(1)}_{j}\equiv 0, 
w^{(2)}_{j}\equiv 0$, if $j<0$. 

For generic values of $\mu_1,\mu_2$ the representation 
$\pi_{(\mu_{1},\mu_{2},-1-\mu_{1})}^{-}$ is irreducible. 
It becomes reducible iff $\mu_{1}-\mu_{2} \in {\mathbb Z}_{\ge -1}$. 
First consider the main case $\mu_{1}-\mu_{2} \in {\mathbb Z}_{\ge 0}$.
In this case $\pi_{(\mu_{1},\mu_{2},-1-\mu_{1})}^{-}$ contains the 
 class-1 atypical finite dimensional representation 
$\pi_{(\mu_{1},\mu_{2},-1-\mu_{1})}^{[0]}$ as an invariant subspace,
 while the factor-module 
\be
\pi_{(\mu_{1},\mu_{2},-1-\mu_{1})}^{-}/\pi_{(\mu_{1},\mu_{2},-1-\mu_{1})}^{[0]}
\simeq \pi_{(\mu_{2}-1,\mu_{1}+1,-\mu_{1}-1)}^{+}, \qquad 
\mu_{1}-\mu_{2} \in {\mathbb Z}_{\ge 0}, \label{cl1-fac1}
\ee
is isomorphic to the infinite-dimensional class-2 representation  
$\pi_{(\mu_{2}-1,\mu_{1}+1,-\mu_{1}-1)}^{+}$, briefly described above in the
beginning of Sect.~\ref{sect-inf-dim} (see Sect.~\ref{cl2-inf-dim} for
more details).
To see this let us write basis vectors in the factor module 
$\pi_{(\mu_{1},\mu_{2},-1-\mu_{1})}^{-}/
\pi_{(\mu_{1},\mu_{2},-1-\mu_{1})}^{[0]} $ as
 $v^{(1)}_{j}=w^{(1)}_{j+2l+1}$ and $v^{(2)}_{j}=w^{(2)}_{j+2l+2}$,
$j=0,1,2,\ldots \infty$.
Then from \eqref{class1-1} one obtains 
\begin{eqnarray}
&&
\begin{array}{rclrrrl}
(E_{11},E_{22},E_{33}) \ v^{(1)}_{j} &=& 
( & \mu_{2}-j-1, & \mu_{1}+j+1, & -1-\mu_{1} & ) \ v^{(1)}_{j},  \\[.25cm]
( E_{11}, E_{22}, E_{33}) \ v^{(2)}_{j} &=& 
( & \mu_{2}-j-2, & \mu_{1}+1+j, & -\mu_{1} & ) \ v^{(2)}_{j}, \\[.2cm]
\end{array}
\nonumber \\
&& 
\begin{array}{rclrcl}
E_{12}v^{(1)}_{j} &=& [j+2l+1]v^{(1)}_{j-1}, &
E_{12}v^{(2)}_{j} &=& [j+2l+2]v^{(2)}_{j-1},  \\[.2cm]
E_{21}v^{(1)}_{j} &=& -[j+1]v^{(1)}_{j+1}, &
E_{21}v^{(2)}_{j} &=& -[j+1]v^{(2)}_{j+1}, \\[.2cm]
E_{23}v^{(2)}_{j} &=& [j+1]v^{(1)}_{j+1}, &
E_{32}v^{(1)}_{j} &=& v^{(2)}_{j-1}, \\[.2cm]
E_{31}v^{(1)}_{j} &=& -q^{1+\mu_{1}}v^{(2)}_{j}, &  
E_{13}v^{(2)}_{j} &=& q^{-1-\mu_{1}}[j+2l+2]v^{(1)}_{j}.
\end{array}
  \label{class1-fac-1}
\end{eqnarray}
$2l=\mu_1-\mu_2$ and all other matrix elements vanish. It is not  difficult
to check that \eqref{class1-fac-1} becomes identical to 
\eqref{class2-1}, provided 
the quantities $\mu_1$, $\mu_2$ and $2l$ in \eqref{class2-1} 
are replaced with $\mu_2-1$, $\mu_1+1$ and $-2-2l$,
respectively, and the basis in \eqref{class2-1} is re-scaled as
\be
w^{(1)}_j={[2l+2]}\,v^{(1)}_j\Big/ {\textstyle {\qbinom{j+2l+1}{j}}}, 
\qquad 
w^{(3)}_j= v^{(2)}_j\Big/{\textstyle {\qbinom{j+2l+2}{j}}} \ .
\ee

In the special case $\mu_1-\mu_2=-1$ the invariant subspace of 
  $\pi_{(\mu_{1},\mu_{2},-1-\mu_{1})}^{-}$ is one-dimensional; the
  analog of 
\eqref{cl1-fac1}
reads 
\be
\pi_{(\mu_{1},\mu_{1}+1,-\mu_{1}-1)}^{-}/
\pi_{(\mu_{1},\mu_{1},-\mu_{1})}^{[1]}
\simeq \pi_{(\mu_{1},\mu_{1}+1,-\mu_{1}-1)}^{+}, \qquad 
\mu_{1}-\mu_{2}=-1, \label{cl1-fac2}
\ee

\ssapp{Class-2 infinite-dimensional representation
  $\pi_{(\mu_{1},\mu_{2},-\mu_{2})}^{+}$}\label{cl2-inf-dim}
This representation is defined by
  \eqref{class2-1}, i.e. by the same formulae as for the finite dimensional
  representation $\pi_{(\mu_{1},\mu_{2},-\mu_{2})}$, except that
  the index $j$ now runs infinitely many values $j\in Z_{\ge 0}$.
The basis consists of the vectors
$\{w^{(1)}_{j}\}_{j=0}^\infty$
and $ \{w^{(3)}_{j}\}_{j=0}^\infty$ with 
the parities $p(w^{(1)}_{j})=0$, $p(w^{(3)}_{j})=1$.
The highest weight vector is $w^{(1)}_{0}$. 
We assume also that $w^{(1)}_{j}\equiv 0, 
w^{(3)}_{j}\equiv 0$, if $j<0$. 

For generic values of $\mu_1,\mu_2$ the representation 
$\pi_{(\mu_{1},\mu_{2},-\mu_{1})}^{+}$ is irreducible. 
It becomes reducible iff $\mu_{1}-\mu_{2} \in {\mathbb Z}_{\ge 0}$. 
In this case it contains the 
 class-2 atypical finite dimensional representation 
$\pi_{(\mu_{1},\mu_{2},-\mu_{1})}^{[0]}$ as an invariant subspace,
 while the factor-module 
\be
\pi_{(\mu_{1},\mu_{2},-\mu_{2})}^{+}/\pi_{(\mu_{1},\mu_{2},-\mu_{2})}^{[0]}
\simeq \pi_{(\mu_{2}-1,\mu_{1}+1,-\mu_{2})}^{-}, \qquad 
2l=\mu_{1}-\mu_{2} \in {\mathbb Z}_{\ge 0}, 
\label{cl2-fac1}
\ee
is isomorphic to the infinite-dimensional class-1 representation  
$\pi_{(\mu_{2}-1,\mu_{1}+1,-\mu_{2})}^{-}$, considered above in 
Sect.\ref{cl1-inf-dim}. 
To see this let us write basis vectors in the factor module 
$\pi_{(\mu_{1},\mu_{2},-\mu_{2})}^{+}/
\pi_{(\mu_{1},\mu_{2},-\mu_{2})}^{[0]} $ as
 $u^{(1)}_{j}=w^{(1)}_{j+2l+1}$ and $u^{(3)}_{j}=w^{(3)}_{j+2l}$,
$j=0,1,2,\ldots \infty$.
Then from \eqref{class2-1} one obtains 
\begin{eqnarray}
&&
\begin{array}{rclrrrl}
(E_{11},E_{22},E_{33}) \ u^{(1)}_{j} &=& 
( & \mu_{2}-j-1, & \mu_{1}+j+1, & -\mu_{2} & ) \ u^{(1)}_{j},  \\[.25cm]
( E_{11}, E_{22}, E_{33}) \ u^{(3)}_{j} &=& 
( & \mu_{2}-1-j, & \mu_{1}+j, & -\mu_{2}+1 & ) \ u^{(3)}_{j}, \\[.2cm]
\end{array}
\nonumber 
\\
&&
\begin{array}{rclrcl}
E_{12}u^{(1)}_{j} &=& [j+2l+1]u^{(1)}_{j-1}, &
E_{12}u^{(3)}_{j} &=& [j+2l]u^{(3)}_{j-1}, \\[.2cm]
E_{21}u^{(1)}_{j} &=& -[j+1]u^{(1)}_{j+1}, &
E_{21}u^{(3)}_{j} &=& -[j+1]u^{(3)}_{j+1}, 
\\[.2cm]
E_{23}u^{(3)}_{j} &=& u^{(1)}_{j},&
E_{32}u^{(1)}_{j} &=& [j+2l+1]u^{(3)}_{j}, \\[.2cm]
E_{31}u^{(1)}_{j} &=& 
-q^{\mu_{2}}[j+1]u^{(3)}_{j+1}, &
E_{13}u^{(3)}_{j} &=& q^{-\mu_{2}}u^{(1)}_{j-1}.
\end{array}
 \label{class2-fac-1}
\end{eqnarray}
$2l=\mu_1-\mu_2$ and all other matrix elements vanish. It is not difficult
to check that \eqref{class2-fac-1} becomes identical to 
\eqref{class1-1}, provided 
the quantities $\mu_1$, $\mu_2$ and $2l$ in \eqref{class1-1} 
are replaced with $\mu_2-1$, $\mu_1+1$ and $-2-2l$,
respectively,
and the basis in \eqref{class1-1} is re-scaled as
\be
w^{(1)}_j=u^{(1)}_j\Big/{\textstyle \qbinom{j+2l+1}{j}}, 
\qquad w^{(2)}_j=[2l+1]\,u^{(3)}_j\Big/{\textstyle \qbinom{j+2l}{j}}\ .
\ee

\app{Representations of the Borel
  subalgebra ${\mathcal B}_{+}(U_q(\widehat{sl}(2|1)))$}
 \label{repsB}
\addcontentsline{toc}{subsection}{B. Representations of the Borel
  subalgebra
${\mathcal
    B}_{+}(U_q(\widehat{sl}(2|1)))$} 
First we will consider  the {\em evaluation representations} of
${\mathcal B}_{+}$, 
based on a composition of the evaluation map (\ref{evmap}) and 
the representations of the (non-affine) algebra $U(gl(2|1))$,
described in the previous subsections. Next we will consider 
representations obtained by the composition of the maps 
\eqref{modmaps} with the Fock representations of the oscillator
algebras \eqref{Hq-def} and \eqref{F-def}. This leads to
the following set of representations (which are all used in this paper)
\begin{itemize}
\item Finite dimensional representations
\begin{itemize}
\item[(I)] typical $\pi_{(\mu_1,\mu_1,\mu_3)}(x)$, \ \ \ \ \ \ \ \ \ \ \ \
\ \ \ \ \ \ $\dim=4$,  
\item[(II)] class-1 atypical 
  $\pi_{(\mu_1,\mu_2,-\mu_1-1)}(x)$, \ \ 
  $\dim=2(\mu_1-\mu_2)+3$,
  $\mu_1-\mu_2\in{\mathbb Z}_{\ge-1}$, 
\item[(III)] class-2 atypical $\pi_{\mu_1,\mu_2,-\mu_2}(x)$,\ \ \  \ \ \
\ \ $\dim=2(\mu_1-\mu_2)+1$,
  $\mu_1-\mu_2\in{\mathbb Z}_{\ge0}$,  
\item[(IV)] oscillator-type $W_3(x)$ and $\overline{W}_3(x)$,  \ \ 
$\dim=4$.
\end{itemize}
\item
infinite-dimensional representations
\begin{itemize}
\item[(V)] class-1 atypical 
  $\pi_{(\mu_1,\mu_2,-\mu_1-1)}^-(x)$, 
\item[(VI)] class-2 atypical $\pi_{\mu_1,\mu_2,-\mu_2}^+(x)$,
\item[(VII)] oscillator-type $W_1(x)$, $\overline{W}_1(x)$,
$W_2(x)$, $\overline{W}_2(x)$.
\end{itemize}
\end{itemize}
{\bf Remark:} There is a trivial isomorphism involving a shift of the 
spectral parameter for any 
evaluation representation $V_{(\nu_{1},\nu_{2},\nu_{3})}(x)$ of 
${\mathcal B}_{+}$  with spectral parameter $x$ and the 
highest weight $(\nu_{1},\nu_{2},\nu_{3})$ 
\begin{equation}
V_{(\nu_{1},\nu_{2},\nu_{3})}(x) \simeq 
V_{(\nu_{1}+\eta,\nu_{2}+\eta,\nu_{3}-\eta)}(xq^{-\eta}).
\label{shift-iso-app}
\end{equation}
 
\sapp{Atypical representations of ${\mathcal B}_+$}\label{app-atypical}
\ssapp{Class 2 representations}
Let us start from the class-2 representations. 
Both the finite dimensional $\pi_{\mu_1,\mu_2,-\mu_2}(x)$,
$\mu_1-\mu_2\in{\mathbb Z}_{\ge0}$, \ and the infinite-dimensional 
$\pi^{+}_{(\mu_{1},\mu_{2},-\mu_{2})}(x)$, $\mu_1-\mu_2\in{\mathbb
  C}$,\ 
representations are defined by the same formulae (which follow from 
\eqref{evmap} and \eqref{class2-1}) 
\begin{eqnarray}
&&
\begin{array}{rclrrrl}
(h_{0},h_{1},h_{2}) \ w^{(1)}_{j} &=& 
( & j-2l, & 2l-2j, & j & ) \ w^{(1)}_{j},  \nonumber 
\\[.25cm]
(h_{0},h_{1},h_{2}) \ w^{(3)}_{j} &=& 
( & j-2l, & 2l-2j-1, & j+1 & ) \ w^{(3)}_{j}, \\[.2cm]
\end{array}
\nonumber \\
&& 
\begin{array}{rclrclrcl}
e_{0}w^{(1)}_{j} &=& -x
q^{\mu_{2}}[2l-j]w^{(3)}_{j}, &
e_{1}w^{(1)}_{j} &=& [j]w^{(1)}_{j-1}, &
e_{2}w^{(1)}_{j} &=& 0, \\[.2cm]
e_{0}w^{(3)}_{j} &=& 0, & 
e_{1}w^{(3)}_{j} &=& [j]w^{(3)}_{j-1}, & 
e_{2}w^{(3)}_{j} &=& w^{(1)}_{j+1},
\end{array} \label{piplus}
\end{eqnarray}
where $2l=\mu_{1}-\mu_{2}$ and $w^{(1)}_{k}=w^{(3)}_{k}\equiv0$ if 
$k <0$. The parity of the vectors is  $p(w^{(1)}_{j})=0$ and  
$p(w^{(3)}_{j})=1$. 
The index $j$
takes any non-negative integer values $j\in
{\mathbb Z}_{\ge0}$ 
in the infinite-dimensional case and a restricted finite set of values in
finite dimensional case, corresponding to the basis vectors \eqref{cl2-bas}. 

\ssapp{Class 1 representations}
Both the finite dimensional 
$\pi_{(\mu_{1},\mu_{2},-\mu_{1}-1)}(x)$, with 
$\mu_1-\mu_2\in{\mathbb Z}_{\ge-1}$, \ and the infinite-dimensional 
$\pi^{-}_{(\mu_{1},\mu_{2},-\mu_{1}-1)}(x)$, with $\mu_1-\mu_2\in{\mathbb
  C}$,\ representations are defined by the same formulae (which follow from 
\eqref{evmap} and \eqref{class1-1}) 
\begin{eqnarray}
&&
\begin{array}{rclrrrl}
(h_{0},h_{1},h_{2}) \ w^{(1)}_{j} &=& 
( & j+1, & 2l-2j, & -2l+j-1 & ) \ w^{(1)}_{j}, \nonumber \\[.25cm]
(h_{0},h_{1},h_{2}) \ w^{(2)}_{j} &=& 
( & j, & 2l-2j+1, & -2l+j-1 & ) \ w^{(2)}_{j}, \\[.2cm]
\end{array}
\nonumber 
\\
&&
\begin{array}{rclrclrcl} 
e_{0}w^{(1)}_{j} &=& xq^{\mu_{1}+1}w^{(2)}_{j+1}, & 
e_{1}w^{(1)}_{j} &=& [j]w^{(1)}_{j-1}, & 
 e_{2}w^{(1)}_{j} &=& 0, \\[.2cm]
e_{0}w^{(2)}_{j} &=& 0, & 
e_{1}w^{(2)}_{j} &=& [j]w^{(2)}_{j-1}, & 
e_{2}w^{(2)}_{j} &=& -[2l+1-j]w^{(1)}_{j}, 
\end{array}
\end{eqnarray}
where $l=(\mu_{1}-\mu_{2})/2$; 
 $w^{(1)}_{k}=w^{(2)}_{k}=0$ if $k \notin {\mathbb Z}_{\ge 0}$. 
The parity of the vectors is 
$p(w^{(1)}_{j})=0$ and  $p(w^{(2)}_{j})=1$.  
The index $j$
takes any non-negative integer values $j\in
{\mathbb Z}_{\ge0}$ 
in the infinite-dimensional case and a restricted finite set of values in
finite dimensional case, corresponding to the basis vectors \eqref{cl1-bas}. 
\ssapp{Reductions} 
For integer values of $\mu_1-\mu_2$ the infinite dimensional
representations $\pi^{+}_{(\mu_1,\mu_2,-\mu_2)}(x)$ and 
$\pi^{-}_{(\mu_1,\mu_2,-\mu_1-1)}(x)$ become reducible.
Namely, \ it follows from 
 \eqref{cl1-fac1}, \eqref{cl1-fac2} and \eqref{cl2-fac1} that
\begin{equation}
\begin{array}{rcl}
\pi^{+}_{(\mu_{1},\mu_{2},-\mu_{2})}(x)/
\pi_{(\mu_{1},\mu_{2},-\mu_{2})}^{[0]}(x) &\simeq& 
\pi^{-}_{(\mu_{2}-1,\mu_{1}+1,-\mu_{2})}(x),
\\[.3cm]
\pi_{(\mu_{1},\mu_{2},-\mu_{1}-1)}^{-} (x)/
\pi_{(\mu_{1},\mu_{2},-\mu_{1}-1)}^{[0]}(x)
&\simeq & \pi_{(\mu_{2}-1,\mu_{1}+1,-\mu_{1}-1)}^{+} (x), \\[.3cm]
\pi_{(\mu_{1},\mu_{1}+1,-\mu_{1}-1)}^{-}(x) /
\pi_{(\mu_{1},\mu_{1},-\mu_{1})}^{[1]}(x)
& \simeq & 
 \pi_{(\mu_{1},\mu_{1}+1,-\mu_{1}-1)}^{+}(x) ,
\end{array}
\label{cl2-inf-iso}
\ee
where $\mu_1,\mu_2\in {\mathbb  C}$, and $\mu_1-\mu_2\in{\mathbb Z}_{\ge0}$.

\ssapp{Symmetry}
There is a correspondence 
\be
\sigma_{02}[\pi^{+[p]}_{(\mu_{1},\mu_{2},-\mu_{2})}(x)] \simeq
\pi^{-[1-p]}_{(-1-\mu_{2},-\mu_{1},\mu_{2})}(q^{2\mu_{2}}x)
\ee 
under the symmetry transformation $\sigma_{02}$ discussed in 
Section~\ref{sect-sym}; $p =0,1 \pmod 2$. 
\bigskip

\sapp{Typical representation of ${\mathcal B}_+$}\label{app-typical}
The 4-dimensional representation
${\pi}_{(\mu_{1},\mu_{1},\mu_{3})}(x)$,
$\mu_{1},\mu_{3}\in {\mathbb C}$, \  
is obtained by the composition of the map \eqref{evmap} with the 
relations \eqref{4dt}. It is spanned by four basis vectors 
$w^{(1)}_{0},w^{(2)}_{0},w^{(2)}_{1},w^{(4)}_{0}$. Explicitly, one obtains
\begin{subequations}\label{typical-eva-app}
\be
\begin{array}{rclrrrl}
(h_{0},h_{1},h_{2}) \ w^{(1)}_{0}&=& 
( & -\mu_{1}-\mu_{3}, & 0, & \mu_{1}+\mu_{3} & ) \ w^{(1)}_{0},  \\[.25cm]
 (h_{0},h_{1},h_{2}) \ w^{(2)}_{0} &=& 
( & -\mu_{1}-\mu_{3}-1, & 1, & \mu_{1}+\mu_{3} & ) \ w^{(2)}_{0}, \\[.25cm]

 (h_{0},h_{1},h_{2}) \ w^{(2)}_{1} &=& 
( & -\mu_{1}-\mu_{3}, & -1, & \mu_{1}+\mu_{3}+1 & ) \ w^{(2)}_{1},\\[.25cm]
(h_{0},h_{1},h_{2}) \ w^{(4)}_{0} &=& 
( & -\mu_{1}-\mu_{3}-1, & 0, & \mu_{1}+\mu_{3}+1 & ) \ w^{(4)}_{0}, \\[.2cm]
\end{array}
\ee
\be
\begin{array}{ll}
e_{0}w^{(1)}_{0}=
xq^{-1-\mu_{1}-2\mu_{3}}w^{(2)}_{1}, 
&e_{0}w^{(2)}_{0}=
-xq^{-1-\mu_{1}-2\mu_{3}}w^{(4)}_{0},\\[.3cm]
e_{1}w^{(2)}_{1} =w^{(2)}_{0},  
&e_{2}w^{(2)}_{0} = [\mu_{1}+\mu_{3}]w^{(1)}_{0}, \qquad
e_{2}w^{(4)}_{0} = [\mu_{1}+\mu_{3}+1]w^{(2)}_{1}, 
\end{array}
\ee
where all omitted matrix elements vanish.
\end{subequations}
The parities of the vectors read  
$p(w^{(1)}_{0})=p(w^{(4)}_{0})=0$, $p(w^{(2)}_{0})=p(w^{(2)}_{1})=1$. 
The representation is irreducible, except when $(\mu_{1}+\mu_{3})= 0,-1 $.
In the latter case one obtains\footnote{%
Note that we are using the symbol $\pi$ instead of 
$\hat{\pi}$, despite the representation is reducible at these special points.}

\begin{equation}
{\pi}_{(\mu_{1},\mu_{1},-\mu_{1})}(x)/
\pi_{(\mu_{1},\mu_{1}-1,1-\mu_{1})}^{[1]}(x)
\simeq \pi_{(\mu_{1},\mu_{1},-\mu_{1})}^{[0]}(x),\qquad 
\mu_1+\mu_3=0,
\end{equation}
\begin{equation}
{\pi}_{(\mu_{1},\mu_{1},-1-\mu_{1})}(x)/
\pi_{(\mu_{1}-1,\mu_{1}-1,1-\mu_{1})}^{[0]}(x)
\simeq \pi_{(\mu_{1},\mu_{1},-1-\mu_{1})}^{[0]}(x),\qquad
\mu_{1}+\mu_{3}=-1.
\end{equation}
%
\sapp{Oscillator representations of ${\mathcal B}_+$}\label{q-osc-appendix} 
Here we list explicit forms of the basis for the Fock representations
introduced  in section \ref{proof},
\begin{enumerate}[(a)]
\item module $\overline{W}_{1}^{++}(x)$ based on
$\overline{\rho}_{1}^{\prime}(x)$, 
\begin{eqnarray}
|m,n>_{++}&=&(b_{1}^{-})^{m}(f_{2}^{-})^{n}|0>_{++} , 
\quad m \in {\mathbb Z}_{\ge 0}, \quad n \in \{ 0,1 \}, \nonumber \\[.2cm]
e_{0}|m,n>_{++}&=&|m,n-1>_{++}, \nonumber \\[.2cm]
e_{1}|m,n>_{++}&=&q^{-n}|m+1,n>_{++}, \label{w1b++}\\[.2cm]
e_{2}|m,n>_{++}&=&-\frac{q^{n+\frac{1}{2}}(1-q^{-2m})x}
{(q-q^{-1})^{2}}|m-1,n+1>_{++},  \nonumber    \\[.2cm]
(h_{0},h_1,h_2)|m,n>_{++}&=&(-m,2m+n,-m-n)|m,n>_{++},\nonumber 
\end{eqnarray}
where $|-1,n>_{++}=|m,-1>_{++}=|m,2>_{++}=0$ and 
$p(|m,n>_{++})= n \pmod 2$.
\item
module $\overline{W}_{2}^{--}(x)$ based on
$\overline{\rho}_{2}^{\prime}(x)$,
\begin{eqnarray}
|m,n>_{--}&=&(f_{1}^{+})^{m}(b_{2}^{+})^{n}|0>_{--} , 
\quad m \in \{ 0,1 \},\quad  n \in {\mathbb Z}_{\ge 0}, \nonumber \\[.2cm]
e_{0}|m,n>_{--}&=&\frac{q^{\frac{1}{2}}(q^{n}-q^{-n})x}
{(q-q^{-1})^{2}}|m+1,n-1>_{--},  \nonumber \\[.2cm]
e_{1}|m,n>_{--}&=&|m,n+1>_{--}, \label{w2b--}\\[.2cm]
e_{2}|m,n>_{--}&=&q^{n}|m-1,n>_{--}, \nonumber \\[.2cm]
(h_{0},h_1,h_2)|m,n>_{--}&=&(-m-n,m+2n,-n)|m,n>_{--},\nonumber 
\end{eqnarray}
where $|-1,n>_{--}=|2,n>_{--}=|m,-1>_{--}=0$ and $p(|m,n>_{--})=m \pmod 2$. 
\item module $\overline{W}_{3}^{-+}(x)$ based on
$\overline{\rho}_{3}^{\prime}(x)$,
\begin{eqnarray}
|m,n>_{-+}&=&(f_{1}^{+})^{m}(f_{2}^{-})^{n}|0>_{-+} , 
\quad m,n \in \{ 0,1 \},\nonumber  \\[.2cm]
e_{0}|m,n>_{-+}&=&q^{n}|m-1,n>_{-+}, \nonumber \\[.2cm]
e_{1}|m,n>_{-+}&=&(-1)^{m+1}\frac{q^{-n-\frac{1}{2}}x}
{q-q^{-1}}|m+1,n+1>_{-+},\label{w3b-+} \\[.2cm]
e_{2}|m,n>_{-+}&=&(-1)^{m}|m,n-1>_{-+}, \nonumber \\[.2cm]
(h_{0},h_1,h_2)|m,n>_{-+}&=&(-n,m+n,-m)|m,n>_{-+}, \nonumber 
\end{eqnarray}
where $|-1,n>_{-+}=|2,n>_{-+}=|m,-1>_{-+}=|m,2>_{-+}=0$ and  
$p(|m,n>_{-+})=m+n \pmod 2$. 
\item module $W_{1}^{--}(x)$ based on $\rho_{1}^{\prime}(x)$,
\begin{eqnarray}
|m,n>_{--}&=&(b_{1}^{+})^{m}(f_{2}^{+})^{n}|0>_{--} , 
\quad m \in {\mathbb Z}_{\ge 0}, \quad n \in \{ 0,1 \},\nonumber \\[.2cm]
e_{0}|m,n>_{--}&=&|m,n-1>_{--}, \nonumber \\[.2cm]
e_{1}|m,n>_{--}&=&q^{n}|m+1,n>_{--}, \label{w1--}\\[.2cm]
e_{2}|m,n>_{--}&=&-\frac{q^{-n-\frac{1}{2}}(1-q^{2m})x}
{(q-q^{-1})^{2}}|m-1,n+1>_{--},\nonumber  \\[.2cm]
(h_{0},h_1,h_2)|m,n>_{--}&=&(-m,2m+n,-m-n)|m,n>_{--},\nonumber  
\end{eqnarray}
where $|-1,n>_{--}=|m,-1>_{--}=|m,2>_{--}=0$ and  
$p(|m,n>_{--})=n \pmod 2$.
\item module $W_{2}^{++}(x)$ based on $\rho_{2}^{\prime}(x)$
\begin{eqnarray}
|m,n>_{++}&=&(f_{1}^{-})^{m}(b_{2}^{-})^{n}|0>_{++}, 
\quad m \in \{ 0,1 \},\quad  n \in {\mathbb Z}_{\ge 0}, \nonumber \\[.2cm]
e_{0}|m,n>_{++}&=&-\frac{q^{-\frac{1}{2}}(q^{n}-q^{-n})x}
{(q-q^{-1})^{2}}|m+1,n-1>_{++},
\nonumber  \\[.2cm]
e_{1}|m,n>_{++}&=&|m,n+1>_{++},\label{w2++}\\[.2cm]
e_{2}|m,n>_{++}&=&q^{-n}|m-1,n>_{++}, \nonumber \\[.2cm]
(h_{0},h_1,h_2)|m,n>_{++}&=&(-m-n,m+2n,-n)|m,n>_{++}, \nonumber 
\end{eqnarray}
where $|-1,n>_{++}=|2,n>_{++}=|m,-1>_{++}=0$ and $p(|m,n>_{++})=m \pmod 2$. 
\item mudule $W_{3}^{+-}(x)$ based on $\rho_{3}^{\prime}(x)$
\begin{eqnarray}
|m,n>_{+-}&=&(f_{1}^{-})^{m}(f_{2}^{+})^{n}|0>_{+-} , 
\quad m,n \in \{ 0,1 \}, \nonumber \\[.2cm]
e_{0}|m,n>_{+-}&=&q^{-n}|m-1,n>_{+-}, \nonumber \\[.2cm]
e_{1}|m,n>_{+-}&=&(-1)^{m}\frac{q^{n+\frac{1}{2}}x}
{q-q^{-1}}|m+1,n+1>_{+-}, \label{w3+-}\\[.2cm]
e_{2}|m,n>_{+-}&=&(-1)^{m}|m,n-1>_{+-}, \nonumber \\[.2cm]
(h_{0},h_1,h_2)|m,n>_{+-}&=&(-n,m+n,-m)|m,n>_{+-}, \nonumber 
\end{eqnarray}
where $|-1,n>_{+-}=|2,n>_{+-}=|m,-1>_{+-}=|m,2>_{+-}=0$ and
$p(|m,n>_{+-})=m+n \pmod 2$.  
\item module $\overline{W}_{1}^{--}(x)$ based on
  $\overline{\rho}_{1}^{\prime}(x)$ 
\begin{eqnarray}
|m,n>_{--}&=&(b_{1}^{+})^{m}(f_{2}^{+})^{n}|0>_{--} , 
\quad m \in {\mathbb Z}_{\ge 0}, \quad n \in \{ 0,1 \},\nonumber \\[.2cm]
e_{0}|m,n>_{--}&=&|m,n+1>_{--}, \nonumber \\[.2cm]
e_{1}|m,n>_{--}&=&-\frac{q^{m+n}[m]}{q-q^{-1}}|m-1,n>_{--}, 
\label{w1b--}\\[.2cm]
e_{2}|m,n>_{--}&=&-q^{-\frac{1}{2}}x|m+1,n-1>_{--}, \nonumber \\[.2cm]
(h_{0},h_1,h_2)|m,n>_{--}&=&(m,-2m-n,m+n)|m,n>_{--}, \nonumber 
\end{eqnarray}
where $|-1,n>_{--}=|m,-1>_{--}=|m,2>_{--}=0$ and $p(|m,n>_{--})=n
\pmod 2$. 
\item module $\overline{W}_{2}^{++}(x)$ based on
$\overline{\rho}_{2}^{\prime}(x)$ 
\begin{eqnarray}
|m,n>_{++}&=&(f_{1}^{-})^{m}(b_{2}^{-})^{n}|0>_{++} , 
\quad m \in \{ 0,1 \},\quad  n \in {\mathbb Z}_{\ge 0}, \nonumber \\[.2cm]
e_{0}|m,n>_{++}&=&-q^{n+\frac{1}{2}}x|m-1,n+1>_{++}, \nonumber \\[.2cm]
e_{1}|m,n>_{++}&=&\frac{q^{-n}[n]}{q-q^{-1}}|m,n-1>_{++}, 
\label{w2b++}\\[.2cm]
e_{2}|m,n>_{++}&=&q^{-n}|m+1,n>_{++}, \nonumber \\[.2cm]
(h_{0},h_1,h_2)|m,n>_{++}&=&(m+n,-m-2n,n)|m,n>_{++}, \nonumber 
\end{eqnarray}
where $|-1,n>_{++}=|2,n>_{++}=|m,-1>_{++}=0$ and $p(|m,n>_{++})=m \pmod 2$. 
\item module $W_{1}^{++}(x)$ based on $\rho_{1}^{\prime}(x)$
\begin{eqnarray}
|m,n>_{++}&=&(b_{1}^{-})^{m}(f_{2}^{-})^{n}|0>_{++} , 
\quad m \in {\mathbb Z}_{\ge 0}, \quad n \in \{ 0,1 \},\nonumber \\[.2cm]
e_{0}|m,n>_{++}&=&|m,n+1>_{++}, \nonumber \\[.2cm]
e_{1}|m,n>_{++}&=&\frac{q^{-m-n}[m]}{q-q^{-1}}|m-1,n>_{++}, \label{w1++}
 \\[.2cm]
e_{2}|m,n>_{++}&=&-q^{\frac{1}{2}}x|m+1,n-1>_{++},\nonumber  \\[.2cm]
(h_{0},h_1h_2)|m,n>_{++}&=&(m,-2m-n,m+n)|m,n>_{++},
\end{eqnarray}
where $|-1,n>_{++}=|m,-1>_{++}=|m,2>_{++}=0$ and 
$p(|m,n>_{++})=n \pmod 2$. 
\item module $W_{2}^{--}(x)$ based on $\rho_{2}^{\prime}(x)$
\begin{eqnarray}
|m,n>_{--}&=&(f_{1}^{+})^{m}(b_{2}^{+})^{n}|0>_{--} , 
\quad m \in \{ 0,1 \},\quad  n \in {\mathbb Z}_{\ge 0}, \nonumber \\[.2cm]
e_{0}|m,n>_{--}&=&-q^{-n-\frac{1}{2}}x|m-1,n+1>_{--}, \nonumber \\[.2cm]
e_{1}|m,n>_{--}&=&-\frac{q^{n}[n]}{q-q^{-1}}|m,n-1>_{--}, \label{w2--}
 \\[.2cm]
e_{2}|m,n>_{--}&=&q^{n}|m+1,n>_{--}, \nonumber \\[.2cm]
(h_{0},h_1,h_2)|m,n>_{--}&=&(m+n,-m-2n,n)|m,n>_{--},\nonumber  
\end{eqnarray}
where $|-1,n>_{--}=|2,n>_{--}=|m,-1>_{--}=0$ and $p(|m,n>_{--})=m \pmod 2$. 
\end{enumerate}

\app{Quantum affine superalgebra analogue of the first and second Weyl
  formulae}\label{app-JTSP}
\addcontentsline{toc}{subsection} {C. Quantum affine superalgebra analogue
  of the first and second Weyl   formulae}  
The so-called Bazhanov-Reshetikhin formula \cite{BR90}
 is a determinant expression of the eigenvalue of the transfer matrix 
for the fusion model for $U_{q}(\widehat{sl}(m))$.
This formula allows a supersymmetric generalization for 
$U_{q}(\widehat{sl}(m|n))$, which may be called the 
"quantum supersymmetric Jacobi-Trudi and Giambelli formula" 
(\ref{superJT1}) and (\ref{superJT2}) \cite{T97,T98,T98-2} 
 (see also \cite{KOS95} for $U_{q}(B^{(1)}_{r})$ case). 
This is a quantum affine superalgebra analogue of the second 
Weyl formula for the transfer matrices.
 It is natural to consider an analogue of the first Weyl formula. 
 Weyl first formula for the superalgebra $gl(m|n)$ is often
called ``Sergeev-Pragacz formula" in mathematical literature 
\cite{Pragacz91,vhkt90} (see (\ref{Sergeev-Pragacz})). 
Eqs.(\ref{atyp-c})-(\ref{app-wrons3}) are quantum affine superalgebra
  analogue of the   Sergeev-Pragacz formula.  
 
\sapp{Partitions, Young diagrams and admissible tableaux.}
Introduce notations for the integer partitions and Young
diagrams (see, e.g., \cite{M95} for additional details). 
A {\em partition} is a sequence
of integers $\mu=(\mu_{1},\mu_{2},\dots) $ such that  
$\mu_{1} \ge \mu_{2} \ge \dots \ge 0$. 
Repeated entries $k,k,\ldots,k$ of the same integer $k$ in the 
partition can be abbreviated as $k^{m_{k}}$, where $m_{k}$ denotes the 
corresponding multiplicity. 
Two partitions are regarded equivalent if all their 
non-zero elements coincide. 
For example, $(3,3,2,1,1,0,0)=(3,3,2,1,1)=(3^2,2,1^2)$. 
Partitions can be visualized by {\em Young diagrams}, formed by
rows of identical square boxes in the plane. 
The Young diagram $\mu$, corresponding to a partition $\mu$, has 
$\mu_{k}$ boxes in the $k$-th 
row, see Fig.~\ref{Young1}.
Individual boxes are referred to by integer coordinates 
$(i,j)\in {\mathbb Z}^{2}$, 
where the row index $i$ increases downwards while the column 
index $j$ increases from left to right. 
The top left corner of $\mu$ has coordinates $(1,1)$.
The partition $\mu^{\prime}=(\mu_{1}^{\prime},\mu_{2}^{\prime},\dots)$ 
is called conjugate of $\mu$, where  
$\mu_{j}^{\prime}$ is defined as the maximal integer $k$ 
such that $\mu_{k} \ge j$. 
The Young diagrams for conjugated partitions are obtained from each
other by the transposition of rows and columns, 
as in example in Fig.~\ref{Young2}.

Let $\lambda =(\lambda_{1},\lambda_{2},\dots)$ and 
$\mu =(\mu_{1},\mu_{2},\dots)$ be two partitions such that
$\mu_{i} \ge \lambda_{i}: i=1,2,\dots$ and 
$\lambda_{\mu_{1}^{\prime}}=\lambda^{\prime}_{\mu_{1}}=0$. 
We denote a skew-Young diagram defined by these two partitions as 
 $\lambda \subset \mu$. This is the domain obtained by the subtraction 
 $\mu-\lambda$ as in the example in Fig.~\ref{Young3}. 
If $\lambda $ is an empty set $\phi$, then 
$\lambda \subset \mu$ coincides with $\mu$. 
Individual boxes on the skew-Young diagram 
$\lambda \subset \mu$ are referred to by their coordinates on $\mu$.
%
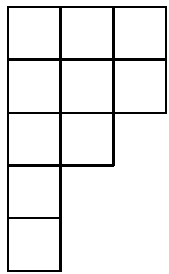
\begin{figure}
  \begin{center}
    \setlength{\unitlength}{2pt}
    \begin{picture}(30,55) 
      \put(0,0){\line(0,1){50}}
      \put(10,0){\line(0,1){50}}
      \put(20,20){\line(0,1){30}}
      \put(30,30){\line(0,1){20}}
      \put(0,0){\line(1,0){10}}      
      \put(0,10){\line(1,0){10}}
      \put(0,20){\line(1,0){20}}
      \put(0,30){\line(1,0){30}}
      \put(0,40){\line(1,0){30}}
      \put(0,50){\line(1,0){30}}
    \end{picture}
  \end{center}
  \caption{The Young diagram with the shape $\mu=(3^2,2,1^2)$.}
  \label{Young1}
\end{figure}
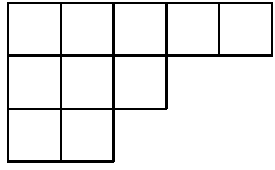
\begin{figure}
  \begin{center}
    \setlength{\unitlength}{2pt}
    \begin{picture}(50,35) 
      \put(0,0){\line(0,1){30}}
      \put(10,0){\line(0,1){30}}
      \put(20,0){\line(0,1){30}}
      \put(30,10){\line(0,1){20}}
      \put(40,20){\line(0,1){10}}  
      \put(50,20){\line(0,1){10}}     
      \put(0,0){\line(1,0){20}}
      \put(0,10){\line(1,0){30}}
      \put(0,20){\line(1,0){50}}
      \put(0,30){\line(1,0){50}}
    \end{picture}
  \end{center}
  \caption{The Young diagram for the partition $\mu^{\prime}=(5,3,2)$, 
conjugated to $\mu=(3^2,2,1^2)$.}
  \label{Young2}
\end{figure}
\begin{figure}
  \begin{center}
    \setlength{\unitlength}{2pt}
    \begin{picture}(30,55) 
      \put(0,0){\line(0,1){30}}
      \put(10,0){\line(0,1){40}}
      \put(20,20){\line(0,1){30}}
      \put(30,30){\line(0,1){20}}
      \put(0,0){\line(1,0){10}}      
      \put(0,10){\line(1,0){10}}
      \put(0,20){\line(1,0){20}}
      \put(0,30){\line(1,0){30}}
      \put(10,40){\line(1,0){20}}
      \put(20,50){\line(1,0){10}}
    \end{picture}
  \end{center}
  \caption{The skew Young diagram $\lambda \subset \mu$ with 
  $\lambda=(2,1)$ and $\mu=(3^2,2,1^2)$.}
  \label{Young3}
\end{figure}
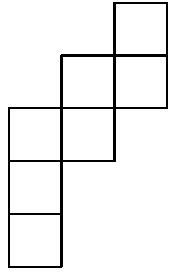
%

Next, define a space of {\em admissible tableaux}  $\mathsf{Tab}(\nu)$
on a (skew) Young diagram $\nu$. 
In each box $(i,j)$ of the diagram
write an integer $t_{ij}$.
An admissible tableau (or,
 simply, a tableau) $t\in\mathsf{Tab}(\nu)$  is 
a set of integers $t=\{t_{jk}\}_{(j,k)\in \nu}$, 
where all $t_{jk} \in B=\{1,2,3\}=B_{+}\cup B_{-}$, $B_{+}=\{1,2 \}$,
$B_{-}=\{3 \}$, satisfy the conditions  
\begin{itemize} 
\item[(i)] $t_{jk}\le t_{j+1,k},t_{j,k+1}$ \\
\item[(ii)]  $t_{jk} < t_{j,k+1}$ if $t_{jk}\in B_{-}$ or
  $t_{j,k+1}\in B_{-}$  \\ 
\item[(iii)]  $t_{jk} < t_{j+1,k}$ if $t_{jk}\in B_{+}$ or
  $t_{j+1,k}\in B_{+}$.  
\end{itemize}
Fig.~\ref{Young4} shows all admissible tableaux for the skew Young diagram
$\lambda \subset \mu$ with $\lambda=(1^2)$ and $\mu=(2^3)$.
\begin{figure}
  \begin{center}
    \setlength{\unitlength}{1.9pt}
    \begin{picture}(230,35) 
      \put(0,0){\line(0,1){10}}
      \put(10,0){\line(0,1){30}}
      \put(20,0){\line(0,1){30}}
      \put(0,0){\line(1,0){20}}
      \put(0,10){\line(1,0){20}}
      \put(10,20){\line(1,0){10}}
      \put(10,30){\line(1,0){10}}
      \put(5,5){\makebox(0,0){$1$}}
      \put(15,25){\makebox(0,0){$1$}}
      \put(15,15){\makebox(0,0){$2$}}
      \put(15,5){\makebox(0,0){$3$}}
      \put(30,0){\line(0,1){10}}
      \put(40,0){\line(0,1){30}}
      \put(50,0){\line(0,1){30}}
      \put(30,0){\line(1,0){20}}
      \put(30,10){\line(1,0){20}}
      \put(40,20){\line(1,0){10}}
      \put(40,30){\line(1,0){10}}
      \put(35,5){\makebox(0,0){$1$}}
      \put(45,25){\makebox(0,0){$1$}}
      \put(45,15){\makebox(0,0){$3$}}
      \put(45,5){\makebox(0,0){$3$}}
      \put(60,0){\line(0,1){10}}
      \put(70,0){\line(0,1){30}}
      \put(80,0){\line(0,1){30}}
      \put(60,0){\line(1,0){20}}
      \put(60,10){\line(1,0){20}}
      \put(70,20){\line(1,0){10}}
      \put(70,30){\line(1,0){10}}
      \put(65,5){\makebox(0,0){$1$}}
      \put(75,25){\makebox(0,0){$2$}}
      \put(75,15){\makebox(0,0){$3$}}
      \put(75,5){\makebox(0,0){$3$}}
      \put(90,0){\line(0,1){10}}
      \put(100,0){\line(0,1){30}}
      \put(110,0){\line(0,1){30}}
      \put(90,0){\line(1,0){20}}
      \put(90,10){\line(1,0){20}}
      \put(100,20){\line(1,0){10}}
      \put(100,30){\line(1,0){10}}
      \put(95,5){\makebox(0,0){$1$}}
      \put(105,25){\makebox(0,0){$3$}}
      \put(105,15){\makebox(0,0){$3$}}
      \put(105,5){\makebox(0,0){$3$}}
      \put(120,0){\line(0,1){10}}
      \put(130,0){\line(0,1){30}}
      \put(140,0){\line(0,1){30}}
      \put(120,0){\line(1,0){20}}
      \put(120,10){\line(1,0){20}}
      \put(130,20){\line(1,0){10}}
      \put(130,30){\line(1,0){10}}
      \put(125,5){\makebox(0,0){$2$}}
      \put(135,25){\makebox(0,0){$1$}}
      \put(135,15){\makebox(0,0){$2$}}
      \put(135,5){\makebox(0,0){$3$}}
      \put(150,0){\line(0,1){10}}
      \put(160,0){\line(0,1){30}}
      \put(170,0){\line(0,1){30}}
      \put(150,0){\line(1,0){20}}
      \put(150,10){\line(1,0){20}}
      \put(160,20){\line(1,0){10}}
      \put(160,30){\line(1,0){10}}
      \put(155,5){\makebox(0,0){$2$}}
      \put(165,25){\makebox(0,0){$1$}}
      \put(165,15){\makebox(0,0){$3$}}
      \put(165,5){\makebox(0,0){$3$}}
      \put(180,0){\line(0,1){10}}
      \put(190,0){\line(0,1){30}}
      \put(200,0){\line(0,1){30}}
      \put(180,0){\line(1,0){20}}
      \put(180,10){\line(1,0){20}}
      \put(190,20){\line(1,0){10}}
      \put(190,30){\line(1,0){10}}
      \put(185,5){\makebox(0,0){$2$}}
      \put(195,25){\makebox(0,0){$2$}}
      \put(195,15){\makebox(0,0){$3$}}
      \put(195,5){\makebox(0,0){$3$}}
      \put(210,0){\line(0,1){10}}
      \put(220,0){\line(0,1){30}}
      \put(230,0){\line(0,1){30}}
      \put(210,0){\line(1,0){20}}
      \put(210,10){\line(1,0){20}}
      \put(220,20){\line(1,0){10}}
      \put(220,30){\line(1,0){10}}
      \put(215,5){\makebox(0,0){$2$}}
      \put(225,25){\makebox(0,0){$3$}}
      \put(225,15){\makebox(0,0){$3$}}
      \put(225,5){\makebox(0,0){$3$}}
    \end{picture}
  \end{center}
  \caption{All admissible tableaux for the 
skew-Young diagram $\lambda \subset \mu$ with 
  $\lambda=(1^2)$ and $\mu=(2^3)$.}
  \label{Young4}
\end{figure}
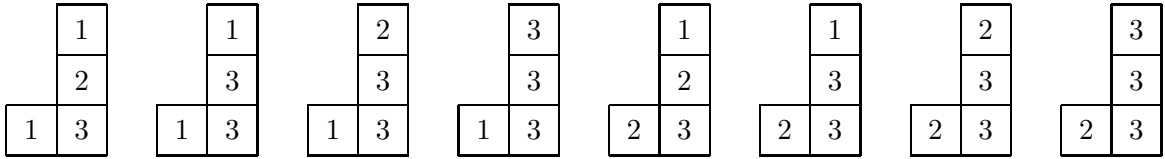

\sapp{{\bf Q}-operators}
Now recall the $\bf Q$-operators $\A_i(x)$ and $\overline{\A}_i(x)$ 
defined in \eqref{Q-def} and \eqref{Qbar-def}. 
They satisfy the functional relations of Sect.\ref{wrtype}. 
When written in terms of $\A_i(x)$ and $\overline{\A}_i(x)$ these
relations have the form
\begin{eqnarray}
 c_{21}{\mathbb A}_{3}(x)&=&\Big(\frac{z_2}{z_1}\Big)^{\frac{1}{2}}\,
\overline{\mathbb A}_{1}(xq)\overline{\mathbb A}_{2}(xq^{-1})-
\Big(\frac{z_1}{z_2}\Big)^{\frac{1}{2}}\,
\overline{\mathbb A}_{1}(xq^{-1})\overline{\mathbb A}_{2}(xq),
 \label{sl21-rel1-app} \\[.2cm]
c_{12}\overline{{\mathbb A}}_{3}(x)&=&\Big(\frac{z_1}{z_2}\Big)^{\frac{1}{2}}\,
{\mathbb A}_{1}(xq){\mathbb A}_{2}(xq^{-1})-
\Big(\frac{z_2}{z_1}\Big)^{\frac{1}{2}}\,{\mathbb
  A}_{1}(xq^{-1}){\mathbb A}_{2}(xq),  
\label{sl21-rel2-app}\\[.2cm]
c_{13}{\mathbb A}_{1}(x){\mathbb A}_{3}(x)&=&
 \left(\frac{z_{1}}{z_{3}}\right)^{\frac{1}{2}} \overline{{\mathbb A}}_{2}(xq)-
 \left(\frac{z_{3}}{z_{1}}\right)^{\frac{1}{2}}
 \overline{\mathbb A}_{2}(xq^{-1}),   
\label{sl21-rel3-app} 
\end{eqnarray} 
\begin{eqnarray} 
c_{23}{\mathbb A}_{2}(x){\mathbb A}_{3}(x)&=&
\left(\frac{z_{2}}{z_{3}}\right)^{\frac{1}{2}} \overline{{\mathbb A}}_{1}(xq)-
 \left(\frac{z_{3}}{z_{2}}\right)^{\frac{1}{2}}
 \overline{\mathbb A}_{1}(xq^{-1}), 
\label{sl21-rel4-app} \\[.2cm]
c_{13}\overline{\mathbb A}_{1}(x)\overline{\mathbb A}_{3}(x)&=&
\left(\frac{z_{1}}{z_{3}}\right)^{\frac{1}{2}} {\mathbb A}_{2}(xq^{-1})-
\left(\frac{z_{3}}{z_{1}}\right)^{\frac{1}{2}} {\mathbb A}_{2}(xq),  
\label{sl21-rel5-app} \\[.2cm]
c_{23}\overline{\mathbb A}_{2}(x)\overline{{\mathbb A}}_{3}(x)&=&
\left(\frac{z_{2}}{z_{3}}\right)^{\frac{1}{2}} {\mathbb A}_{1}(xq^{-1})-
\left(\frac{z_{3}}{z_{2}}\right)^{\frac{1}{2}} {\mathbb A}_{1}(xq).
\label{sl21-rel6-app}
\end{eqnarray}
where the quantities $z_{1},z_{2},
z_{3}$, defined in \eqref{zs-def},
are constant ($x$-independent) operators, satisfying 
the relation $z_{1}z_{2}z_{3}^{-1}=1$. 

\sapp{Higher transfer matrices and sums over tableaux} 
Define operator-valued functions 
\begin{eqnarray} 
{\mathcal X}(1,x)
 =z_{1}\frac{\overline{\mathbb A}_{1}(xq^{-\frac{3}{2}})}
{\overline{\mathbb A}_{1}(xq^{\frac{1}{2}})}, \quad 
{\mathcal X}(2,x)
=z_{2}\frac{\overline{\mathbb A}_{1}(xq^{\frac{5}{2}})
{\mathbb A}_{3}(xq^{-\frac{1}{2}})}
{\overline{\mathbb A}_{1}(xq^{\frac{1}{2}})
{\mathbb A}_{3}(xq^{\frac{3}{2}})},
 \quad  
{\mathcal X}(3,x)
=z_{3}\frac{{\mathbb A}_{3}(xq^{-\frac{1}{2}})}
{{\mathbb A}_{3}(xq^{\frac{3}{2}})}, 
\label{fun-tab-def}
\end{eqnarray}
of the spectral variable $x\in {\mathbb C}$. 
Remind, that all the operators, appearing in (\ref{sl21-rel1-app})-
(\ref{fun-tab-def}), are commutative.  
For any skew Young diagram $\nu$ 
define an operator  
\begin{equation}
{\mathcal F}_{\nu}(x)=
\sum_{t\in\mathsf{Tab}(\nu)}
\prod_{(i,j) \in \nu}
(-1)^{p(t_{i,j})}
{\mathcal X}(t_{i,j} ,xq^{-\nu_{1}+\nu_{1}^{\prime}-2i+2j}),
\label{DVF-tab}
\end{equation}
where the sum is taken over all admissible tableaux, and 
the product is taken over all boxes of the
Young diagram $\nu$. The parities are  
$p(1)=p(2)=0$, $p(3)=1$ and the integers $\nu_1,\nu'_1$ for a skew
 diagram $\nu=\lambda\subset \mu$ are defined as $\nu_1=\mu_1$,
 $\nu_1'=\mu_1'$. 
For example for the diagram in Fig.~\ref{Young4}, one has from \eqref{DVF-tab}
\be
\begin{array}{l} 
{\mathcal F}_{(1^2) \subset (2^3)}(x)=\\[.3cm] 
-{\mathcal X}(1,xq^{-3}){\mathcal X}(1,xq^{3})
{\mathcal X}(2,xq){\mathcal X}(3,xq^{-1})
+{\mathcal X}(1,xq^{-3}){\mathcal X}(1,xq^{3})
{\mathcal X}(3,xq){\mathcal X}(3,xq^{-1})  \\[0.3cm]
+{\mathcal X}(1,xq^{-3}){\mathcal X}(2,xq^{3})
{\mathcal X}(3,xq){\mathcal X}(3,xq^{-1})
-{\mathcal X}(1,xq^{-3}){\mathcal X}(3,xq^{3})
{\mathcal X}(3,xq){\mathcal X}(3,xq^{-1})  \\[0.3cm]
-{\mathcal X}(2,xq^{-3}){\mathcal X}(1,xq^{3})
{\mathcal X}(2,xq){\mathcal X}(3,xq^{-1})
+{\mathcal X}(2,xq^{-3}){\mathcal X}(1,xq^{3})
{\mathcal X}(3,xq){\mathcal X}(3,xq^{-1})  \\[0.3cm]
+{\mathcal X}(2,xq^{-3}){\mathcal X}(2,xq^{3})
{\mathcal X}(3,xq){\mathcal X}(3,xq^{-1})
-{\mathcal X}(2,xq^{-3}){\mathcal X}(3,xq^{3})
{\mathcal X}(3,xq){\mathcal X}(3,xq^{-1}).
\end{array}\label{examp}
\ee
Note, that since 
 $\overline{\mathbb A}_{1}(0)={\mathbb A}_{3}(0)=1$,
Eq.\eqref{DVF-tab} with $x=0$ reduces to the
 super-character formula (the supersymmetric Shur function) 
\begin{equation}
S_{\nu}(z_1,z_2|z_3) =
\sum_{t\in\mathsf{Tab}(\nu)}
\prod_{(i,j) \in \nu}
(-1)^{p(t_{i,j})}
z_{t_{i,j}}\label{shur1}
\end{equation}
for the classical super-algebra $gl(2|1)$.

An important feature of Eq.\eqref{DVF-tab} is that
 it defines an entire function of $x$ (the pole terms cancel out due
 to the Bethe Ansatz equations \eqref{BAijk}, see \cite{T97,T98,T98-2}).
It is expected that this formula defines the most
general (higher) transfer matrix, associated with $U_q(\widehat{sl}(2|1))$.
For example, the quantity ${\cal
  F}_{(1)}(x)$ (for the simplest diagram
 $\nu=(1)$, consisting of one square) exactly coincides with one of
 the six expressions  \eqref{six2} for the fundamental transfer matrix
 $\TT(x)$.  
The general statement will be formulated in the 
Conjecture \ref{JT-SP-conj} below, 
but before that let us discuss some other properties of  
\eqref{DVF-tab}.
 
Note that sum over the tableaux expressions, like 
\eqref{DVF-tab}, arise also in the theory of character for quantum
 affine algebras \cite{FR99,FM01}.

\sapp{Bazhanov-Reshetikhin formulae}
The definition \eqref{DVF-tab}  implies the following determinant identities
\begin{eqnarray}
{\mathcal F}_{\lambda \subset \mu}(x)&=& \det_{1 \le i,j \le \mu_{1}}
    ({\mathcal F}_{(1^{\mu_{i}^{\prime}-\lambda_{j}^{\prime}-i+j})}
(xq^{-\mu_{1}+\mu_{1}^{\prime}-\mu_{i}^{\prime}-\lambda_{j}^{\prime}+i+j-1}))	
 \label{superJT1}
	\\[.3cm] 
& =&\det_{1 \le i,j \le \mu_{1}^{\prime}}
    ({\mathcal F}_{(\mu_{j}-\lambda_{i}+i-j)}
    (xq^{-\mu_{1}+\mu_{1}^{\prime}+\mu_{j}+\lambda_{i}-i-j+1})),
     \label{superJT2}
\end{eqnarray}
where  ${\mathcal F}_{(1^{0})}={\mathcal F}_{(0)}\equiv1$ and 
${\mathcal F}_{(1^{a})}={\mathcal F}_{(a)}\equiv 0$ for $a <0$. 
For example, for the same operator as in \eqref{examp} one obtains
\begin{eqnarray}
{\mathcal F}_{(1^2) \subset (2^3)}(x)&=&
\det 
\left(
\begin{array}{cc}
{\mathcal F}_{(1)}(xq^{-3}) & {\mathcal F}_{(1^4)}(x) \\
1 & {\mathcal F}_{(1^3)}(xq)
\end{array}
\right) \\[0.3cm]
&=&
\det 
\left(
\begin{array}{ccc}
{\mathcal F}_{(1)}(xq^{3}) & 1 & 0 \\
{\mathcal F}_{(2)}(xq^2) &{\mathcal F}_{(1)}(xq) & 1 \\
{\mathcal F}_{(4)}(x) & {\mathcal F}_{(3)}(xq^{-1}) & {\mathcal
  F}_{(2)}(xq^{-2}) 
\end{array}
\right) .
\end{eqnarray}

The determinant representations (\ref{superJT1}), (\ref{superJT2}) for
the algebra $U_{q}(\widehat{sl}(m))$ with $\lambda=\phi$ were first
obtained by Bazhanov and Reshetikhin in \cite{BR90}.  For the relevant
here case of $U_{q}(\widehat{sl}(m|n))$ these representations were
generalized in \cite{T97} (see, also \cite{T98,T98-2}).  

The fact that (\ref{DVF-tab}) implies the relations 
(\ref{superJT1})-(\ref{superJT2}) is a combinatorial theorem which can be
proven by induction on the size of the determinants $\mu_{1}$ or
$\mu_{1}^{\prime}$ (this theorem mentioned in \cite{KOS95} in the same
context for $U_{q}(B^{(1)}_{r})$ case).  In the case $x=0$ the above
determinant expressions reduce to the super-symmetric Jacobi-Trudi and
Giambelli formulae for the characters of $gl(2|1)$ (or the second Weyl
formula).
On the other hand,  the fact that the tableau sum  
(or the related determinant formulae \eqref{superJT1} or \eqref{superJT2}) 
exactly coincides with a higher transfer matrix is a non-trivial 
statement. Obviously, to prove this one needs to connect
\eqref{DVF-tab} with 
the definition of transfer matrices by using algebraic properties 
of representations of the quantum affine algebra (or the corresponding Yangian in the
rational case $q=1$) in the auxiliary space. 

So far the formulae \eqref{superJT1} or \eqref{superJT2} 
have been proven only for a few cases:
(i) for $U_q(\widehat{sl}(2))$ \cite{KR87}, (ii) for $U_q(\widehat{sl}(3))$  
 with an arbitrary quantum space \cite{BHK}
and (iii) for the Yangian case $Y(gl(m|n))$, corresponding
to $q=1$, but when the quantum space is a tensor product of the fundamental
$(m+n)$-dimensional representations.
In all cases only non-skew\footnote{%
The evaluation representations considered here obviously connected with
non-skew diagrams as well.} diagrams (with $\lambda=0$) were considered.

\sapp{Quantum affine analogue of the first Weyl formula}
Here we consider quantum affine
analogues of the {\em first Weyl formula} for the characters. 
Such representations for the ${\bf T}$-operators were first introduced in
\cite{BLZ97} for  $U_q(\widehat{sl}(2))$ and in \cite{BHK} for 
$U_q(\widehat{sl}(3))$.
Examples of such representations 
for $U_q(\widehat{sl}(2|1))$ are 
the Wronskian-like expressions \eqref{atyp} and \eqref{dim4},
considered in the main text. Here it
is convenient to 
rewrite them in terms of of $\A_i(x)$ and $\overline{\A}_i(x)$,
\begin{eqnarray}
{\mathbb T}_{m}^{(1)}(x)&=&
\frac{c_{13}}{c_{12}}
z_{1}^{m+\frac{1}{2}}
{\mathbb A}_{1}(xq^{m+\frac{1}{2}})
\overline{{\mathbb A}}_{1}(xq^{-m-\frac{1}{2}})
-\frac{c_{23}}{c_{12}}
z_{2}^{m+\frac{1}{2}}
{\mathbb A}_{2}(xq^{m+\frac{1}{2}})
\overline{{\mathbb A}}_{2}(xq^{-m-\frac{1}{2}}), 
\label{atyp-c} \\[.3cm]
{{\mathbb T}}_m^{(2)}(x)&=&c_{13}\, c_{23}\, z_{3}^{m+\frac{1}{2}}\,
{\mathbb A}_{3}(xq^{-m-\frac{1}{2}})
\overline{{\mathbb A}}_{3}(xq^{m+\frac{1}{2}}). \label{dim4-c}
\end{eqnarray}
For any $m_{1},m_{2},m_{3} \in {\mathbb C}$, 
define the following functions
\begin{eqnarray}
&& \overline{{\mathcal T}}_{(m_{1},m_{2},m_{3})}(x)=
(-1)^{m_{3}+1}\frac{c_{31}c_{23}}{c_{12}}
z_{3}^{\frac{m_{1}+m_{3}}{2}} \nonumber \\[.2cm] 
&& \hspace{30pt} \times \Bigl(
 z_{1}^{\frac{m_{1}+m_{3}}{2}+1}z_{2}^{\frac{-m_{1}+2m_{2}+m_{3}}{2}}
\overline{{\mathbb A}}_{1}(xq^{-m_{1}-m_{3}-\frac{3}{2}})
 \overline{{\mathbb A}}_{2}(xq^{m_{1}-2m_{2}-m_{3}+\frac{1}{2}}) 
 \label{app-wrons2} \\[.2cm] 
&& \hspace{30pt}
 -z_{1}^{\frac{-m_{1}+m_{3}+2m_{2}}{2}}z_{2}^{\frac{m_{1}+m_{3}}{2}+1}
\overline{{\mathbb A}}_{1}(xq^{m_{1}-2m_{2}-m_{3}+\frac{1}{2}})
\overline{{\mathbb A}}_{2}(xq^{-m_{1}-m_{3}-\frac{3}{2}})
\Bigr) \overline{{\mathbb A}}_{3}(xq^{m_{1}+m_{3}+\frac{1}{2}}),
 \nonumber \\[.3cm] 
&& {\mathcal T}_{(m_{1},m_{2},m_{3})}(x)=
(-1)^{m_{3}+1}\frac{c_{31}c_{23}}{c_{12}}
z_{3}^{\frac{m_{1}+m_{3}}{2}} \nonumber \\[.2cm] 
&& \hspace{30pt} \times \Bigl(
z_{1}^{\frac{m_{1}+m_{3}}{2}+1}
z_{2}^{\frac{-m_{1}+2m_{2}+m_{3}}{2}}
{\mathbb A}_{1}(xq^{m_{1}+m_{3}+\frac{3}{2}})
{\mathbb A}_{2}(xq^{-m_{1}+2m_{2}+m_{3}-\frac{1}{2}}) 
\label{app-wrons3} \\[.2cm] 
&& \hspace{30pt} 
-z_{1}^{\frac{-m_{1}+2m_{2}+m_{3}}{2}}z_{2}^{\frac{m_{1}+m_{3}}{2}+1}
{\mathbb A}_{1}(xq^{-m_{1}+2m_{2}+m_{3}-\frac{1}{2}}) 
{\mathbb A}_{2}(xq^{m_{1}+m_{3}+\frac{3}{2}}) 
\Bigr) {\mathbb A}_{3}(xq^{-m_{1}-m_{3}-\frac{1}{2}}), \nonumber
\end{eqnarray}
where $c_{ij}=(z_{i}-z_{j})/(z_{i}z_{j})^{\frac{1}{2}}$.
They obey  the following symmetry relations 
\begin{eqnarray}
\overline{{\mathcal T}}_{(m_{1},m_{2},m_{3})}(x)&=&(-1)^{m_{3}}
\overline{{\mathcal T}}_{(m_{1}+m_{3},m_{2}+m_{3},0)}(x), 
\label{app-dual1} \\[.3cm] 
{\mathcal T}_{(m_{1},m_{2},m_{3})}(x)&
=&(-1)^{m_{3}}{\mathcal T}_{(m_{1}+m_{3},m_{2}+m_{3},0)}(x).
\label{app-dual2}
\end{eqnarray}
From (\ref{sl21-rel1-app})--(\ref{sl21-rel6-app}) it follows, that
in a particular case, when $m_{1}=m_{2}=m \in {\mathbb C}$
the expressions (\ref{app-wrons2}) and (\ref{app-wrons3}) reduce to 
\begin{eqnarray}
\hspace{-15pt}
\overline{{\mathcal T}}_{(m,m,m_{3})}(x)={\mathcal T}_{(m,m,m_{3})}(x)
 =(-1)^{m_{3}+1}c_{23}c_{31}z_{3}^{m+m_{3}+\frac{1}{2}}
{\mathbb A}_{3}(xq^{-m-m_{3}-\frac{1}{2}})
\overline{{\mathbb A}}_{3}(xq^{m+m_{3}+\frac{1}{2}}). \label{app-wrons4}
\end{eqnarray}
Comparing this with \eqref{dim4} one concludes, 
\be
{\mathbb T}_{m}^{(2)}(x)={\mathcal T}_{(m,m,0)}(x)\ .\label{tdva}
\ee
Similarly, 
\begin{eqnarray}
{\mathbb T}^{(1)}_{m}(xq^{-1})-{\mathbb T}^{(1)}_{m+1}(x)&=&
\overline{{\mathcal T}}_{(m,0,0)}(x)
\label{app-aty1}\\[.3cm]
{\mathbb T}^{(1)}_{m}(xq)
-{\mathbb T}^{(1)}_{m+1}(x)&=&
{\mathcal T}_{(m,0,0)}(x)\ .
\label{app-aty2}
\end{eqnarray}

For the general case we have the following
\begin{conj} \label{JT-SP-conj}
{\it For \ $m_{1},m_{2},m_{3},m_{1}-m_{2} \in {\mathbb Z}_{\ge 0}$\
  and \  
$(m_{2}+m_{3})\ne 0 $, \ the expressions (\ref{app-wrons2}) and
  (\ref{app-wrons3}) can be written as tableau sums 
\begin{eqnarray}
\overline{\mathcal T}_{(m_{1},m_{2},m_{3})}(x)&
=&{\mathcal F}_{((m_{1}-1)^{m_{3}},m_{1}-m_{2}) 
\subset (m_{1}^{m_{3}+2})}(x), \label{app-tf1} \\[.3cm] 
{\mathcal T}_{(m_{1},m_{2},m_{3})}(x)
&=&{\mathcal F}_{(m_{1},m_{2},1^{m_{3}})}(x), \label{app-tf2}\\[.3cm] 
\nonumber 
\end{eqnarray}
and, similarly, 
\begin{eqnarray}
{\mathbb T}^{(1)}_{m}(x)&=&{\mathcal F}_{(m)}(x),\qquad m\in {\mathbb
  Z}_{\ge0}.  
\label{t1f}
\end{eqnarray}
}
\end{conj}
We conviced ourselves in the validity of the these relations by
numerous checks for particular values of $m,m_1,m_2,m_3$. Apparently 
there exist an elementary general proof, which we postpone for the
future work. 
For example for $(m_{1},m_{2},m_{3})=(2,1,1)$, one has
\begin{eqnarray}
\overline{\mathcal T}_{(2,1,1)}(x)&=&{\mathcal F}_{(1^{1},1) 
\subset (2^{3})}(x)
={\mathcal F}_{(1^2) \subset (2^{3})}(x), \\[.3cm] 
{\mathcal T}_{(2,1,1)}(x)
&=&{\mathcal F}_{(2,1,1^{1})}(x)={\mathcal F}_{(2,1^{2})}(x).
\end{eqnarray}
From \eqref{tdva} and \eqref{app-tf2} it follows that 
\be
{\mathbb T}^{(2)}_m(x)={\mathcal F}_{(m,m,0)}(x),
\qquad m \in {\mathbb Z}_{\ge0} \ . \label{t2f}
\ee

The relations 
and \eqref{t1f} and \eqref{t2f} provide one with the  
tableau sum (and, thus, the determinant expressions \eqref{superJT1},
\eqref{superJT2}) for the ${\bf T}$-operators \eqref{atyp-c} and
\eqref{dim4-c}. 
Remind that the later were derived form the {\em ab initio} definition
of Sect.~\ref{tmatsec}. 
This is the main result in this section. 

The $gl(m|n)$ analog  of the first Weyl formula is called 
the Sergeev-Pragacz formula  \cite{Pragacz91,vhkt90}. 
It gives an alternative representation of the supersymmetric Shur function, in
addition to the tableau sum formula of the type \eqref{shur1}.  
Let $\mu$ be a (non-skew)
partition, then
\footnote{The sign of the  variables $y_{1},\dots,y_{n}$ 
in \cite{Pragacz91,vhkt90} is opposite to our definition.}
\begin{eqnarray}
S_{\mu}(x_1,\ldots,x_m|y_1,\ldots,y_n)
=\frac{\ds \sum_{\sigma \in S_{m}\times S_{n}}
{\rm sgn}(\sigma) \ \sigma    
\left[\ds
\prod_{i=1}^{m-1}x_{i}^{m-i}\prod_{j=1}^{n-1}y_{j}^{n-j}
\prod_{(i,j)\in \mu}(x_{i}-y_{j}) 
\right]
}{\ds \prod_{i<j}(x_{i}-x_{j})\prod_{i<j}(y_{i}-y_{j})},
\label{Sergeev-Pragacz}
\end{eqnarray}
where the third product in the numerator is taken over all boxes 
${(i,j)\in \mu}$ of  
the Young diagram $\mu $.
It is assumed that $x_{i}\equiv0$ 
if $i \ge m+1$ and $y_{j}\equiv0$ if $j \ge n+1 $. 
The symbol $S_{m}$ (resp. $S_{n}$) denotes the symmetric group of order $m$
(resp. $n$). The notation $\sigma[\ldots]$ stands for the action of the 
permutation $\sigma$ on the variables $x_1,\ldots,x_m,y_1,\ldots,y_n$
inside the square brackets. 

When the spectral parameter vanishes, $x=0$, the Wronskian type 
relations \eqref{atyp-c} and \eqref{dim4-c}, connecting 
${\bf T}$- and ${\bf Q}$-operators,  
reduce to the $gl(2|1)$ Sergeev-Pragacz formula 
with $x_{1}=z_{1}$, $x_{2}=z_{2}$ and $y_{1}=z_{3}$. 
Note that a similar statement \cite{workinprogress} 
holds also for any quantum affine superalgebra $U_{q}(\hat{sl}(m|n))$ with  
arbitrary $m$ and $n$.  

\newpage
\app{Expansion of the universal R-matrix}\label{app-R-expan}
\addcontentsline{toc}{subsection}{D. Expansion of the universal R-matrix}
The third order in $e_j$ term in the expansion \eqref{R-expan}
of the reduced universal $R$-matrix read 
\begin{eqnarray}
\begin{array}{lll}
O(e_j^3)\mbox{ term in \eqref{R-expan}}=\\[.3cm]
\phantom{=}
\ds  -\frac{q^{-3}(q-q^{-1})^{3}}{[2][3]}e_{1}^{3}\otimes_{s}f_{1}^{3}
+q(q-q^{-1})^{2}
 (e_{0}e_{1}e_{0} 
\otimes_{s} f_{0}f_{1}f_{0}+e_{2}e_{1}e_{2}\otimes_{s}f_{2}f_{1}f_{2}
 )
  \\[.3cm] 
\phantom{=}\ds
 +q^{-1}(q-q^{-1})^{2}
 (e_{0}e_{2}e_{0} 
\otimes_{s} f_{0}f_{2}f_{0}+e_{2}e_{0}e_{2}\otimes_{s}f_{2}f_{0}f_{2}
 ) 
 \\[.4cm]
\phantom{=}\ds 
 +\frac{(q-q^{-1})^{2}}{q[2]}
\sum_{j\in
  \{0,2\}}(qe_{j}e_{1}^{2}
\otimes_{s}f_{j}f_{1}^{2}+qe_{1}^{2}e_{j}\otimes_{s}f_{1}^{2}f_{j}  
-[2]e_{1}e_{j}e_{1}\otimes_{s}f_{1}f_{j}f_{1})  \\[.3cm]
\phantom{=}\ds
 +(q-q^{-1})
\Bigl\{
e_{0}e_{1}e_{2}\otimes_{s}
\big(q[2]f_{0}f_{1}f_{2}-qf_{0}f_{2}f_{1}-qf_{1}f_{0}f_{2}+
f_{1}f_{2}f_{0}+f_{2}f_{0}f_{1}-[2]f_{2}f_{1}f_{0}\big) \\[.3cm]
\phantom{========}\ds
 +e_{2}e_{1}e_{0}\otimes_{s}
\big(q[2]f_{2}f_{1}f_{0}-qf_{2}f_{0}f_{1}-qf_{1}f_{2}f_{0}+
f_{1}f_{0}f_{2}+f_{0}f_{2}f_{1}-[2]f_{0}f_{1}f_{2}\big)  \\[.3cm]
\phantom{========}\ds
 +e_{0}e_{2}e_{1}\otimes_{s}
\big(-qf_{0}f_{1}f_{2}+f_{1}f_{0}f_{2}-qf_{2}f_{0}f_{1}+f_{2}f_{1}f_{0}\big)
 \\[.3cm] 
\phantom{========}\ds
 +e_{2}e_{0}e_{1}\otimes_{s}
\big(-qf_{2}f_{1}f_{0}+f_{1}f_{2}f_{0}-qf_{0}f_{2}f_{1}+f_{0}f_{1}f_{2}\big)
 \\[.3cm] 
\phantom{========}\ds
 +e_{1}e_{0}e_{2}\otimes_{s}
(-qf_{0}f_{1}f_{2}+f_{0}f_{2}f_{1}-qf_{1}f_{2}f_{0}+f_{2}f_{1}f_{0})
 \\[.3cm] 
\phantom{========}\ds
 +e_{1}e_{2}e_{0}\otimes_{s}
(-qf_{2}f_{1}f_{0}+f_{2}f_{0}f_{1}-qf_{1}f_{0}f_{2}+f_{0}f_{1}f_{2}) 
\Bigr\}.
\end{array}\label{3term}
\end{eqnarray}
On the other hand the expansion of the CFT ${\bf L}$-operator
\eqref{zz-def} reads 
\begin{eqnarray}
\overline{\mathcal L}&=&{\mathcal P}\exp \left( \int {\cal Z}(u) du
\right)
\\[.2cm]
&=&1+\int {\cal Z}(u) du+ 
\int_{u_{1}\ge u_{2}} {\cal Z}(u_{1}){\cal Z}(u_{2}) du_{1}du_{2}+\cdots, 
\\[.2cm]
{\cal Z}(u)&=&e_{0}\otimes_{s} V_{0}(u)+e_{1}\otimes_{s}
V_{1}(u)+e_{2}\otimes_{s} V_{2}(u)\label{L-expan}
\end{eqnarray}
where ${\cal Z}(u)$ is defined in \eqref{zz-def}. Introduce the
ordered integrals 
\begin{eqnarray}
J(i_{1},i_{2},\cdots,i_{n})=
\int_{u_{1}\ge u_{2} \ge \cdots \ge u_{n}}
 V_{i_{1}}(u_{1}) V_{i_{2}}(u_{2}) \cdots V_{i_{n}}(u_{n})
du_{1}du_{2} \cdots du_{n}.
\end{eqnarray}
Then the products of the vertex operators \eqref{ff-def} can be written as 
\begin{eqnarray}
&& f_{1}=-\frac{1}{q-q^{-1}}J(1), \\
&& f_{1}^{2}=\frac{q[2]}{(q-q^{-1})^{2}}J(1,1), \\
&& f_{1}^{3}=-\frac{q^{3}[2][3]}{(q-q^{-1})^{3}}J(1,1,1), 
\end{eqnarray}
\begin{eqnarray}
&& f_{0}f_{1}f_{0}=-\frac{q^{-1}}{(q-q^{-1})^{2}}J(0,1,0), \\
&& f_{2}f_{1}f_{2}=-\frac{q^{-1}}{(q-q^{-1})^{2}}J(2,1,2), \\
&& f_{0}f_{2}f_{0}=-\frac{q}{(q-q^{-1})^{2}}J(0,2,0), \\
&& f_{2}f_{0}f_{2}=-\frac{q}{(q-q^{-1})^{2}}J(2,0,2)
\end{eqnarray}
\begin{eqnarray}
&& \left(
\begin{array}{c}
f_{0}f_{1}^{2} \\
f_{1}f_{0}f_{1} \\
f_{1}^{2}f_{0}
\end{array}
\right)
=
\frac{[2]}{(q-q^{-1})^{3}}
\left(
\begin{array}{ccc}
q & 1 & q^{-1} \\
1 & \frac{2}{[2]} & 1 \\
q^{-1} & 1 & q
\end{array}
\right)
\left(
\begin{array}{c}
J(0,1,1) \\
J(1,0,1) \\
J(1,1,0)
\end{array}
\right), 
\\
&& \left(
\begin{array}{c}
f_{2}f_{1}^{2} \\
f_{1}f_{2}f_{1} \\
f_{1}^{2}f_{2}
\end{array}
\right)
=
\frac{[2]}{(q-q^{-1})^{3}}
\left(
\begin{array}{ccc}
q & 1 & q^{-1} \\
1 & \frac{2}{[2]} & 1 \\
q^{-1} & 1 & q
\end{array}
\right)
\left(
\begin{array}{c}
J(2,1,1) \\
J(1,2,1) \\
J(1,1,2)
\end{array}
\right), 
\end{eqnarray}
Determinant of the above matrix is 0. The Serre relations 
\eqref{aser1}  for $\{f_{j}\}$ follows immediately. Then one can derive
\begin{eqnarray}
&& qf_{0}f_{1}^{2}-f_{1}f_{0}f_{1}
= \frac{q[2]}{(q-q^{-1})^{2}}
\left(J(0,1,1)+\frac{J(1,0,1)}{[2]}\right), \\
&& qf_{1}^{2}f_{0}-f_{1}f_{0}f_{1}
= \frac{q[2]}{(q-q^{-1})^{2}}
\left(J(1,1,0)+\frac{J(1,0,1)}{[2]}\right), \\
&& qf_{2}f_{1}^{2}-f_{1}f_{2}f_{1}
= \frac{q[2]}{(q-q^{-1})^{2}}
\left(J(2,1,1)+\frac{J(1,2,1)}{[2]}\right), \\
&& qf_{1}^{2}f_{2}-f_{1}f_{2}f_{1}
= \frac{q[2]}{(q-q^{-1})^{2}}
\left(J(1,1,2)+\frac{J(1,2,1)}{[2]}\right).
\end{eqnarray}
Thanks to the Serre relations \eqref{aser1} for $\{e_j\}$ 
one can derive 
\begin{eqnarray}
&&e_{j}e_{1}^{2}\otimes_{s}J(j,1,1)
+e_{1}e_{j}e_{1}\otimes_{s}J(1,j,1)+e_{1}^{2}e_{j}\otimes_{s}J(1,1,j) 
\nonumber \\
&& = e_{j}e_{1}^{2}\otimes_{s}\left(J(j,1,1)+\frac{J(1,j,1)}{[2]}\right)+
e_{1}^{2}e_{j}\otimes_{s}\left(J(1,1,j)+\frac{J(1,j,1)}{[2]}\right)
\nonumber \\ 
&& = \frac{(q-q^{-1})^{2}}{q[2]}
\left\{ e_{j}e_{1}^{2}\otimes_{s}(qf_{j}f_{1}^{2}-f_{1}f_{j}f_{1}) +
e_{1}^{2}e_{j}\otimes_{s} (qf_{1}^{2}f_{j}-f_{1}f_{j}f_{1}) \right\}
\nonumber \\ 
&&= 
 \frac{(q-q^{-1})^{2}}{q[2]}
(qe_{j}e_{1}^{2}\otimes_{s}f_{j}f_{1}^{2}
+qe_{1}^{2}e_{j}\otimes_{s}f_{1}^{2}f_{j} 
-[2]e_{1}e_{j}e_{1}\otimes_{s}f_{1}f_{j}f_{1}),
\end{eqnarray}
where $j \in \{0,2 \}$.
We also have
\begin{eqnarray}
&& \hspace{-40pt} \left(
\begin{array}{c}
f_{0}f_{1}f_{2} \\
f_{0}f_{2}f_{1} \\
f_{1}f_{0}f_{2} \\
f_{1}f_{2}f_{0} \\
f_{2}f_{0}f_{1} \\
f_{2}f_{1}f_{0}
\end{array}
\right)
=
\frac{1}{q^{2}(q-q^{-1})^{3}}
\left(
\begin{array}{cccccc}
-q^{2} & -q     & -q     & q^{2} & q^{2}  & q \\
-q     & -q^{2} & -1     & q     & q^{3}  & q^{2} \\
-q     & -1     & -q^{2} & q^{3} & q      & q^{2} \\
q^{2}  & q      & q^{3}  & -q^{2}&-1      & -q \\
q^{2}  & q^{3}  & q      & -1    & -q^{2} & -q \\
q      & q^{2}  & q^{2} & -q     & -q     & -q^{2}
\end{array}
\right)
\left(
\begin{array}{c}
J(0,1,2) \\
J(0,2,1) \\
J(1,0,2) \\
J(1,2,0) \\
J(2,0,1) \\
J(2,1,0) 
\end{array}
\right).
\end{eqnarray}
This relation is invertible. Now one can show that the third order
term in \eqref{L-expan}
\begin{eqnarray}
 && \int_{u_{1}\ge u_{2} \ge u_{3}}{\cal Z}(u_{1})
{\cal Z}(u_{2}){\cal Z}(u_{3})du_{1}du_{2}du_{3} 
=\sum_{j_{1}=0}^{2}\sum_{j_{2}=0}^{2}\sum_{j_{3}=0}^{2}
(-1)^{p(j_{1})p(j_{2})+p(j_{1})p(j_{3})+p(j_{2})p(j_{3})}\nonumber \\
&& \hspace{90pt}
 \times e_{j_{1}}e_{j_{2}}e_{j_{3}} \otimes_{s} 
\int_{u_{1}\ge u_{2} \ge u_{3}}
 V_{j_{1}}(u_{1})V_{j_{2}}(u_{2})V_{j_{3}}(u_{3})du_{1}du_{2}du_{3}, 
\end{eqnarray}
exactly coincide with \eqref{3term}.

\addcontentsline{toc}{section}{References}

\end{document}